\documentclass[aps,twocolumn,showpacs,amsmath,amssymb,prc]{revtex4-2}
\usepackage{graphicx,here}
\usepackage{multirow}
\usepackage{tikz}
\usepackage{epstopdf}
\usepackage[percent]{overpic}
\usepackage{scrextend}
\usepackage{xcolor,xspace}
\usepackage[ulem=normalem]{changes}
\usepackage{hyperref}
\hypersetup{colorlinks=True,urlcolor=blue,linkcolor=blue,citecolor=blue,filecolor=black}

\colorlet{Changes@Color}{red}
\usepackage{dcolumn,color,footnote,bm,braket}
\usepackage{url,longtable,tabularx}
\usepackage{threeparttable}

\usepackage{fancybox}
\usetikzlibrary{matrix}

\newcommand\+{\dagger}
\newcommand\tbeta{\tilde{\beta}}

\begin{document}

\title{Octupole correlations in collective excitations of neutron-rich $N\approx56$ nuclei}

\author{Kosuke Nomura}
\email{knomura@phy.hr}
\affiliation{Department of Physics, Faculty of Science, 
University of Zagreb, HR-10000 Zagreb, Croatia}

\date{\today}

\begin{abstract}
Octupole correlations in the low-energy collective states  
of neutron-rich nuclei with the neutron number $N\approx56$ are 
studied within the interacting boson model (IBM) that is 
based on the nuclear density functional theory. 
The constrained self-consistent mean-field (SCMF) 
calculations using a universal energy 
density functional and a pairing interaction provide the 
potential energy surfaces in terms of the axially-symmetric 
quadrupole and octupole deformations for the even-even nuclei 
$^{86-94}$Se, $^{88-96}$Kr, $^{90-98}$Sr, $^{92-100}$Zr, 
and $^{94-102}$Mo. 
The SCMF energy surface is then mapped onto the energy 
expectation value of a version of the IBM in the boson 
condensate state, which consists of the neutron 
and proton monopole $s$, quadrupole $d$, and octupole $f$ 
bosons. 
This procedure determines the strength parameters of the 
IBM Hamiltonian, which is used to 
compute relevant spectroscopic properties. 
At the SCMF level, no octupole deformed ground state is obtained, 
while the energy surface is generally soft in the octupole deformation 
at $N\approx56$. 
The predicted negative-parity yrast bands with the bandhead state 
$3^-_1$ weakly depend on $N$ and become lowest in energy at $N=56$ 
in each of the considered isotopic chains. 
The model further predicts finite electric octupole transition 
rates between the lowest negative- and the positive-parity 
ground-state bands. 
\end{abstract}

\maketitle

\section{Introduction}

The ground-state shape and the related spectroscopic properties 
of those neutron-rich nuclei in the mass $A\approx100$ region 
have been of considerable interest. 
The low-energy structure of these nuclei 
depends on the underlying shell structures and nuclear forces, 
and is characterized by a subtle interplay between the single-particle 
and collective degrees of freedom. 
A number of nuclei belonging to this mass region have been suggested 
to demonstrate intriguing nuclear shape-related features 
that include the triaxial deformation, the 
proton $Z=40$ and neutron $N=56$ subshell gaps, the coexistence 
of several different shapes in the vicinity of the ground state 
\cite{heyde2011}, the quantum phase 
transitions in their shapes \cite{cejnar2010} 
represented by a sudden onset of 
deformation near the neutron number $N\approx60$. 

Of particular interest is the octupole correlations 
and the related low-energy negative-parity states. 
In a simple spherical shell model, 
the octupole correlations become enhanced in particular mass 
regions in the nuclear mass table, in which the coupling 
occurs between the normal and unique parity single-particle orbitals 
in a given major oscillator shell that satisfy the conditions 
$\Delta\ell=\Delta j=3\hbar$, 
with $\ell$ and $j$ the quantum numbers for a single-particle state. 
The so-called ``octupole magic numbers'' at which the above 
conditions are met and stable octupole deformations are supposed 
to emerge are the neutron and/or proton numbers 
34, 56, 88, 134, $\ldots$ \cite{butler1996,butler2016}. 
Experimental evidence for the permanent octupole deformations were 
found at CERN in the light actinides with $(N,Z)\approx(134,88)$ 
such as $^{220}$Rn and $^{224}$Ra 
\cite{gaffney2013}, and $^{228}$Th \cite{chishti2020} 
and at ANL in the lanthanides with $(N,Z)\approx(88,56)$ 
such as $^{144}$Ba \cite{bucher2016} and $^{146}$Ba \cite{bucher2017}. 
On the other hand, the possible octupolarity 
in the lighter-mass regions, such as those with $(N,Z)\approx(56,34)$, 
$(56,56)$ and $(34,34)$, has not been as extensively investigated, 
both experimentally and theoretically, as in the 
cases of the actinides and lanthanides. 

In some of the neutron-rich even-even 
Mo, Zr, Sr, and Kr isotopes that are in the vicinity of the 
neutron number $N\approx56$, low-lying negative-parity 
states $3^-$ at the excitation energy $E_x\approx2$ MeV, 
as well as the bands built on them, have been suggested experimentally 
(see, e.g., Refs.~\cite{rzacaurban2000,lalkovski2007,scheck2010,li2011,gregor2017,dudouet2017,gregor2019,gerst2022}). 
In this case, the negative-parity states 
are supposed to appear as a consequence of the coupling 
between the neutron $h_{11/2}$ and $d_{5/2}$ orbitals. 
In the neutron-rich nuclei $^{94,96}$Kr, in particular, the $3^-$ 
states have been newly obtained in an experiment at RIKEN 
\cite{gerst2022}. With the recurrent interests in the studies 
of the octupole shapes and the new data 
on the negative-parity states in radioactive nuclei, 
it would be meaningful to 
pursue timely theoretical investigations to address 
the relevance of the octupole degrees of freedom in the description 
of low-lying nuclear structure in the neutron-rich 
$A\approx100$ region. 

Theoretical approaches to the octupole deformations and 
collective excitations include the self-consistent mean-field 
(SCMF) framework based on the macroscopic-microscopic method 
with Strutinski shell correction \cite{naza1984b,leander1985,moeller2008} 
and the energy density functionals (EDFs) 
\cite{bonche1986,heenen1994,robledo1987,egido1992,robledo2010,robledo2011,
li2013,robledo2013,yao2015,bernard2016,xia2017,agbemava2017,ebata2017,
marevic2018,robledo2019,cao2020,rayner2020oct,nomura2021qoch}, 
the interacting boson model (IBM) 
\cite{engel1985,engel1987,otsuka1986,otsuka1988,sugita1996,kusnezov1988,
yoshinaga1993,zamfir2001,SMIRNOVA2000,pietralla2003,nomura2013oct,
nomura2014,nomura2015,nomura2020oct,nomura2021oct-u,nomura2021oct-ba,
nomura2021oct-zn,hennig2014,vallejos2021}, 
the geometrical collective model 
\cite{bonatsos2005,lenis2006,bizzeti2013,bonatsos2015}, 
the cluster models \cite{shneidman2002,shneidman2003,jolos2012}, 
and the nuclear shell model 
\cite{brown2000,kaneko2002,yoshinaga2018,vanisacker2020}. 
In particular, the SCMF methods using a given universal nonrelativistic 
\cite{bender2003,robledo2019}
or relativistic \cite{vretenar2005,niksic2011} 
EDF provides a global description of the bulk nuclear 
matter and intrinsic properties, 
as well as collective excitations, 
over the entire region of the nuclear mass table. 
In recent years, the EDF-based approaches have been extensively 
used for the studies of the octupole deformations and collectivity. 
A straightforward approach is 
the beyond SCMF calculations with symmetry projections and 
configuration mixing within the generator coordinate method (GCM) 
\cite{RS,bonche1986,heenen1994,robledo1987,egido1992,robledo2010,robledo2011,robledo2013,yao2015,bernard2016,marevic2018,robledo2019,rayner2020oct}. 
The full GCM calculations are, however, computationally so demanding 
that some alternative approaches have also been considered, 
such as the mapping onto the quadrupole-octupole collective 
Hamiltonian \cite{li2013,xia2017,nomura2021qoch} 
and onto the interacting-boson Hamiltonian 
\cite{nomura2013oct,nomura2014,nomura2015}.

In this article, the octupole correlations 
and the relevant spectroscopic properties of 
the low-lying positive- and negative-parity states in the neutron-rich 
nuclei near the octupole magic number $N=56$ are investigated 
within the framework of the nuclear EDF and the mapped IBM. 
The starting point is the axially-symmetric 
quadrupole and octupole constrained SCMF calculations of the 
potential energy surfaces for those nuclei in which octupolarity 
is expected to be enhanced, i.e., 
$^{86-94}$Se, $^{88-96}$Kr, $^{90-98}$Sr, $^{92-100}$Zr, 
and $^{94-102}$Mo. 
Then, by using the method of Ref.~\cite{nomura2014}, the quadrupole-octupole 
SCMF energy surface is mapped onto the equivalent one 
in the system of the interacting monopole $s$, quadrupole $d$, 
and octupole $f$ bosons. 
The mapping procedure specifies the strength parameters for the $sdf$-IBM 
Hamiltonian, which in turn gives excitation spectra and 
electromagnetic transition rates of both the positive- and negative-parity 
states. Previously, the mapped $sdf$-IBM framework has been 
applied to the studies of octupole shape phase transitions in 
light actinides and rare-earth nuclei using the relativistic 
EDF \cite{nomura2013oct,nomura2014}. A number of spectroscopic 
calculations have been carried out based on the 
Gogny-type EDF \cite{robledo2019}, and revealed the onset of 
octupole deformations and collectivity in various mass regions 
characteristic of the octupole deformations 
\cite{nomura2015,nomura2020oct,nomura2021oct-u,nomura2021oct-ba,nomura2021oct-zn}.

In the present study, a version of the $sdf$-IBM that makes 
distinction between the neutron and proton boson degrees of freedom, 
denoted hereafter by $sdf$-IBM-2, is considered 
since it is more realistic than a simpler version of the 
IBM ($sdf$-, or $spdf$-IBM-1) that does not distinguish 
between the protons and neutrons. 
In most of the previous IBM calculations for the octupole collective 
states, the simple IBM-1 framework has been employed. 
There are a few instances in which the $sdf$-IBM-2 
has been considered on phenomenological grounds, e.g., 
in Refs.~\cite{yoshinaga1993,SMIRNOVA2000,pietralla2003,vallejos2021}. 
In this study, all the strength parameters for the $sdf$-IBM-2 are 
determined by using the microscopic input from 
the constrained SCMF calculations. 
As a basis of the SCMF method, the 
relativistic density-dependent point-coupling 
(DD-PC1) EDF \cite{DDPC1} is employed. 
The universal functional DD-PC1 has been 
successfully employed in the spectroscopic studies 
on the octupole collective modes both at the static and beyond 
SCMF levels \cite{agbemava2017,marevic2018,nomura2021qoch}.

The paper is structured as follows. 
In Sec.~\ref{sec:theory} the theoretical procedure in the mapped 
$sdf$-IBM-2 calculations is illustrated. 
Results of the quadrupole-octupole constrained potential 
energy surfaces, the systematics of the low-energy spectra for the 
positive- and negative-parity states, electric quadrupole, octupole 
and dipole transition rates that characterize the octupole correlations 
in the considered Se, Kr, Sr, Zr, and Mo 
nuclei are discussed in Sec.~\ref{sec:results}. 
In the same section detailed level structures for the octupole 
magic nuclei, i.e., the $N=56$ isotones, are analyzed, 
with a special attention to the neutron-rich Kr isotopes where 
some new experimental data are available. 
Finally, Sec.~\ref{sec:summary} gives a summary of the main results 
and conclusions.

\begin{figure*}
\begin{center}
\includegraphics[width=\linewidth]{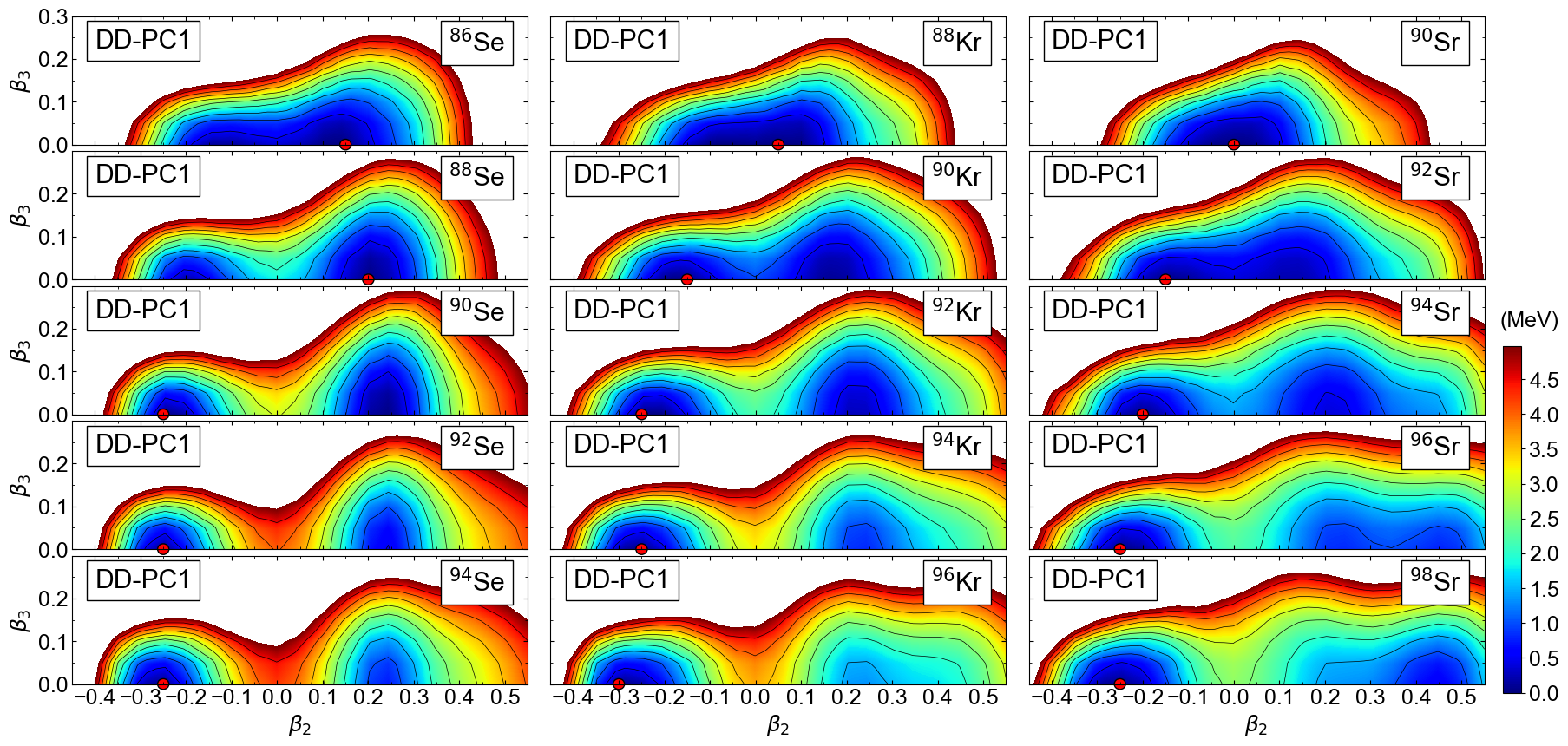}
\caption{Axially-symmetric quadrupole and octupole SCMF 
potential energy surfaces for the $^{86-94}$Se, $^{88-96}$Kr, 
and $^{90-98}$Sr isotopes 
as functions of the $\beta_2$ and $\beta_3$ deformations, 
computed by the constrained RHB method with the DD-PC1 EDF and 
the separable pairing force of finite range. The energy 
difference between neighboring contours is 0.5 MeV, and 
the global minimum within the $\beta_2-\beta_3$ plane is indicated 
by the solid circle. 
}
\label{fig:pesdft1}
\end{center}
\end{figure*}

\begin{figure*}
\begin{center}
\includegraphics[width=.7\linewidth]{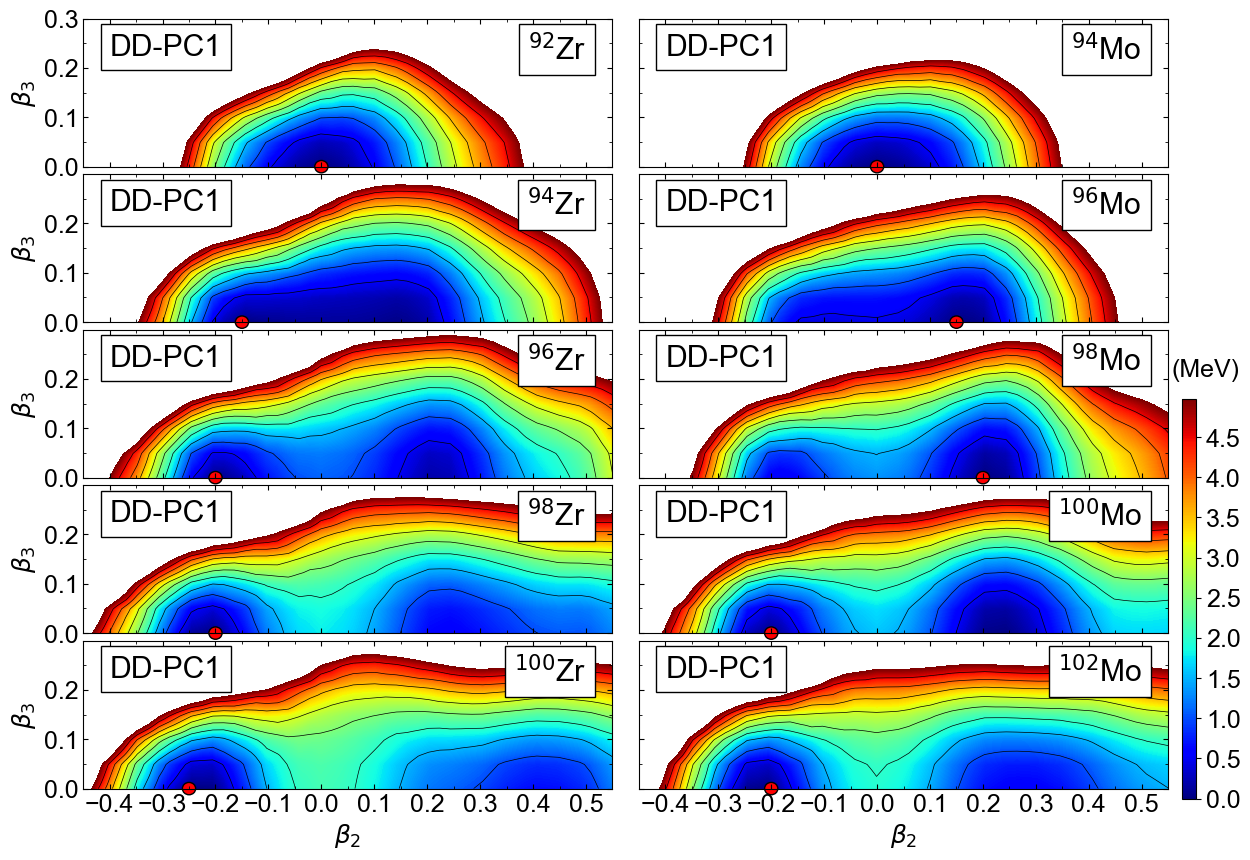}
\caption{Same as Fig.~\ref{fig:pesdft1}, but for the 
$^{92-100}$Zr, and $^{94-102}$Mo isotopes. }
\label{fig:pesdft2}
\end{center}
\end{figure*}

\section{Theoretical framework\label{sec:theory}}

\subsection{Self-consistent mean-field calculations\label{sec:scmf}}

The constrained SCMF calculations are performed within the 
relativistic Hartree-Bogoliubov (RHB) framework
\cite{vretenar2005,niksic2011,DIRHB} using the DD-PC1 
interaction for the particle-hole channel and a separable pairing 
force of finite range \cite{tian2009} for the particle-particle channel. 
The constraints imposed in the SCMF calculations are on the 
expectation values of the axially-symmetric quadrupole 
$\hat Q_{20}$ and octupole $\hat Q_{30}$ moments, 
\begin{align}
 &\hat Q_{20}=2z^2-x^2-y^2,\\
 &\hat Q_{30}=2z^3-3z(x^2+y^2),
\end{align}
which are associated with the dimensionless quadrupole $\beta_2$ 
and octupole $\beta_3$ deformations through the relations
\begin{align}
 &\beta_2=\frac{\sqrt{5\pi}}{3r_0^2A^{5/3}}\braket{\hat Q_{20}}\\
 &\beta_3=\frac{\sqrt{7\pi}}{3r_0^3A^{2}}\braket{\hat Q_{30}}
\end{align}
with $r_0=1.2$ fm. The RHB equation is solved in a harmonic oscillator 
basis with the number of oscillator shells $N_F=12$. 
Here the strength of the pairing force 
$V_0=837$ MeV\,fm$^3$, corresponding to an 
increase of the original value 728 MeV\,fm$^3$ 
of Ref.~\cite{tian2009} by 15 \%, is used for 
both the proton and neutron pairs. 
An effect of increasing the 
pairing strength in the SCMF calculations is such that 
the potential energy surface becomes less steep. 
The mapping procedure then provides the IBM 
parameters closer to those  
that would be obtained in the conventional IBM fit, 
and is expected to give a better description of the experimental 
excitation energies, especially, those for nonyrast states.  
As a further support for the enhancement of the pairing strength, 
a global analysis of the separable pairing force 
over the entire region of the mass chart 
based on the covariant density functional theory \cite{teeti2021} 
indicated that, in order to account for the empirical odd-even 
mass staggering, the strength of the pairing force 
needs to be scaled with nucleon number dependent 
factors. In the case of the $N\approx56$ nuclei considered 
here, such scaling factors 
are calculated to be 
approximately within the range 1.1$-$1.2. 
which is consistent with the 15 \% increase of the pairing 
strength in the present study. 
The constrained calculations provide the 
potential energy surfaces $E(\beta_2,\beta_3)$ 
as functions of the $\beta_2$ and $\beta_3$ deformations. 
The corresponding results are shown in Figs.~\ref{fig:pesdft1} 
and \ref{fig:pesdft2}, 
and will be discussed in detail in Sec.~\ref{sec:pes}. 
The $\beta_2-\beta_3$ energy surface has a property that 
it is symmetric with respect to the $\beta_3=0$ axis, 
i.e., $E(\beta_2,\beta_3)=E(\beta_2,-\beta_3)$, and it is therefore 
enough to consider the $\beta_3\geq0$ sector of the energy surface.

\subsection{Mapping onto the $sdf$-IBM-2 Hamiltonian}

The $sdf$-IBM-2 is comprised of the neutron and proton 
monopole $s_\nu$ and $s_\pi$, quadrupole $d_\nu$ and $d_\pi$, 
and octupole $f_\nu$ and $f_\pi$ bosons, which are, from a 
microscopic point of view \cite{OAIT,OAI}, associated with 
the correlated pairs of valence neutrons and protons 
with spin and parity $J=0^+$, $2^+$, and $3^-$, respectively. 
The calculations are carried out within the neutron $N=50-82$ 
and proton $Z=28-50$ major shells and, therefore, the doubly-magic 
nucleus $^{78}$Ni is taken as the inert core. 
The number of the neutron (or proton) bosons $n_{\nu}$ (or $n_{\pi}$) 
is conserved for each nucleus, and is equal to half the number 
of valence neutron (or proton) pairs. 
The neutron (or proton) $s_\nu$ ($s_\pi$), $d_\nu$ ($d_\pi$), 
and $f_\nu$ ($f_\pi$) bosons, denoted by $n_{s_{\nu}}$, 
$n_{d_{\nu}}$, and $n_{f_{\nu}}$ ($n_{s_{\pi}}$, 
$n_{d_{\pi}}$, and $n_{f_{\pi}}$), respectively, 
satisfy the condition that 
$n_{s_{\nu}}+n_{d_{\nu}}+n_{f_{\nu}}=n_\nu$ 
(or $n_{s_{\pi}}+n_{d_{\pi}}+n_{f_{\pi}}=n_\pi$). 

Based on the earlier microscopic considerations \cite{OAIT} 
that the low-lying 
structure of medium-heavy and heavy nuclei is mainly determined 
by the pairing-like correlations between identical nucleons 
and the quadrupole-quadrupole interactions between nonidentical 
nucleons, the following form of the $sdf$-IBM-2 Hamiltonian 
is adopted: 
\begin{align}
\label{eq:ham}
 \hat H = 
\epsilon_d \hat n_d + \epsilon_f \hat n_f
+ \kappa_2 \hat Q_{\nu} \cdot \hat Q_{\pi}
+ \kappa_3 \hat O_{\nu} \cdot \hat O_{\pi}.
\end{align}
The first (second) term stands for the $d$ ($f$) boson number 
operator $\hat n_d = \hat n_{d_\nu} + \hat n_{d_\pi}$ 
($\hat n_f = \hat n_{f_\nu} + \hat n_{f_\pi}$) with the 
single $d$ ($f$) boson energy $\epsilon_d$ ($\epsilon_f$) 
relative to the $s$-boson one. 
The single boson energies, $\epsilon_d$ and $\epsilon_f$, 
could in principle be different, 
but are here assumed to be the same for simplicity, between 
the neutron and proton bosons. 
The third term is the quadrupole-quadrupole interaction with 
the strength $\kappa_2$, with the quadrupole operator given by
\begin{align}
 \hat Q_\rho = s^\+_\rho\tilde d_\rho + d^\+_\rho s_\rho 
+ \chi_{\rho}(d^\+_{\rho}\times\tilde d_{\rho})^{(2)}
+ \chi_{\rho}'(f^\+_{\rho}\times\tilde f_{\rho})^{(2)}
\end{align}
for neutron $\rho=\nu$ and proton $\rho=\pi$. 
$\chi_\rho$ and $\chi'_\rho$ are dimensionless parameters. 
The last term in (\ref{eq:ham}) represents the octupole-octupole interaction 
between the neutron and proton bosons with the strength $\kappa_3$ 
and the octupole operator is given as 
\begin{align}
 \hat O_\rho = s^\+_\rho\tilde f_\rho + f^\+_\rho s_\rho 
+ \chi''_{\rho}(d^\+_{\rho}\times\tilde f_{\rho}
+f^\+_{\rho}\times\tilde d_{\rho})^{(3)},
\end{align}
with another dimensionless parameter $\chi''_\rho$. 

To make a connection between the geometrical structure of a 
multifermion systems specified by the $\beta_2$ and $\beta_3$ 
deformations and the boson Hamiltonian (\ref{eq:ham}), 
the following wave function for the condensate state \cite{ginocchio1980}
of the $s_\nu$, $s_\pi$, 
$d_\nu$, $d_\pi$, $f_\nu$, and $f_\pi$ bosons is considered:
\begin{align}
\label{eq:coherent}
 \ket{\Psi}
=\prod_{\rho=\nu,\pi}\frac{1}{\sqrt{n_\rho!}}
\left(\lambda_{\rho}^\+\right)^{n_{\rho}}
\ket{0}
\end{align}
with
\begin{align}
\label{eq:coherent2}
 \lambda_\rho^\+ = s^\+_{\rho} + \tbeta_{2\rho} d^\+_{\rho,0} 
+ \tbeta_{3\rho} f^\+_{\rho,0},
\end{align}
and  $\ket{0}$ the boson vacuum, i.e., the inert core.   
The amplitudes $\tbeta_{2\rho}$ and $\tbeta_{3\rho}$ in 
(\ref{eq:coherent2}) are 
boson analogs of the quadrupole $\beta_2$ and octupole $\beta_3$
deformations, respectively. The expectation value of 
the $sdf$-IBM-2 Hamiltonian gives energy surface 
of the boson system specified by the four variables 
$\tbeta_{2\nu}$, $\tbeta_{2\pi}$, $\tbeta_{3\nu}$, and $\tbeta_{3\pi}$. 
However, to treat the energy surface in the full four-dimensional space 
is practically so complicated that, in this study, it is assumed that 
the neutron and proton bosons should have the same deformations, 
i.e., 
$\tbeta_{2\nu}=\tbeta_{2\pi}\equiv\tbeta_2$ 
and $\tbeta_{3\nu}=\tbeta_{3\pi}\equiv\tbeta_3$. 
The equal deformations for the neutron and proton bosons 
may seem practically equivalent to the mapping onto the 
$sdf$-IBM-1 system. An advantage of using the $sdf$-IBM-2 
is that it produces more eigenstates than the $sdf$-IBM-1, 
for the latter represents the fully-symmetric states 
of the former. Therefore, the $sdf$-IBM-2 is expected to 
give a better description of the spectroscopic 
properties of the low-lying states, 
especially the $M1$ transitions, than in the $sdf$-IBM-1. 
Under the assumption of the equal proton and 
neutron deformations, 
the bosonic potential energy surface 
in the $\tbeta_2-\tbeta_3$ space is calculated as
\begin{widetext}
 \begin{align}
\label{eq:pesibm}
 \frac{\braket{\Psi|\hat H|\Psi}}{\braket{\Psi|\Psi}}
=&\frac{(n_\nu+n_\pi)(\epsilon_d\,\tbeta_{2}^2+\epsilon_f\,\tbeta_{3}^2)}
{1+\tbeta_{2}^2+\tbeta_{3}^2}
+\frac{n_\nu n_\pi}
{(1+\tbeta_{2}^2+\tbeta_{3}^2)^2}
\nonumber\\
&\times
\Biggl[
\kappa_2
\left(
2\tbeta_{2}-\sqrt{\frac{2}{7}}\chi_{\nu}\tbeta^2_{2}
-\frac{2}{\sqrt{21}}\chi_\nu'\tbeta_{3}^2
\right)
\left(
2\tbeta_{2}-\sqrt{\frac{2}{7}}\chi_{\pi}\tbeta^2_{2}
-\frac{2}{\sqrt{21}}\chi_\pi'\tbeta_{3}^2
\right)
\nonumber\\
&+\kappa_3
\left(
2\tbeta_{3}-\frac{4}{\sqrt{15}}\chi''_{\nu}\tbeta_{2}\tbeta_{3}
\right)
\left(
2\tbeta_{3}-\frac{4}{\sqrt{15}}\chi''_{\pi}\tbeta_{2}\tbeta_{3}
\right)
\Biggr].
\end{align}
\end{widetext}
The strength parameters of the boson Hamiltonian (\ref{eq:ham}) 
to be determined are 
$\{\epsilon_d,\epsilon_f,\kappa_2,\chi_\nu,\chi_\pi,\chi'_\nu,\chi'_\pi,\kappa_3,\chi''_\nu,\chi''_\pi\}$. 
To reduce the number of parameters, it is assumed that 
the dimensionless parameters 
$\chi'_\nu$, $\chi'_\pi$, $\chi''_\nu$, and $\chi''_\pi$ 
are all equal, $\chi'_\nu=\chi'_\pi=\chi''_\nu=\chi''_\pi\equiv\chi$.  
It is further assumed that the bosonic deformation $\tbeta_\lambda$ 
($\lambda=2$ or 3) is proportional to the fermionic counterparts, 
i.e., $\tbeta\equiv C_\lambda\beta_{\lambda}$. $C_\lambda$ 
is a constant of proportionality, and is also determined by the 
procedure described below.

The seven parameters for the $sdf$-IBM-2 Hamiltonian, i.e., 
$\{\epsilon_d,\epsilon_f,\kappa_2,\chi_\nu,\chi_\pi,\kappa_3,\chi\}$ 
and the two constants $C_2$ and $C_3$, are determined for each nucleus 
by mapping the axially-symmetric $\beta_2-\beta_3$ SCMF 
potential energy surface onto the corresponding bosonic 
energy surface given by Eq.~(\ref{eq:pesibm}). 
Or equivalently, these strength parameters are 
fine tuned so that the $sdf$-IBM-2 energy surface of (\ref{eq:pesibm}) 
becomes similar in topology to the SCMF one and that the approximate 
equality
\begin{align}
\label{eq:mapping}
 E_\text{SCMF}(\beta_2,\beta_3) \approx E_\text{IBM}(\beta_2,\beta_3)
\end{align}
should be satisfied in the neighborhood of the 
global minimum. Limiting the range within which 
the mapping of (\ref{eq:mapping}) is to be made 
to the vicinity of the 
global mean-field minimum is due to the fact that in the 
SCMF framework, the configurations 
near the global minimum make the most relevant contributions 
to determining the low-energy collective states. 
On the other hand, the regions very far from the minimum 
with large $\beta_2$ and $\beta_3$ deformations are more dominated 
by the quasiparticle degrees of freedom, which are, by construction, 
not taken into account in the present $sdf$-IBM-2 framework. 
For further details of the mapping procedure, the reader 
is referred to Refs.~\cite{nomura2008,nomura2010,nomura2013oct,nomura2014}. 

Diagonalization of the mapped $sdf$-IBM-2 Hamiltonian 
with the strength parameters determined by the aforementioned 
procedure produces excitation spectra for both parities and 
electromagnetic transition rates. For the numerical diagonalization 
of the Hamiltonian, the computer program ARBMODEL \cite{arbmodel} 
is used.

\begin{figure*}
\begin{center}
\includegraphics[width=\linewidth]{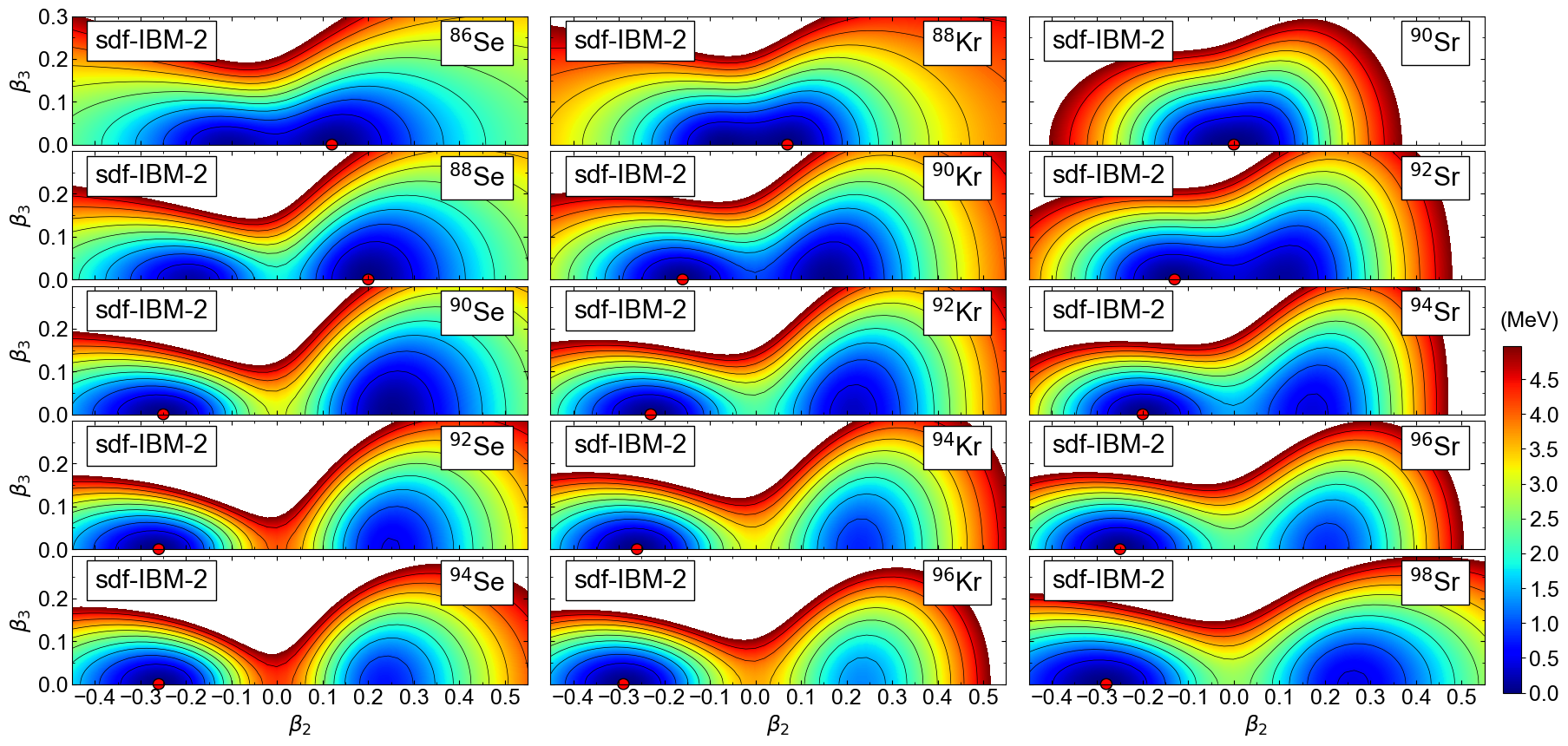}
\caption{
Same as Fig.~\ref{fig:pesdft1}, but for the mapped 
$sdf$-IBM-2 potential energy surfaces for the $^{86-94}$Se, $^{88-96}$Kr, 
and $^{90-98}$Sr isotopes. 
}
\label{fig:pesibm1}
\end{center}
\end{figure*}

\begin{figure*}
\begin{center}
\includegraphics[width=.7\linewidth]{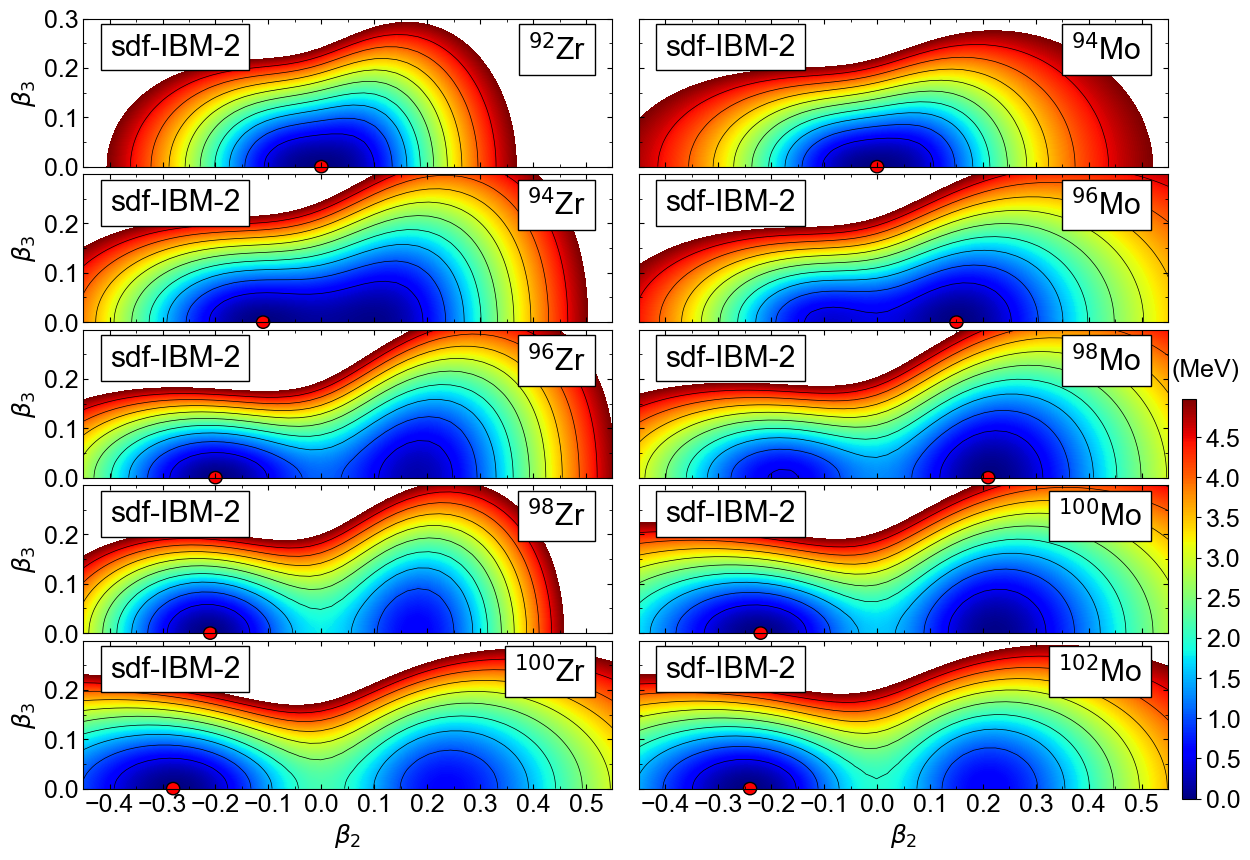}
\caption{
Same as Fig.~\ref{fig:pesibm1}, but for 
the $^{92-100}$Zr, and $^{94-102}$Mo isotopes. 
}
\label{fig:pesibm2}
\end{center}
\end{figure*}

\section{Results and discussion\label{sec:results}}

\subsection{Potential energy surfaces\label{sec:pes}}

Figures~\ref{fig:pesdft1} and \ref{fig:pesdft2} 
show contour plots of the axially-symmetric 
quadrupole and octupole deformation energy surfaces for the 
even-even nuclei $^{86-94}$Se, $^{88-96}$Kr, 
$^{90-98}$Sr, $^{92-100}$Zr, and $^{94-102}$Mo as functions of the 
$\beta_2$ and $\beta_3$ deformations, computed by the SCMF 
method described in Sec.~\ref{sec:scmf}.  
For any of the nuclei under study, 
one finds no octupole deformed ground state minimum 
with $\beta_3\neq0$. 
On closer inspection, the potential energy surface 
appears to be most rigid in the $\beta_2$ deformation 
at the $N=56$ nucleus among 
each isotopic chain, the neutron number corresponding to the 
empirical octupole magic number. 
The potential only gradually becomes softer along the 
$\beta_3$ direction from $N=52$ to 56. 
The $\beta_3$-softest potential is obtained for those nuclei 
with the neutron number approximately equal to 54 or 56, 
for which nuclei the curvature of the energy surface at the global 
minimum on the $\beta_3=0$ axis is indeed smallest. 
These features are most notably observed in the Se and Kr 
isotopes, with the proton number of the former isotopes, $Z=34$, 
identified as the proton octupole magic number. 
For the $N>56$ nuclei, the energy surfaces 
become more rigid, especially in the $\beta_3$ deformation, 
and the octupole correlations are expected to be much 
less pronounced. 

There are some other properties worth remarking 
of the SCMF energy surfaces. 
For the $N=52$ isotones, a nearly spherical minimum is suggested: 
$\beta_2\approx0.05$ for $^{86}$Se and $^{88}$Kr, 
and $\beta_2\approx0.0$ for $^{90}$Sr, $^{92}$Zr, and $^{94}$Mo, 
the last three nuclei being in the immediate vicinity of the proton 
subshell closure $Z=40$. 
All the $N=54$ isotones are here suggested to be soft in $\beta_2$ 
deformation in the interval $|\beta_2|\lesssim0.2$. 
In all the isotopic chains, the prolate-to-oblate shape transition 
is observed: in Figs.~\ref{fig:pesdft1} and \ref{fig:pesdft2} 
a change in the location of the minimum from the prolate, 
or nearly spherical, to oblate sides at $N=54$ (Kr and Zr), 
$N=56$ (Se), and $N=58$ (Mo). 
In $^{96,98}$Sr and $^{100}$Zr,  
another prolate minimum with quite a large 
$\beta_2$ deformation ($\beta_2\gtrsim0.4$) appears. 
These behaviours are considered a signature of the onset of intruder 
deformed configuration that is suggested to occur around $N=60$. 

It can be also shown that the SCMF calculations with the 
reduced pairing strength $V_0=728$ MeV\,fm$^3$ produce the energy surfaces 
that are slightly steeper in both $\beta_2$ and $\beta_3$ deformations but 
that are qualitatively similar to those shown in Figs.~\ref{fig:pesdft1} 
and \ref{fig:pesdft2}, 
obtained with the increased pairing strength $V_0=837$ MeV\,fm$^3$. 
In addition, the same constrained RHB calculations, but employing the 
density-dependent meson-exchange (DD-ME2) functional \cite{DDME2}, 
another representative effective interaction in the 
relativistic EDF framework, give strikingly similar mean-field 
results to those in the case of the DD-PC1 EDF. 

Figures \ref{fig:pesibm1} and \ref{fig:pesibm2} show the mapped $sdf$-IBM-2 
potential energy surfaces. As compared with the SCMF energy surfaces 
in Figs.~\ref{fig:pesdft1} and \ref{fig:pesdft2}, one notices 
that the basic topology of the SCMF energy surface, 
up to typically 2 MeV excitation 
from the minimum, is reproduced by the bosonic ones. 
One also finds that the $sdf$-IBM-2 energy surfaces are flat for higher 
excitation energies associated with large $\beta_2$ and $\beta_3$ 
deformations, in comparison to the SCMF counterparts. 
This difference illustrates that the $sdf$-IBM-2 space consists 
of only limited number of valence nucleons, while the SCMF model 
includes all nucleon degrees of freedom. 

\begin{figure}[ht]
\begin{center}
\includegraphics[width=\linewidth]{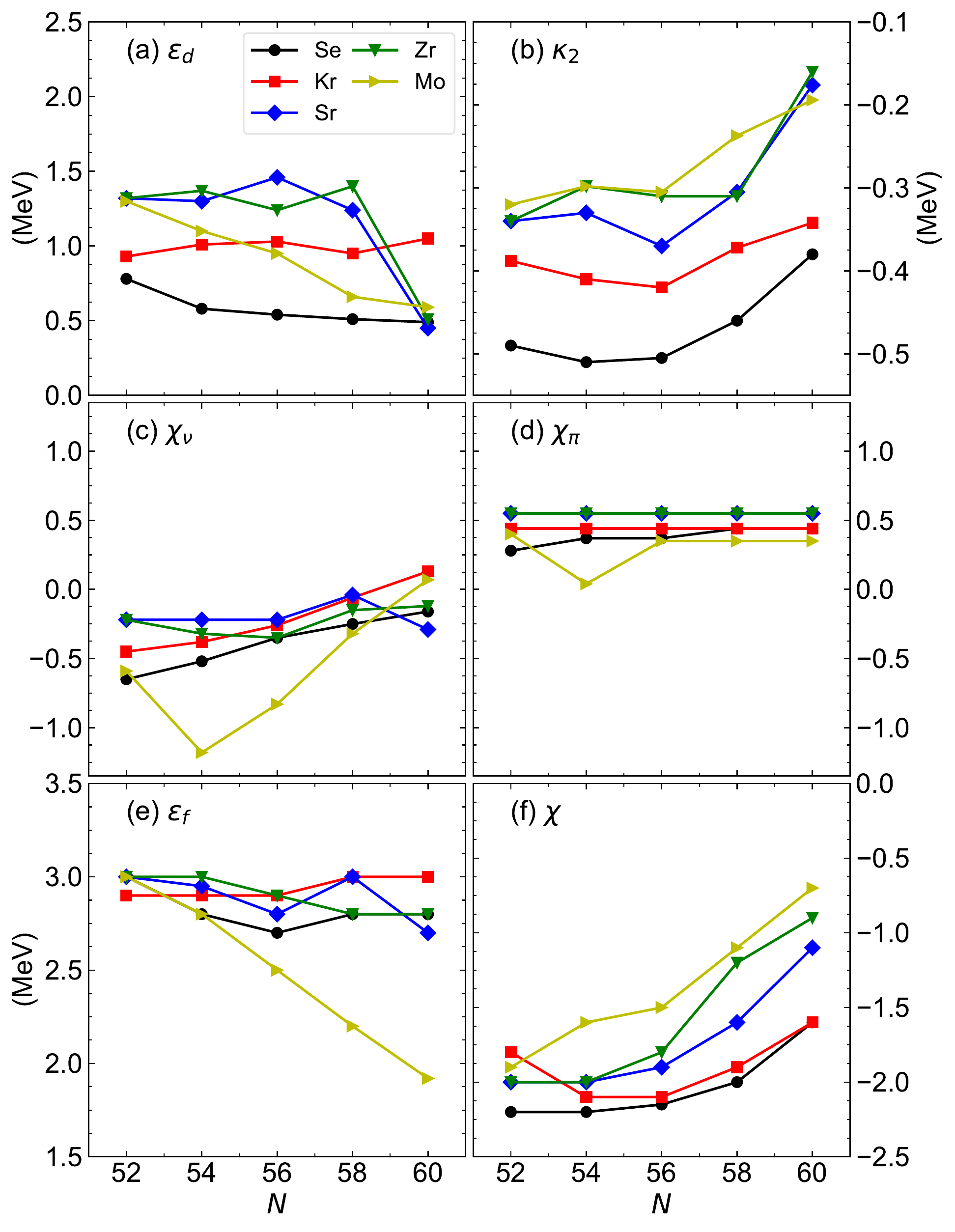}
\caption{Values of the strength parameters for the $sdf$-IBM-2 
Hamiltonian, plotted as functions of $N$. 
Note that the parameter $\chi$ in panel (f) is defined as 
$\chi=\chi'_\nu=\chi'_\pi=\chi''_{\nu}=\chi''_{\pi}$. 
The octupole-octupole interaction strength $\kappa_3$ 
is kept constant as $\kappa_3=0.12$ MeV, and is not plotted.}
\label{fig:para}
\end{center}
\end{figure}

\begin{figure*}[ht]
\begin{center}
\includegraphics[width=.8\linewidth]{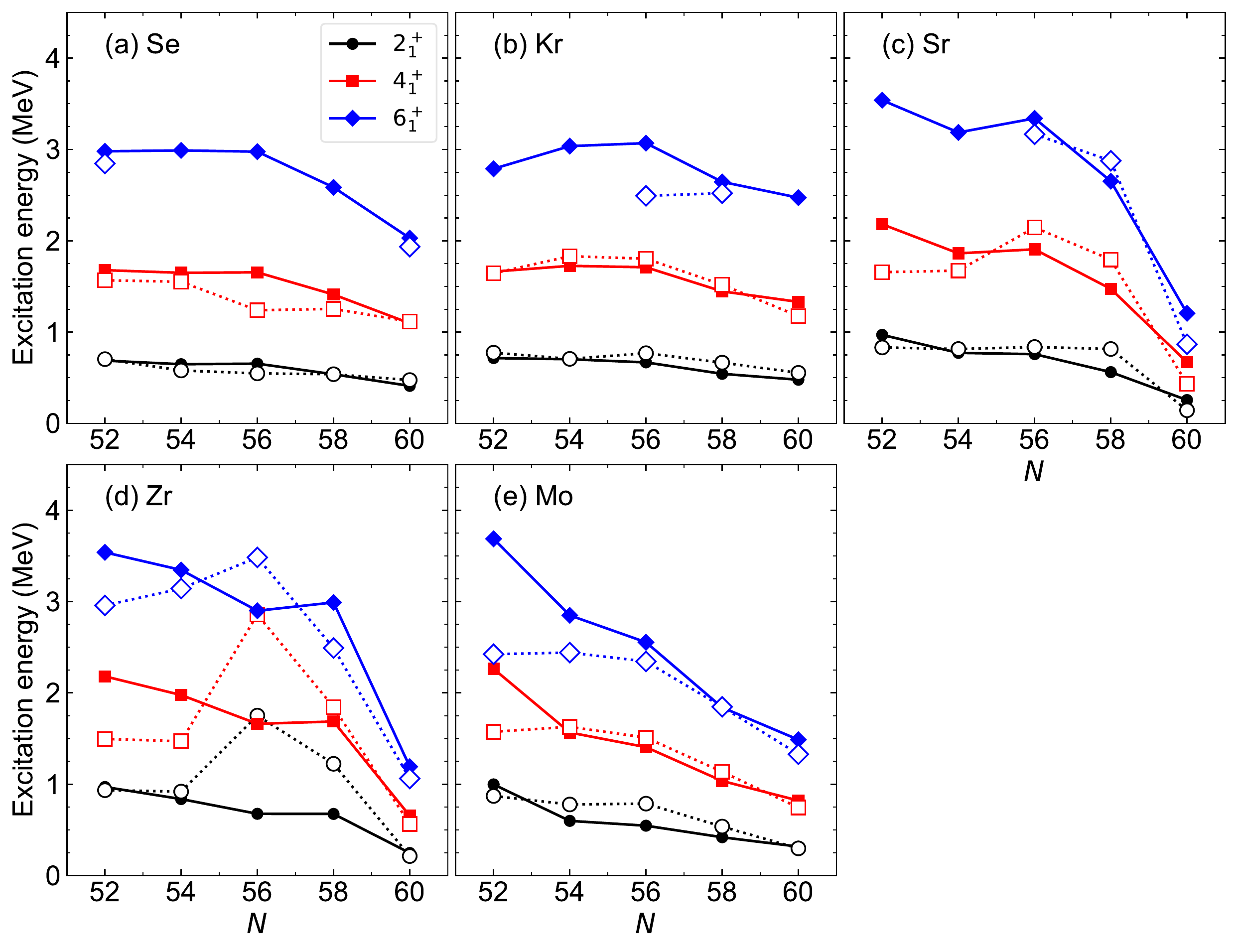}
\caption{Low-energy spectra 
for positive-parity even-spin states 
$2^+_1$, $4^+_1$, and $6^+_1$ 
of the considered even-even nuclei $^{86-94}$Se, $^{88-96}$Kr, 
$^{90-98}$Sr, $^{92-100}$Zr, and $^{94-102}$Mo. 
Theoretical and experimental values are 
represented by the solid and open symbols, which are 
connected by the solid and dotted lines, respectively. 
The experimental data are taken from 
Refs.~\cite{data,rzacaurban2000,lalkovski2007,rzacaurban2009,chen2017,lizarazo2020,gerst2020,gerst2022}.}
\label{fig:pos}
\end{center}
\end{figure*}

\begin{figure*}[ht]
\begin{center}
\includegraphics[width=.8\linewidth]{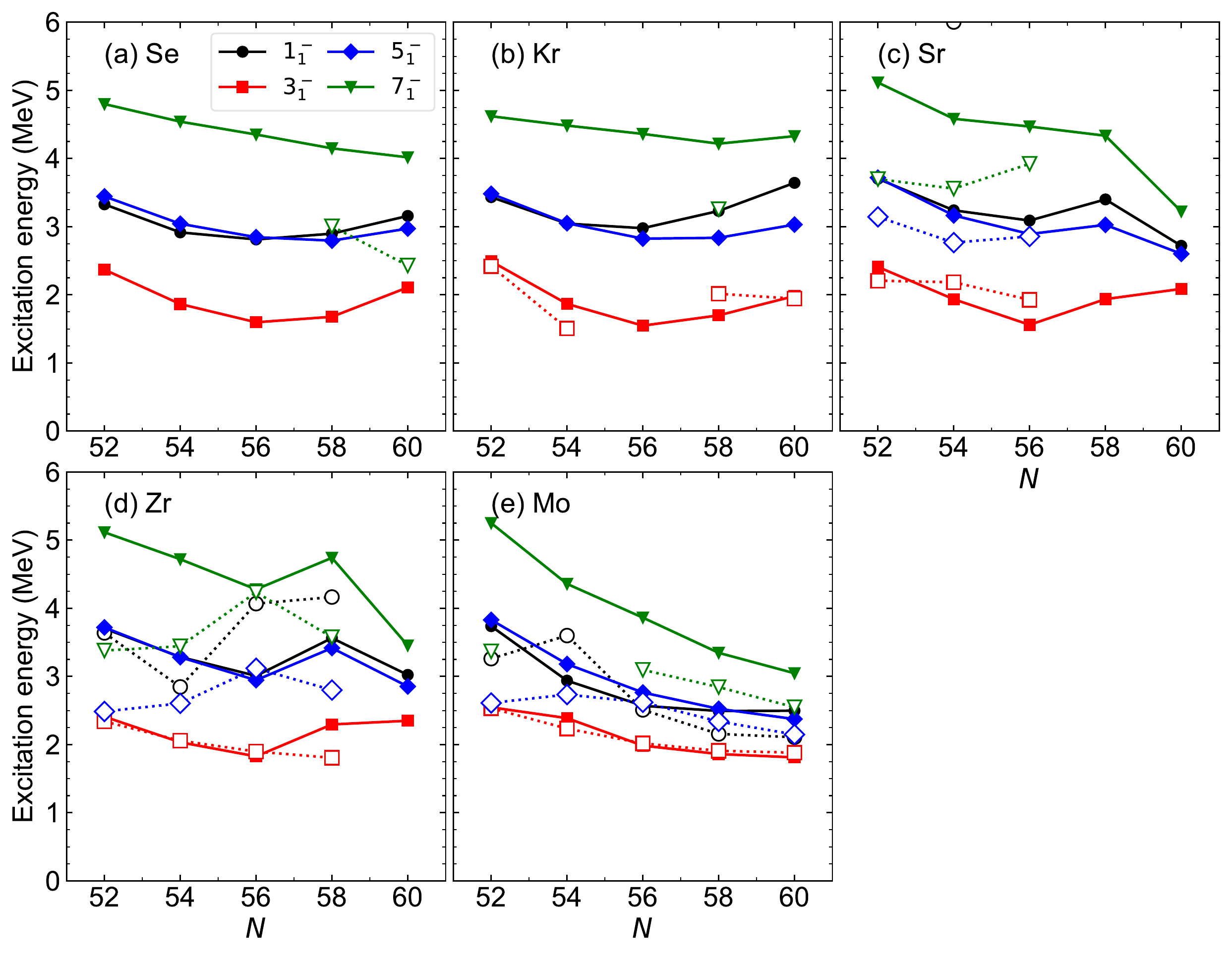}
\caption{Same as Fig.~\ref{fig:pos}, but for 
negative-parity odd-spin states $1^-_1$, $3^-_1$, $5^-_1$, and $7^-_1$. 
}
\label{fig:neg}
\end{center}
\end{figure*}

\subsection{Derived strength parameters}

The derived $sdf$-IBM-2 strength parameters used for the spectroscopic 
calculations on the considered Se, Kr, Sr, Zr, and Mo nuclei 
are shown as functions of $N$ in Fig.~\ref{fig:para}. 
In Fig.~\ref{fig:para}(a) one sees that 
the single-$d$ boson energy $\epsilon_d$ is mostly stable 
for $52\lesssim N\lesssim 58$, but gradually decreases 
along the Mo chain. A sudden drop from $N=58$ to 60 in the derived $\epsilon_d$ 
for the Sr and Zr isotopes is partly to account for 
the rapid structural change, i.e., onset of strong deformation, 
suggested empirically in these isotopic chains. 
The quadrupole-quadrupole interaction strength $\kappa_2$ 
generally decreases in magnitude as the 
number of valence neutrons increases [cf. Fig.~\ref{fig:para}(b)]. 
The parameters $\chi_\nu$ [Fig.~\ref{fig:para}(c)] 
and $\chi_\pi$ [Fig.~\ref{fig:para}(d)] do not show any strong 
dependence on $N$, except for the $\chi_\nu$ value for the 
Mo isotopes. The positive (negative) sign of the sum $\chi_\nu+\chi_\pi$ 
determines whether a nucleus is oblate (prolate) deformed. 
In many of the considered nuclei, a weakly-deformed mean-field 
minimum occurs on the oblate side, hence the value of the sum 
$\chi_\nu+\chi_\pi$ here has positive sign and small magnitude. 
The single $f$-boson energy $\epsilon_f$ is here basically kept 
constant or made only gradually change within the range 
$\epsilon_f\simeq 2.7-3.0$ MeV, 
while it significantly 
decreases with $N$ for the Mo isotopic chain [Fig.~\ref{fig:para}(e)]. 
The value of the common parameter 
$\chi(=\chi'_\nu=\chi'_\pi=\chi''_{\nu}=\chi''_{\pi})$ 
in Fig.~\ref{fig:para}(f) is determined according to the degree 
of the $\beta_3$, as well as $\beta_2$, softness of the potential. 
One sees that the derived $\chi$ value is indeed 
large in magnitude for $52\lesssim N\lesssim 56$ as compared with 
those nuclei with $N>56$ for most of the studied isotopic chains. 
A fixed value of the octupole-octupole interaction 
strength $\kappa_3=0.12$ MeV, determined for a particular 
nucleus $^{92}$Kr, is here used for all the nuclei under study. 
The use of the constant $\kappa_3$ value is not only for the 
sake of simplicity to reduce the number of parameters, 
but is also to take into account 
the facts that the topology of the SCMF energy surface 
varies only gradually in the $\beta_3$ deformation 
as a function of the nucleon number, 
and that the observed low-lying negative-parity levels 
also do not show a strong nucleon-number dependence. 

\subsection{Systematics of low-energy spectra}

Figure~\ref{fig:pos} compares the calculated low-energy spectra for the 
positive-parity even-spin states $2^+_1$, $4^+_1$, and $6^+_1$ 
of the considered nuclei with the experimental data \cite{data}. 
One observes an overall reasonable agreement with the data, 
except perhaps for the Zr isotopes. 
For the Se, Kr, and Mo isotopes, both the theoretical and experimental 
energy levels are gradually lowered with the increasing $N$. 
The modest decrease of the 
calculated yrast spectra, from $^{94}$Kr to $^{96}$Kr 
in particular, suggests a smooth onset of deformation 
in agreement with the experiment \cite{albers2012}. 
For the Sr and Zr isotopic chains the present calculation 
gives a more rapid decrease of these states. 
One notices, in the corresponding experimental spectra in 
Figs.~\ref{fig:pos}(c) and \ref{fig:pos}(d), a pronounced 
peak at the neutron number $N=56$. 
This indicates the effect of the neutron $N=56$ subshell 
gap due to the filling in the $\nu d_{5/2}$ orbital, which is 
even more enhanced for the Zr nuclei corresponding to the 
proton subshell closure $Z=40$. 
The mapped $sdf$-IBM-2 does not reproduce this trend 
in Zr, mainly because no spherical minimum occurs 
in the axially-symmetric quadrupole-octupole SCMF 
energy surface for $^{96}$Zr (cf. Fig.~\ref{fig:pesdft2}). 
The SCMF calculation gives for this nucleus only a weakly deformed 
oblate minimum at $\beta_2\approx-0.2$, and consequently 
the mapped $sdf$-IBM-2 yields the collective energy spectrum, 
characterized by the low-lying 
$2^+_1$ energy level and by the ratio of the $4^+_1$ to $2^+_1$ 
excitation energies $R_{4/2}=2.46$. 
Another notable feature observed in the Sr and Zr isotopes 
is a sudden drop of the energy levels from $N=58$ to 60. 
The present calculation reproduces well this systematic. 
As for the $N=52$ isotones, the $sdf$-IBM-2 generally does not 
give a very good description, because these nuclei are close 
to the neutron $N=50$ major shell closure and the model space 
of the calculation including only one neutron boson may not be large 
enough to reproduce the positive-parity levels of the nearly spherical nuclei. 
It is worth noting that the low-lying nuclear 
structure in the $N\approx60$ region is often characterized 
by the occurrence of the competing intrinsic shapes, 
and that a more refined calculation 
of the relevant spectroscopic properties 
would need to include the configuration mixing between the 
normal and intruder states \cite{duval1981} within the $sdf$-IBM. 

In Fig.~\ref{fig:neg}, the calculated energy spectra for the 
low-lying negative-parity odd-spin states 
$1^-_1$, $3^-_1$, $5^-_1$, and $7^-_1$ are shown as functions of $N$. 
Of particular interest regarding the predicted negative-parity 
energy spectra is their parabolic dependence with $N$ with 
the minimum values at $N=56$ for all the considered isotopic chains, 
except for Mo. This result is consistent with the systematic behaviour 
of the quadrupole-octupole potential energy surfaces, which exhibit 
the $\beta_3$ softest potential for those nuclei with $N\approx56$. 
As seen in Fig.~\ref{fig:neg}, in accordance with the experimental data, 
the lowest negative-parity state is here predicted to 
be $I=3^-$ at the excitation energy $E_x\approx2$ MeV. 
Within the calculation, the $1^-$ energy level is 
rather high and is close to the $5^-_1$ one. 
This level structure is at variance with those observed, 
e.g., in the actinide nuclei with $N\approx134$. 
In many of the nuclei in the latter mass region, the low-lying 
negative-parity band with the bandhead $1^-$ state appears, and 
forms an approximate alternating-parity rotational band with 
the positive-parity ground-state band. 
One may notice that the predicted higher-spin negative-parity states, 
especially the $7^-_1$ one, is higher than the 
experimental value, most notably, for $^{94,96}$Se and $^{94}$Kr. 
It is noted, however, that the observed low-lying $7^-$ state in these 
nuclei could be attributed to isomeric state based on the neutron 
two-quasiparticle excitations (see, e.g., Ref.~\cite{gerst2020}), 
which are beyond the present $sdf$-IBM-2 model space.

\begin{figure*}
\begin{center}
\includegraphics[width=\linewidth]{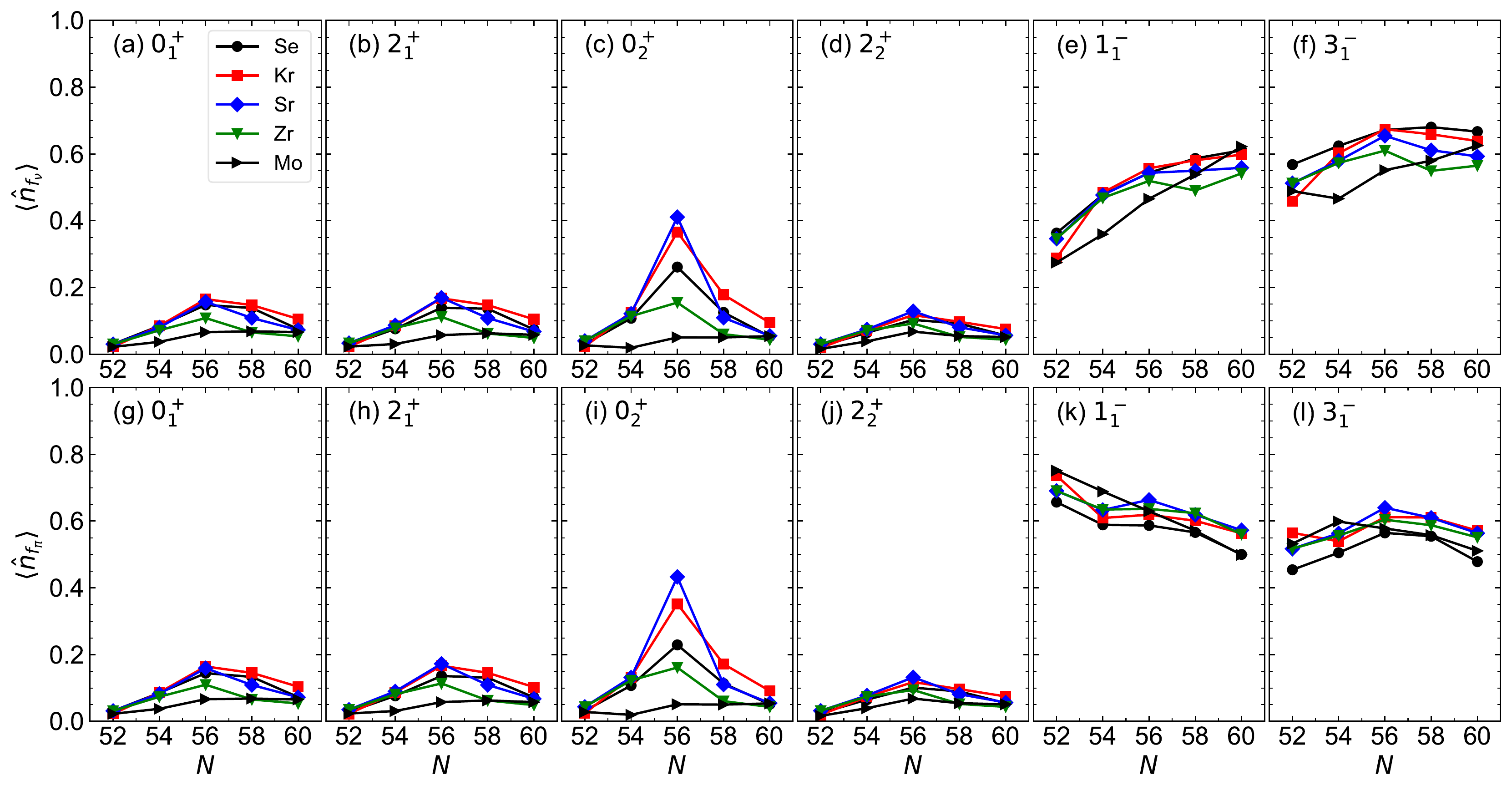}
\caption{Expectation values of 
neutron $\hat n_{f_\nu}$ (upper row) and proton 
$\hat n_{f_\pi}$ (lower row) number operators 
calculated for the (a), (g) $0^+_1$, (b), (h)
$2^+_1$, (c), (i) $0^+_2$, 
(d), (j) $2^+_2$, (e), (k) $1^-_1$, and 
(f), (l) $3^-_1$ states for the 
considered Se, Kr, Sr, Zr, and Mo nuclei.}
\label{fig:wf}
\end{center}
\end{figure*}

\subsection{$f$-boson contributions to wave functions}

To interpret the nature of the low-lying states, contributions 
of the $f$ bosons to the corresponding $sdf$-IBM-2 wave functions 
are studied. As an illustrative example, Fig.~\ref{fig:wf} shows 
the expectation values of the 
neutron $\hat n_{f_\nu}$ and proton $\hat n_{f_\pi}$ boson 
number operators calculated by using the wave functions for 
the low-spin positive-parity states $0^+_1$, $2^+_1$, $0^+_2$, and 
$2^+_2$, and negative-parity states $1^-_1$, and $3^-_1$. 
As one can see in panels (a) to (d) 
and (g) to (j) of Fig.~\ref{fig:wf}, 
there is only a small admixture of the $f_\nu$ and $f_\pi$ bosons 
into the low-energy positive-parity states, with the expectation 
value $\hat n_{f_\rho}$ typically $\braket{\hat n_{f_\rho}}\lesssim0.2$. 
In the $0^+_{2}$ wave functions there appears to be a 
relatively large $f$-boson contributions at $N=56$, in particular, 
for $^{90}$Se, $^{92}$Kr and $^{94}$Sr, in which nuclei the 
octupole correlations are expected to be most pronounced. 
For the negative-parity states $1^-_1$ and $3^-_1$, 
the calculated expectation value $\braket{\hat n_{f_\rho}}\approx0.5$ 
for both proton and neutron bosons, and hence 
$\braket{\hat n_{f_\rho} + \hat n_{f_\rho}}\approx1$. 
Therefore, the wave functions for the above 
negative-parity states contain approximately one-$f$ boson components.

\begin{figure}[ht]
\begin{center}
\includegraphics[width=\linewidth]{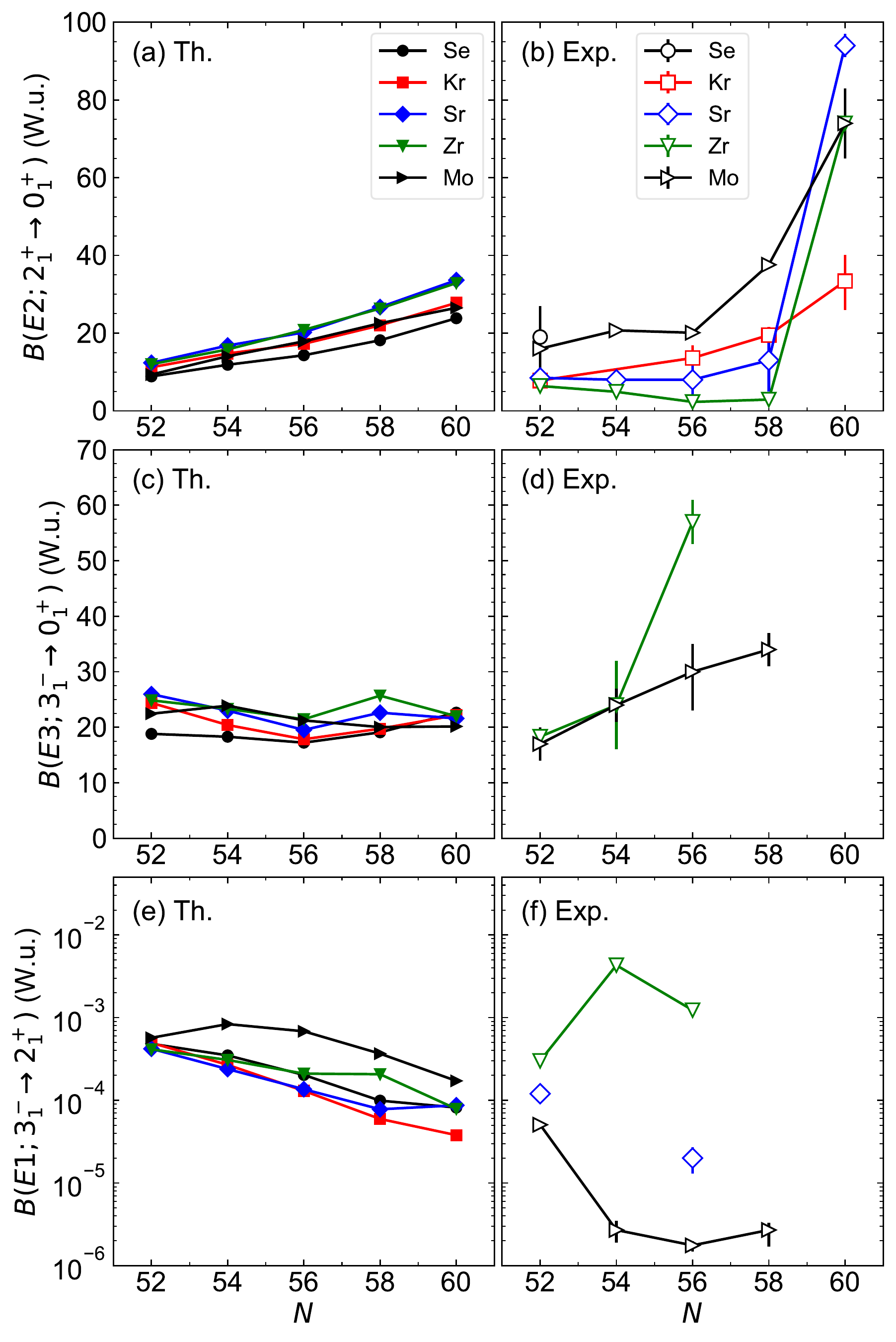}
\caption{Evolution of calculated and experimental 
(a), (b) $B(E2;2^+_1\to0^+_1)$, 
(c), (d) $B(E3;3^-_1\to0^+_1)$, 
and (e), (f) $B(E1;3^-_1\to2^+_1)$ 
transition strengths in Weisskopf units (W.u.) 
for the Se, Kr, Sr, Zr, and Mo isotopes 
with $52\leq N \leq 60$. The experimental data are adopted from 
Refs.~\cite{data,kibedi2002,elhami2008,albers2013,kremer2016,gregor2017,gregor2019}. 
Note that a lower limit is shown for 
the experimental $B(E1)$ value for $^{90}$Sr in panel (f).}
\label{fig:trans}
\end{center}
\end{figure}

\subsection{Electromagnetic transition properties}

Electromagnetic properties considered in the present 
theoretical analysis 
are those of the electric quadrupole $E2$, octupole $E3$, 
dipole $E1$, and magnetic dipole $M1$ transitions. 
The corresponding operators read: 
\begin{align}
\label{eq:e2}
&\hat T^{(E2)} = \sum_{\rho=\nu,\pi} e_{2,\rho} \hat Q_\rho,\\
\label{eq:e3}
&\hat T^{(E3)} = \sum_{\rho=\nu,\pi} e_{3,\rho} \hat O_\rho,\\
\label{eq:e1}
&\hat T^{(E1)} = \sum_{\rho=\nu,\pi} e_{1,\rho} \hat D_\rho,\\
\label{eq:m1}
&\hat T^{(M1)} = \sqrt{\frac{3}{4\pi}}\sum_{\rho=\nu,\pi} g_{\rho}\hat L_\rho,
\end{align}
where the operators $\hat Q_\rho$ and $\hat O_\rho$ are the same quadrupole 
and octupole operators as in the Hamiltonian (\ref{eq:ham}) 
with the same values of the parameters $\chi_\rho$, $\chi'_{\rho}$, 
and $\chi_\rho''$, and 
\begin{align}
 &\hat D_\rho = 
(d^\+_\rho\times\tilde f_\rho + f^\+_\rho\times\tilde d_\rho)^{(1)}\\
 &\hat L_\rho = 
\sqrt{10}(d^\+_\rho\times\tilde d_\rho)^{(1)} 
+\sqrt{28}(f^\+_\rho\times\tilde f_\rho)^{(1)}
\end{align}
are electric and magnetic dipole transition operators, respectively. 
$e_{\lambda,\rho}$ ($\lambda=1,2,3$) in Eqs.~(\ref{eq:e2})--(\ref{eq:e1}) 
are effective boson charges, 
and $g_\rho$ in (\ref{eq:m1}) is the bosonic gyromagnetic ($g$) factor. 
The neutron and proton effective charges are assumed to be 
equal, i.e., $e_{1,\nu}=e_{1,\pi}\equiv e_1$, 
$e_{2,\nu}=e_{2,\pi}\equiv e_2$, and $e_{3,\nu}=e_{3,\pi}\equiv e_3$, 
and the fixed effective charges $e_{1}=0.005$ $e$b$^{1/2}$, 
$e_2=0.06$ $e$b, and $e_3=0.06$ $e$b$^{3/2}$ are used so as to 
reasonably reproduce the experimental data. 
The empirical boson $g$ factors $g_\nu=0$ $\mu_N$ and $g_\pi=1$ $\mu_N$ are 
adopted by following the microscopic calculations in the previous 
($sd$-)IBM-2 studies \cite{IBM}.

Systematic behaviors of the reduced $E2$, $E3$, and $E1$ transition 
probabilities between the lowest positive- and negative-parity states, 
$B(E2;2^+_1\to0^+_1)$, $B(E3;3^-_1\to0^+_1)$, and $B(E1;3^-_1\to2^+_1)$, 
are shown in Fig.~\ref{fig:trans}. Results for the $M1$ properties 
are discussed in the next section for individual nuclei. 
Both the calculated and experimental $2^+_1\to0^+_1$ $E2$ transition 
rates are weak for $52\leq N \leq 58$, with the $B(E2)$ values 
typically lower than 20 Weisskopf units (W.u.). 
The experimental data show a significant rise of the 
$B(E2;2^+_1\to0^+_1)$ values from $N=58$ to 60 in the Sr, Zr, 
and Mo isotopes, as a consequence of the onset of  
strong quadrupole deformation. 
The current model calculation is not able to reproduce 
this sharp increase of the $B(E2;2^+_1\to0^+_1)$ values. 
The inconsistency could have arisen mainly due to the fact that 
the present IBM framework does not include 
the configuration mixing, which would be required for 
dealing with the phenomenon of shape coexistence. 
Indeed, the SCMF potential energy surface, 
e.g., for $^{98}$Sr, indicates a development of 
a strongly-deformed prolate local minimum at $\beta_2\approx0.45$ 
in addition to the oblate global minimum at $\beta_2\approx-0.25$ 
(see Fig.~\ref{fig:pesdft1}). 
For the Zr and Mo isotopes, 
on the other hand, the SCMF $\beta_2-\beta_3$ 
energy surface shown in Fig.~\ref{fig:pesdft2} does not exhibit 
a substantial variation from $N=58$ to 60 along the $\beta_2$ 
deformation, or a pronounced competition between different 
mean-field minima. 
It is, therefore, not straightforward to uniquely identify the 
major source of the discrepancy between the predicted 
and experimental $B(E2)$ systematics in the Zr and Mo chains. 
It could be attributed to the lack 
of the configuration mixing in the IBM, but may also indicate 
a deficiency of the employed EDF or particular choice of 
the pairing strength. In addition, only the 
axially-symmetric shape degrees of freedom are considered here 
as relevant collective coordinates, while the triaxial deformation 
could also play an important 
role in the considered mass region. 

The $B(E3;3^-_1\to0^+_1)$ rates are a direct measure of the octupole 
collectivity, and are generally large for those nuclei that are 
expected to have an enhanced octupole deformation. 
Experimental data for the $B(E3;3^-_1\to0^+_1)$ rates are available 
\cite{kibedi2002,gregor2017,gregor2019} for Zr and Mo nuclei. 
One finds in Fig.~\ref{fig:trans}(d) that experimental 
$B(E3;3^-_1\to0^+_1)$ value is particularly large for $^{96}$Zr, 
$57\pm4$ W.u., and slowly increases toward the middle of 
the neutron major shell in the Mo isotopes. 
The $B(E3;3^-_1\to0^+_1)$ rates predicted by the $sdf$-IBM-2 calculation 
are approximately within the range $20-30$ W.u. 
for all the studied isotopic chains. 
Note, however, that the calculated values 
for the Zr and Mo chains are virtually constant against $N$, 
being at variance with the observed systematic. 

The $B(E1;3^-_1\to2^+_1)$ transition rate is also a relevant 
quantity to the octupole deformation. There are some data 
available. Especially, recent measurements for $^{94}$Kr and $^{96}$Kr 
suggested the 1350 and 1390 keV decays, respectively, of the 
$3^-$ to $2^+_1$ states \cite{gerst2022}. 
The calculated $B(E1;3^-_1\to2^+_1)$ values shown in 
Fig.~\ref{fig:trans}(e) are large near the neutron $N=50$ major 
shell gap, and decrease rather rapidly with $N$ toward $N=60$. 
The measured $B(E1;3^-_1\to2^+_1)$ values seem to be quite different 
from one isotopic chain to another, whereas the present calculation 
gives rather similar values and systematic behaviours among 
the five isotopic chains. 
One should also note that, since the $E1$ mode is more of 
single-particle nature than the $E2$ and $E3$ ones, the $sdf$-IBM 
descriptions in general, which consist only of collective 
degrees of freedom, may not give meaningful predictions for the 
$E1$ transition properties. 
For a more reliable IBM description of the $E1$ transition properties, 
the dipole $p$ bosons with spin and parity $J=1^-$ could 
be introduced as additional 
degrees of freedom to the boson model space 
\cite{otsuka1986,otsuka1988,sugita1996}. This extension, 
however, lies beyond the scope of the present theoretical analysis.

\begin{figure}[ht]
\begin{center}
\includegraphics[width=\linewidth]{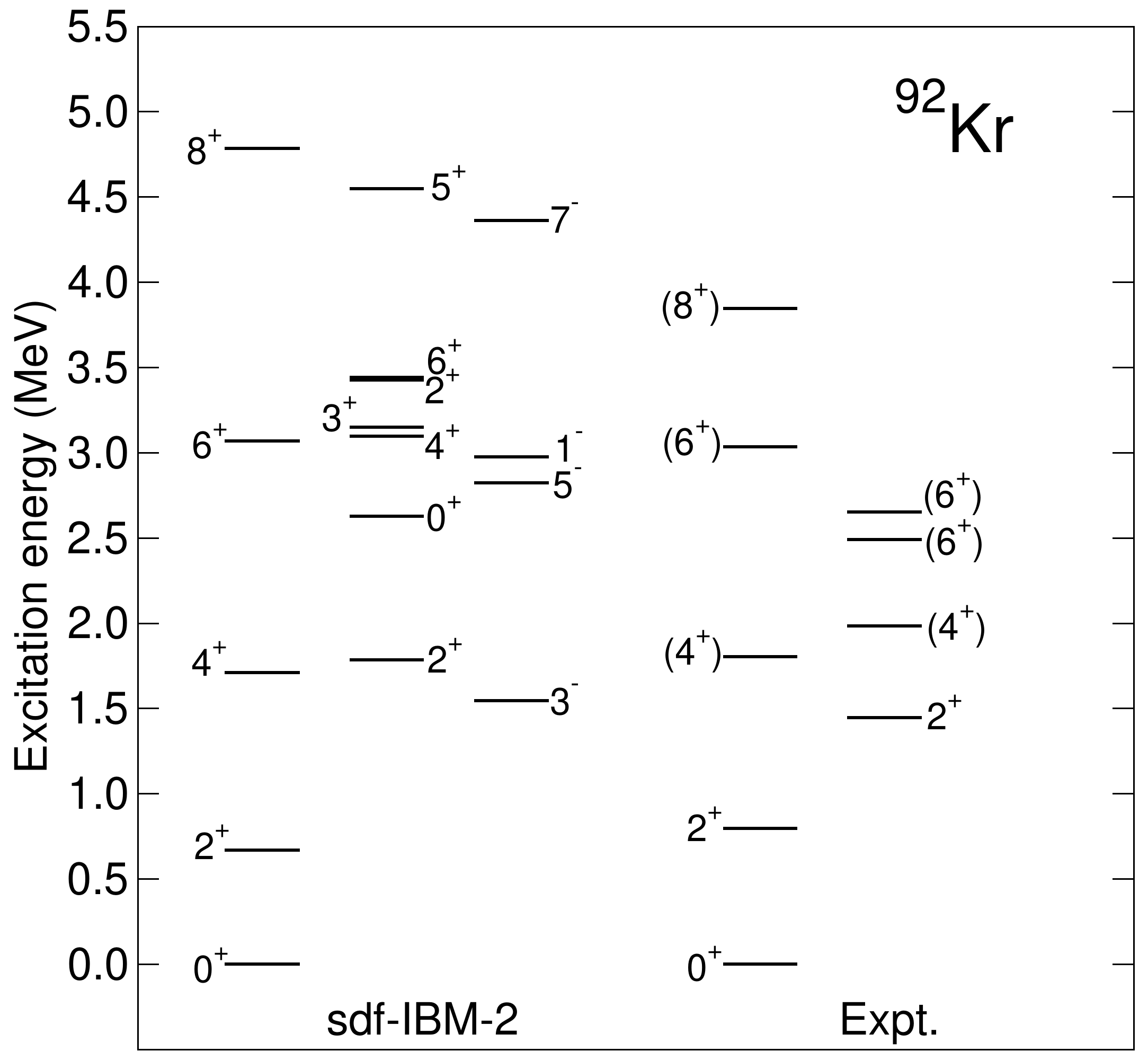}
\caption{Predicted and experimental low-energy positive- 
and negative-parity spectra for $^{92}$Kr. The experimental data 
are taken from Ref.~\cite{li2011}.
}
\label{fig:kr92}
\end{center}
\end{figure}

\begin{figure}[ht]
\begin{center}
\includegraphics[width=\linewidth]{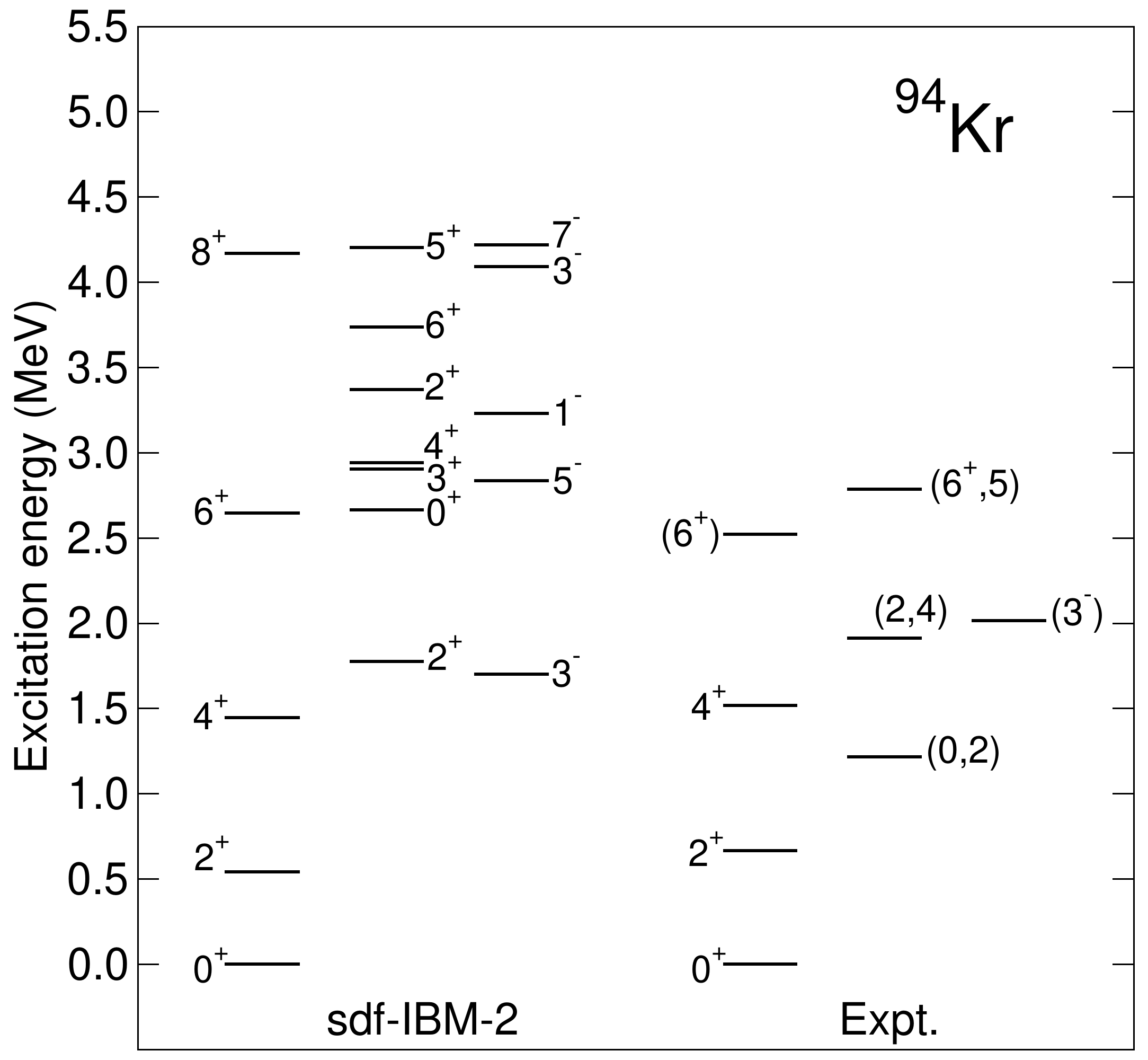}
\caption{Same as Fig.~\ref{fig:kr92}, but for $^{94}$Kr. 
The experimental data are taken from Ref.~\cite{gerst2022}.
}
\label{fig:kr94}
\end{center}
\end{figure}

\begin{figure}[ht]
\begin{center}
\includegraphics[width=\linewidth]{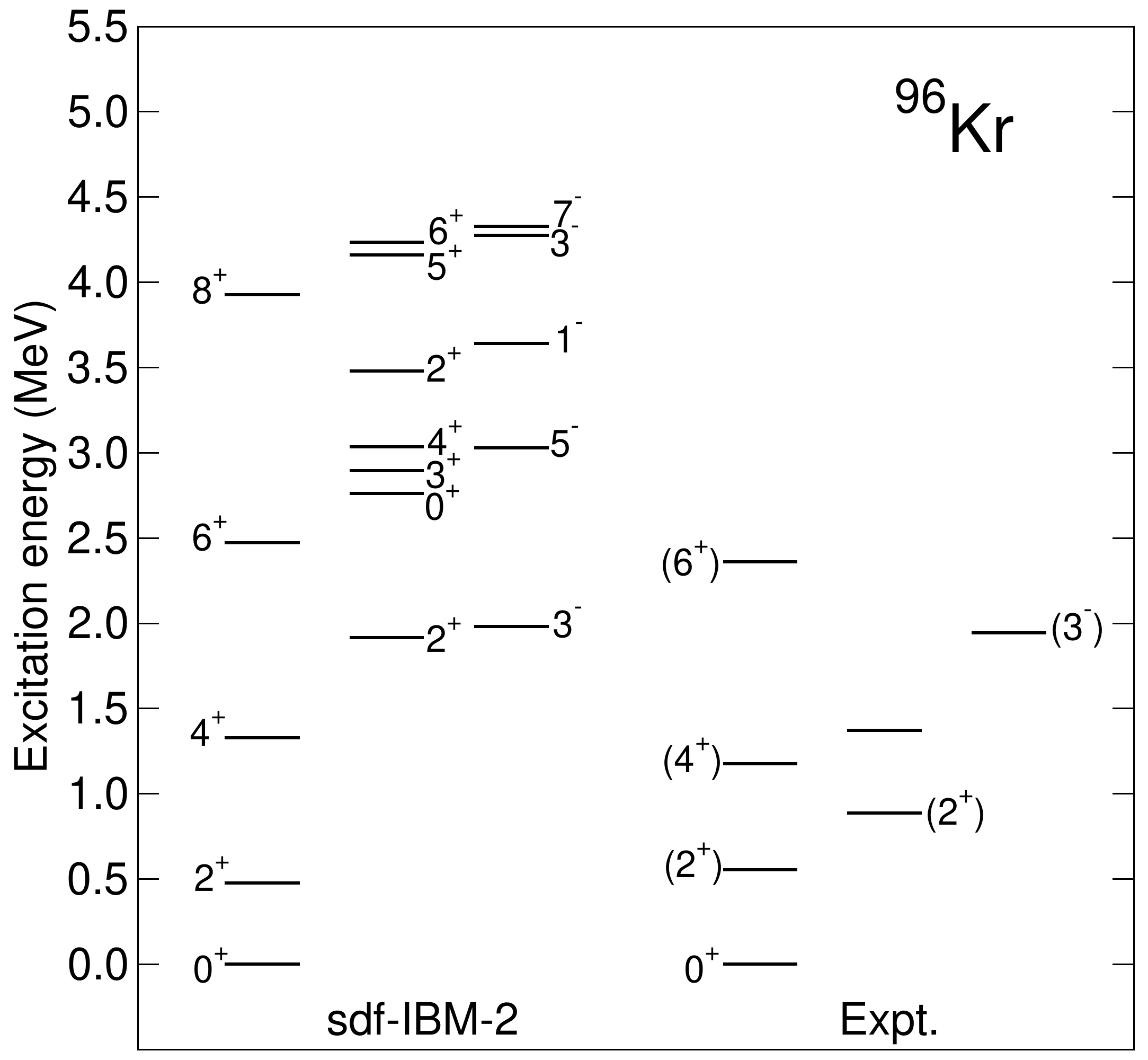}
\caption{Same as Fig.~\ref{fig:kr92}, but for $^{96}$Kr. 
The experimental data are taken from Ref.~\cite{gerst2022}.
}
\label{fig:kr96}
\end{center}
\end{figure}

\subsection{Low-energy spectra for Kr nuclei}

Figures~\ref{fig:kr92}, \ref{fig:kr94}, and \ref{fig:kr96} 
show detailed low-energy spectra for the neutron-rich isotopes 
$^{92}$Kr, $^{94}$Kr, and $^{96}$Kr, respectively. 
The positive-parity part of the level scheme consist of the states in the 
$K=0^+_1$ ground-state band and states not classified into bands. 
For the negative-parity part only the spectra for the odd-spin states 
are shown. 

The $^{92}$Kr nucleus corresponds to the octupole neutron magic 
number $N=56$, and the corresponding $\beta_2-\beta_3$ potential 
energy surface is soft in $\beta_3$ deformation 
(cf. Figs.~\ref{fig:pesdft1} and \ref{fig:pesibm1}). 
As one sees in Fig.~\ref{fig:kr92}, the calculation 
reproduces well the observed $K=0^+_1$ band, and predicts the $2^+_2$ 
states near the $4^+_1$ state, which is considered as the bandhead 
of the quasi-$\gamma$ or $K=2^+_2$ band that comprises the 
$4^+_2$, $3^+_1$, and $5^+_1$ states. The near degeneracy of the 
members of the quasi-$\gamma$ band, i.e., $3^+_1$ and $4^+_2$, 
indicates the O(6)-like level structure \cite{IBM}. 
The $0^+_2$ and $2^+_3$ states predicted by the 
calculation are part of the $K=0^+_2$ band. 
The mapped $sdf$-IBM-2 gives the low-lying $3^-_1$ state at 
around the excitation energy $E_x=1.5$ MeV, below the $4^+_1$ state. 
Remember that, among the considered Kr isotopes, the lowest $3^-_1$ level 
is obtained for $^{92}$Kr in the present calculation 
[cf. Fig.~\ref{fig:neg}(b)]. 

Similar results to $^{92}$Kr are obtained for the $^{94}$Kr 
(Fig.~\ref{fig:kr94}) and $^{96}$Kr (Fig.~\ref{fig:kr96}). 
In these two nuclei, the $sdf$-IBM-2 energy spectra for the 
$K=0^+_1$ band looks rather rotational as compared with the 
experimental counterparts, in such a way that the calculated 
$2^+_1$ energy level is more compressed. 
The calculated spectra for the positive-parity 
nonyrast bands in $^{94,96}$Kr are also rather high with respect to 
the $K=0^+_1$ band, and especially for $^{96}$Kr, overestimate 
the observed nonyrast levels. 
The calculated $3^-_1$ excitation energies for both $^{94}$Kr 
and $^{96}$Kr are, however, in a good agreement with the 
experimental values \cite{gerst2022}, 2015 keV and 1944 keV, respectively.

\begin{table}
\caption{
\label{tab:kr}
Calculated $B(E2)$, $B(M1)$, $B(E1)$, and $B(E3)$ values (in W.u.), and 
electric quadrupole $Q(I)$ (in $e$b) and magnetic 
dipole $\mu(I)$ (in $\mu_N$) moments for $^{92,94,96}$Kr.}
 \begin{center}
 \begin{ruledtabular}
  \begin{tabular}{lccc}
 & $^{92}$Kr & $^{94}$Kr & $^{96}$Kr \\
\hline
$B(E2;2^{+}_{1}\to0^{+}_{1})$ & 17 & 22 & 28 \\
$B(E2;4^{+}_{1}\to2^{+}_{1})$ & 24 & 31 & 40 \\
$B(E2;6^{+}_{1}\to4^{+}_{1})$ & 25 & 34 & 43 \\
$B(E2;0^{+}_{2}\to2^{+}_{1})$ & 0.021 & 0.46 & 2 \\
$B(E2;0^{+}_{2}\to2^{+}_{2})$ & 19 & 29 & 32 \\
$B(E2;2^{+}_{2}\to2^{+}_{1})$ & 19 & 16 & 14 \\
$B(E2;3^{+}_{1}\to2^{+}_{2})$ & 18 & 23 & 30 \\
$B(E2;4^{+}_{2}\to2^{+}_{1})$ & 0.11 & 0.08 & 0.084 \\
$B(E2;4^{+}_{2}\to2^{+}_{2})$ & 14 & 18 & 22 \\
$B(E2;4^{+}_{2}\to3^{+}_{1})$ & 1.5 & 4.6 & 10 \\
$B(E2;4^{+}_{2}\to4^{+}_{1})$ & 10 & 10 & 9.9 \\
$B(E2;3^{-}_{2}\to3^{-}_{1})$ & 0.16 & 0.036 & 0.011 \\
$B(E2;5^{-}_{1}\to3^{-}_{1})$ & 7.2 & 8.5 & 11 \\
$B(E2;7^{-}_{1}\to5^{-}_{1})$ & 12 & 15 & 21 \\
$B(E2;9^{-}_{1}\to7^{-}_{1})$ & 0.031 & 18 & 25 \\
\hline
$B(M1;2^{+}_{2}\to2^{+}_{1})$ & 0.00046 & 0.00028 & 0.00015 \\
$B(M1;3^{+}_{1}\to2^{+}_{2})$ & 0.00086 & 0.0004 & 0.00032 \\
$B(M1;4^{+}_{2}\to4^{+}_{1})$ & 0.0018 & 0.0011 & 0.0006 \\
$B(M1;4^{+}_{2}\to3^{+}_{1})$ & $1.8\times10^{-5}$ & $9.7\times10^{-6}$ & $6.0\times10^{-5}$ \\
$B(M1;3^{-}_{2}\to3^{-}_{1})$ & 0.015 & 0.011 & 0.011 \\
\hline
$B(E1;1^{-}_{1}\to0^{+}_{1})$ & $7.8\times10^{-5}$ & $7.6\times10^{-5}$ & $9.9\times10^{-5}$ \\
$B(E1;1^{-}_{1}\to2^{+}_{1})$ & $1.0\times10^{-5}$ & $6.9\times10^{-6}$ & $3.4\times10^{-6}$ \\
$B(E1;3^{-}_{1}\to2^{+}_{1})$ & $1.3\times10^{-4}$ & $5.9\times10^{-5}$ & $3.8\times10^{-5}$ \\
$B(E1;3^{-}_{1}\to4^{+}_{1})$ & $1.2\times10^{-7}$ & $5.1\times10^{-7}$ & $1.0\times10^{-6}$ \\
$B(E1;5^{-}_{1}\to4^{+}_{1})$ & $3.7\times10^{-4}$ & $2.1\times10^{-4}$ & $1.6\times10^{-4}$ \\
$B(E1;5^{-}_{1}\to6^{+}_{1})$ & $1.0\times10^{-6}$ & $1.3\times10^{-6}$ & $4.3\times10^{-6}$ \\
$B(E1;7^{-}_{1}\to6^{+}_{1})$ & $7.9\times10^{-4}$ & $5.3\times10^{-4}$ & $4.5\times10^{-4}$ \\
\hline
$B(E3;3^{-}_{1}\to0^{+}_{1})$ & 18 & 20 & 22 \\
$B(E3;3^{-}_{1}\to2^{+}_{1})$ & 41 & 43 & 44 \\
$B(E3;5^{-}_{1}\to2^{+}_{1})$ & 11 & 10 & 11 \\
$B(E3;5^{-}_{1}\to4^{+}_{1})$ & 34 & 39 & 42 \\
$B(E3;7^{-}_{1}\to4^{+}_{1})$ & 9.8 & 8.6 & 9.9 \\
$B(E3;7^{-}_{1}\to6^{+}_{1})$ & 29 & 34 & 37 \\
\hline
$Q(2^{+}_{1})$ & 0.19 & 0.34 & 0.45 \\
$\mu(2^{+}_{1})$ & 0.37 & 0.32 & 0.32 \\
  \end{tabular}
 \end{ruledtabular}
 \end{center}
\end{table}

Table~\ref{tab:kr} lists the predicted $B(E2)$, $B(M1)$, 
$B(E1)$, and $B(E3)$ transition strengths, and 
electric quadrupole $Q(I)$ and magnetic dipole $\mu(I)$ moments 
for the $^{92,94,96}$Kr isotopes. One can see from the table that the 
calculated values for these 
electromagnetic properties are basically similar among the 
three Kr nuclei. The inband $\Delta I=2$ $E2$ transitions, i.e., 
$2^{+}_{1}\to0^{+}_{1}$, $4^{+}_{1}\to2^{+}_{1}$, 
$6^{+}_{1}\to4^{+}_{1}$, $5^{-}_{1}\to3^{-}_{1}$, and 
$7^{-}_{1}\to5^{-}_{1}$, as well as the $Q(2^+_1)$ values, 
gradually increase from $^{92}$Kr to $^{96}$Kr. 
These systematics illustrate gradual evolution 
of quadrupole collectivity. 
The available experimental data \cite{albers2013}, 
$B(E2;2^+_1\to0^+_1)=13.6^{+2.8}_{-3.3}$ W.u., 
$19.5^{+2.2}_{-2.1}$ W.u., and $33.4^{+7.4}_{-6.7}$ W.u. 
for $^{92}$Kr, $^{94}$Kr, 
and $^{96}$Kr, respectively, are reasonably reproduced 
by the present $sdf$-IBM-2 calculation. 
The data are also available from Ref.~\cite{albers2013} 
for the $Q(2^+_1)$ values, $-0.61^{+0.53}_{-0.45}$ $e$b 
and $-0.26^{+0.19}_{-0.16}$ $e$b, for $^{92}$Kr and $^{94}$Kr, respectively, 
are rather at variance with the 
present theoretical values both in sign and magnitude. 
The calculated $Q(2^+_1)$ value for $^{96}$Kr is, however, 
within error bar of the experimental one \cite{albers2013}, 
$0.15\pm0.53$ $e$b. 
The considered $B(E3)$ values 
also become gradually larger for heavier Kr as 
the neutron number increases. This is a consequence 
of the fact that almost constant strength parameters are 
used for the Kr isotopes (cf. Fig.~\ref{fig:para}), 
and thus these transition rates increase only monotonically 
with the boson number.

For completeness, Table~\ref{tab:kr} shows some calculated 
$B(M1)$ and $B(E1)$ dipole transitions. 
An interesting feature is that the calculated $B(M1;3^-_2\to3^-_1)$ 
transition rate is large $\gtrsim0.01$ W.u., while 
the corresponding $B(E2)$ value  
$B(E2;3^-_2\to3^-_1)<1$ W.u. is small. 
Experimentally, strong $M1$ transitions from the $3^-_i$ 
($i=1$ or 2) excited state at $E_x\approx3.1$ MeV to the $3^-_1$ state 
have been suggested in some of the nearly spherical 
$N=52$ isotones, $^{92}$Zr and $^{94,96}$Mo \cite{scheck2010}, 
and have been identified as the isovector octupole excitation modes. 
As for the $B(E1)$ rates, the transitions from the higher-spin 
negative-parity to lower-spin positive-parity states 
are systematically larger than those 
in the opposite direction, i.e., from the lower-spin 
negative-parity to higher-spin positive-parity 
states. Similar tendency has been obtained in the previous 
$sdf$-IBM calculations in other mass regions \cite{zamfir2001,nomura2014}, 
and this staggering pattern in the $B(E1)$ rates can be partly 
attributed to the fact that the model space and the $E1$ transition 
operator do not include the effect of $p$-boson degree of freedom. 
At any rate, experimental data on the transition properties 
is so scarce for the neutron-rich Kr isotopes that an extensive 
assessments of the model description and of the quality of the 
$sdf$-IBM-2 wave functions remain to be done.

\begin{figure*}[ht]
\begin{center}
\begin{tabular}{cc}
\includegraphics[width=0.5\linewidth]{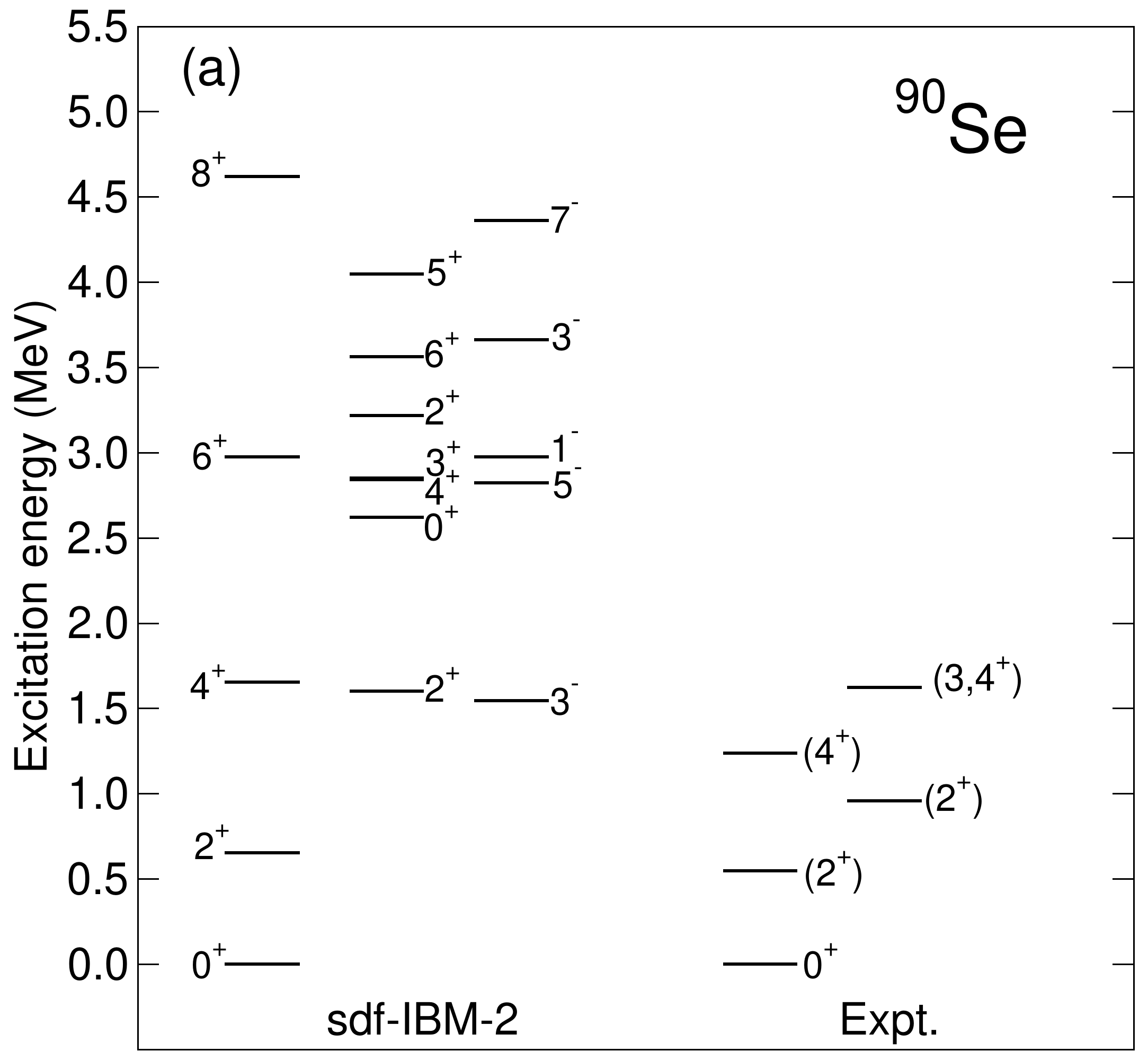} &
\includegraphics[width=0.5\linewidth]{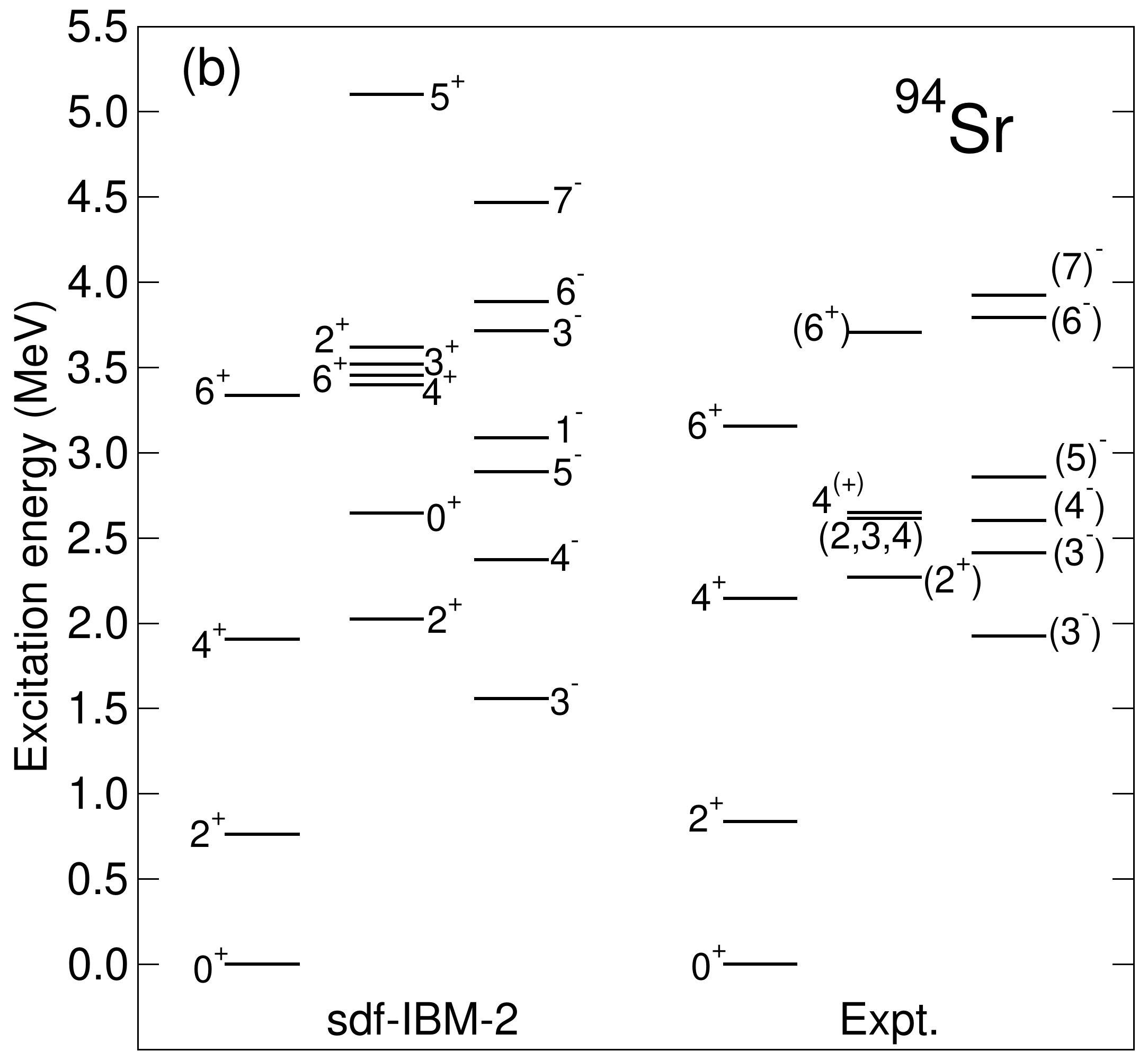}\\
\includegraphics[width=0.5\linewidth]{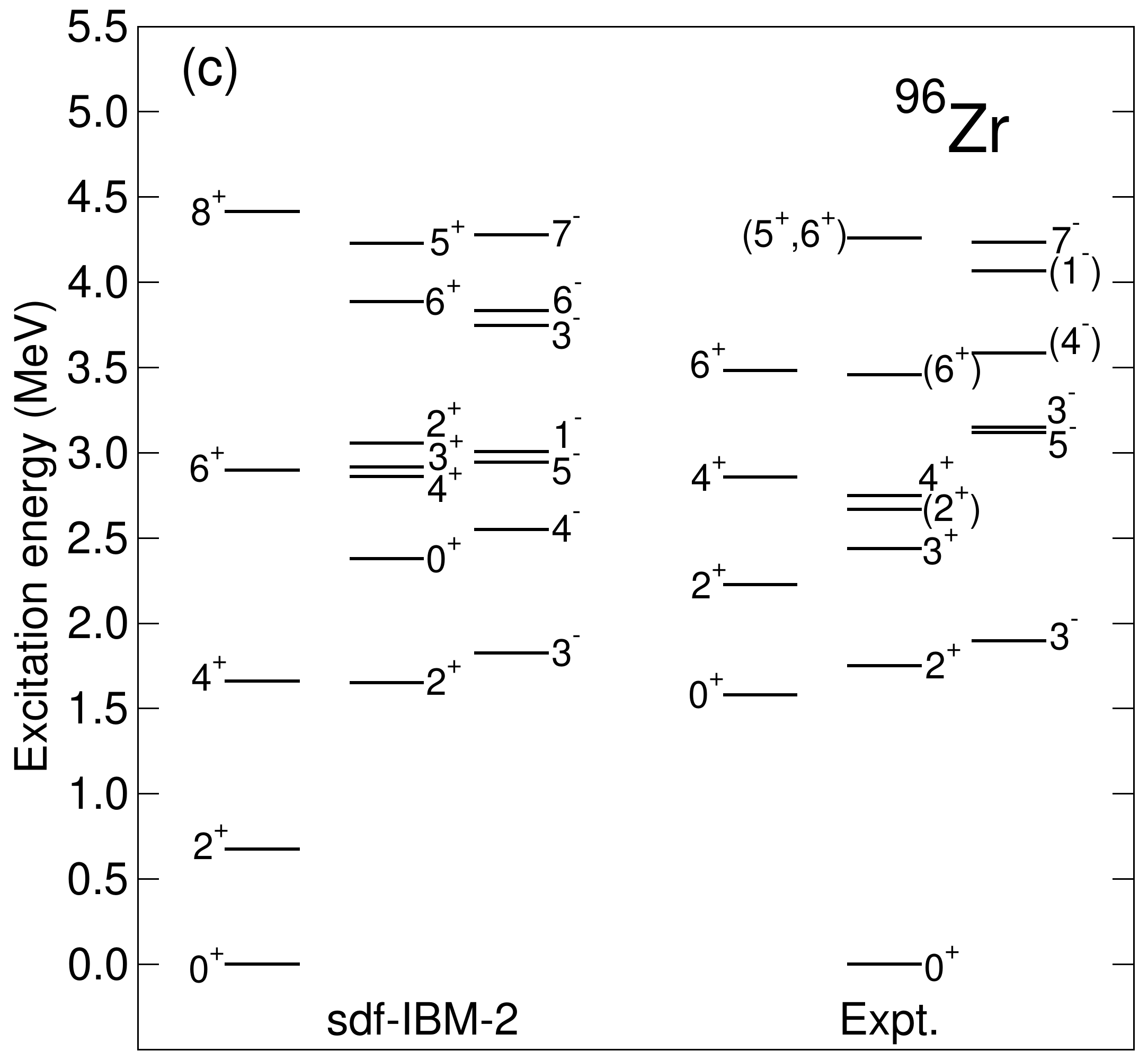} &
\includegraphics[width=0.5\linewidth]{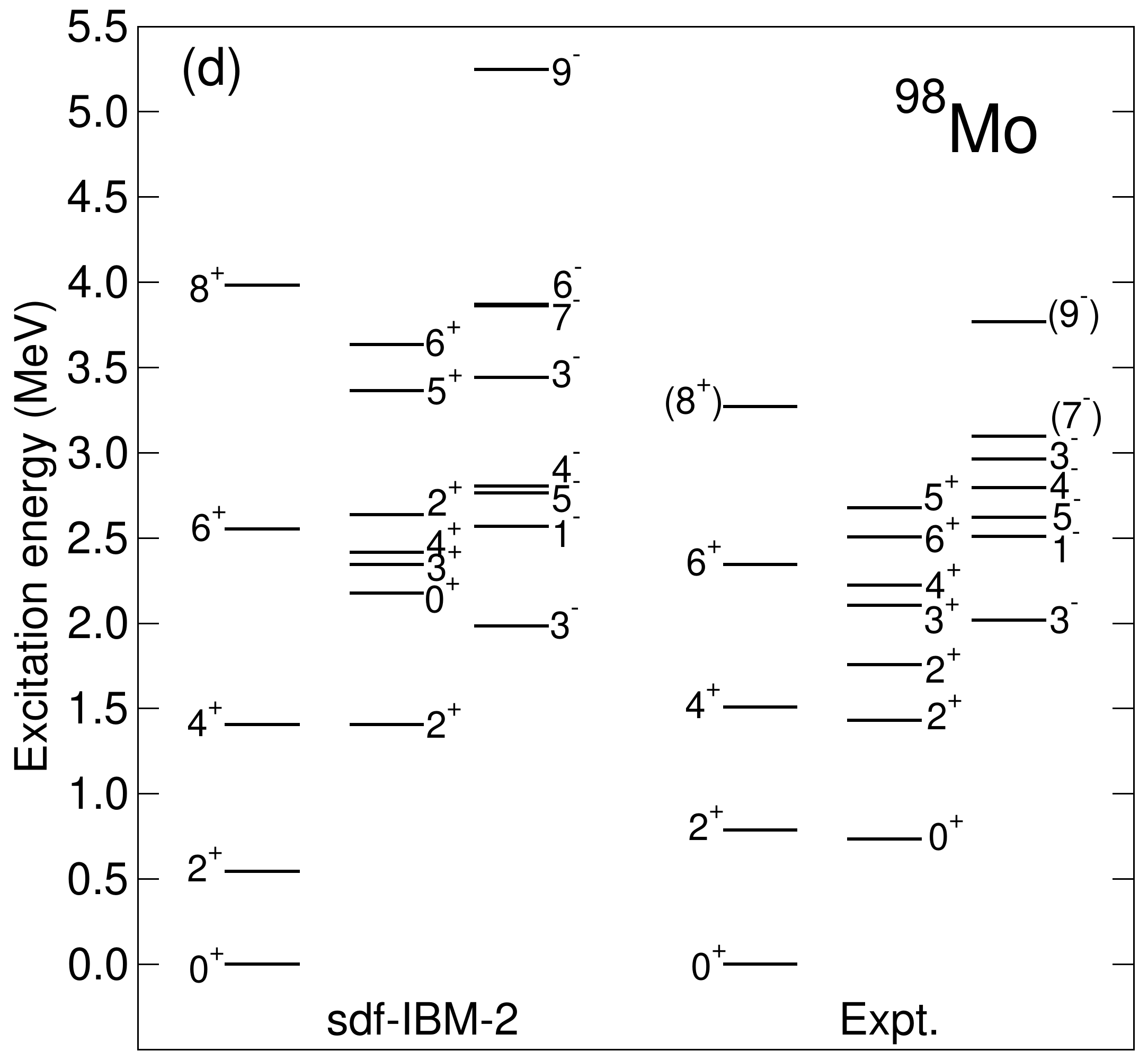}\\
\end{tabular}
\caption{Same as Fig.~\ref{fig:kr92}, but for the $N=56$ isotones 
(a) $^{90}$Se, (b) $^{94}$Sr, (c) $^{96}$Zr, and (d) $^{98}$Mo. 
The experimental data are taken from 
Refs.~\cite{chen2017} ($^{90}$Se), \cite{rzacaurban2009} ($^{94}$Sr), 
\cite{data} ($^{96}$Zr), and \cite{lalkovski2007,thomas2013} ($^{98}$Mo).}
\label{fig:n56}
\end{center}
\end{figure*}

\subsection{Low-energy spectra for $N=56$ isotones}

Figure~\ref{fig:n56} shows the low-energy level schemes for the 
$N=56$ isotones other than $^{92}$Kr, that is, $^{90}$Se, $^{94}$Sr, 
$^{96}$Zr, and $^{98}$Mo. 

The nucleus $^{90}$Se corresponds to both the neutron $N=56$ and 
proton $Z=34$ empirical octupole magic numbers. 
The SCMF result for this nucleus suggests that it is 
more or less soft in the octupole $\beta_3$ deformation 
(see, Fig.~\ref{fig:pesdft1}). 
The predicted low-energy spectrum in Fig.~\ref{fig:n56} for $^{90}$Se 
resembles the one for the neighboring isotone $^{92}$Kr. 
For instance,  the $4^+_1$, $2^+_2$, 
and $3^-_1$ states are nearly degenerate at $E_x\approx1.5$ MeV. 
In addition, the calculation predicts the $3^-_1$ excitation energy 
to be $E_x=1597$ keV, while, experimentally \cite{chen2017}, 
there is a level at $E_x=1600$ keV 
with tentatively assigned spin and parity $(3,4^+)$. 
As compared with the experimental data 
\cite{chen2017}, the predicted 
$K=0^+_1$ band is rather stretched in energy. 
Some calculated $B(E2)$, $B(M1)$, $B(E1)$, and 
$B(E3)$ transition rates and the $Q(I)$ and $\mu(I)$ moments are 
listed in Table~\ref{tab:se90sr94}. 
Note that the experimental data for the electromagnetic transition 
properties are not available for this nucleus. 

For $^{94}$Sr, an agreement between the $sdf$-IBM-2 and 
the experimental \cite{rzacaurban2009,data} low-energy spectra is 
generally satisfactory, particularly 
for the $K=0^+_1$ band, the bandhead energy 
of the quasi-$\gamma$ band, i.e., $2^+_2$, 
and many of the negative-parity states. 
Note that the $6^+_2$ state is calculated to be particularly 
low in energy, being close to the $6^+_1$ one. 
The approximate degeneracy arises probably because 
a large fraction of the $f$-boson 
components is admixed into its wave function. 
The calculation gives the low-lying $3^-_1$ state below the $4^+_1$ state, 
and also reproduces the even-spin negative-parity states $4^-$ and $6^-$. 
The predicted $3^-_2$ energy is, however, much higher than the 
experimental one. Note that the spin and parity of the $3^-_1$ 
and $3^-_2$ states are not firmly established experimentally. 
Predicted $B(E2)$, $B(M1)$, $B(E1)$, and 
$B(E3)$ transition rates and the $Q(I)$ and $\mu(I)$ moments for $^{94}$Sr 
are summarized in Table~\ref{tab:se90sr94}. 
The available experimental data \cite{data} are: 
$B(E2;2^+_1\to0^+_1)=8\pm4$ W.u., 
$B(E1;3^-_2\to2^+_1)=(2.0\pm0.7)\times10^{-5}$ W.u., 
and $B(E1;5^-_1\to4^+_1)=(1.5\pm0.7)\times10^{-5}$ W.u.

\begin{table}
\caption{
\label{tab:se90sr94}
Same as Table~\ref{tab:kr}, but for $^{90}$Se and $^{94}$Sr. 
}
 \begin{center}
 \begin{ruledtabular}
  \begin{tabular}{lcc}
 & $^{90}$Se & $^{94}$Sr \\
\hline
$B(E2;2^{+}_{1}\to0^{+}_{1})$ & 14 & 20 \\
$B(E2;2^{+}_{2}\to0^{+}_{1})$ & 0.29 & 0.68 \\
$B(E2;2^{+}_{2}\to2^{+}_{1})$ & 18 & 21 \\
$B(E2;4^{+}_{1}\to2^{+}_{1})$ & 19 & 29 \\
$B(E2;4^{+}_{2}\to2^{+}_{1})$ & 0.047 & 0.14 \\
$B(E2;4^{+}_{2}\to2^{+}_{2})$ & 11 & 17 \\
$B(E2;4^{+}_{2}\to4^{+}_{1})$ & 9.3 & 12 \\
$B(E2;3^{+}_{1}\to2^{+}_{1})$ & 0.24 & 0.86 \\
$B(E2;3^{+}_{1}\to2^{+}_{2})$ & 14 & 21 \\
$B(E2;3^{-}_{2}\to3^{-}_{1})$ & 0.015 & 0.54 \\
\hline
$B(M1;2^{+}_{2}\to2^{+}_{1})$ & 0.001 & 0.00039 \\
$B(M1;3^{+}_{1}\to2^{+}_{1})$ & 0.00015 & 0.00012 \\
$B(M1;3^{+}_{1}\to2^{+}_{2})$ & 0.0032 & 0.00037 \\
$B(M1;4^{+}_{2}\to4^{+}_{1})$ & 0.0025 & 0.0017 \\
$B(M1;3^{-}_{2}\to3^{-}_{1})$ & 0.014 & 0.017 \\
\hline
$B(E1;1^{-}_{1}\to0^{+}_{1})$ & $1.0\times10^{-4}$ & $7.6\times10^{-5}$ \\
$B(E1;3^{-}_{2}\to2^{+}_{1})$ & $1.7\times10^{-5}$ & $8.8\times10^{-5}$ \\
$B(E1;3^{-}_{1}\to2^{+}_{1})$ & $2.0\times10^{-4}$ & $1.4\times10^{-4}$ \\
$B(E1;3^{-}_{1}\to2^{+}_{2})$ & $5.6\times10^{-4}$ & $5.5\times10^{-4}$ \\
$B(E1;3^{-}_{1}\to4^{+}_{1})$ & $4.5\times10^{-7}$ & $2.2\times10^{-8}$ \\
$B(E1;5^{-}_{1}\to4^{+}_{1})$ & $5.3\times10^{-4}$ & $3.8\times10^{-4}$ \\
$B(E1;7^{-}_{1}\to6^{+}_{1})$ & $1.1\times10^{-3}$ & $6.9\times10^{-4}$ \\
\hline
$B(E3;3^{-}_{1}\to0^{+}_{1})$ & 17 & 20 \\
$B(E3;3^{-}_{1}\to2^{+}_{1})$ & 42 & 36 \\
$B(E3;3^{-}_{1}\to4^{+}_{1})$ & 12 & 6.6 \\
$B(E3;5^{-}_{1}\to2^{+}_{1})$ & 12 & 13 \\
$B(E3;5^{-}_{1}\to4^{+}_{1})$ & 33 & 31 \\
$B(E3;7^{-}_{1}\to4^{+}_{1})$ & 11 & 12 \\
$B(E3;7^{-}_{1}\to6^{+}_{1})$ & 30 & 19 \\
\hline
$Q(2^{+}_{1})$ & 0.087 & 0.24 \\
$\mu(2^{+}_{1})$ & 0.31 & 0.39 \\
  \end{tabular}
 \end{ruledtabular}
 \end{center}
\end{table}

The $^{96}$Zr nucleus corresponds to the doubly subshell closure 
of $N=56$ and $Z=40$. As already shown in Fig.~\ref{fig:pos}(d), 
the $sdf$-IBM-2 only produces a collective band structure for 
the positive-parity yrast band, 
whereas experimentally \cite{kremer2016}, the first excited state 
is the $0^+_2$ state at $E_x=1582$ keV and the high-lying $2^+_1$ state 
is found at 1750 keV. 
It was shown by the Monte Carlo Shell Model calculation 
\cite{togashi2016,kremer2016} that this $0^+_2$ is likely 
to arise from the intruder deformed configuration. The $0^+_2$ state 
obtained by the present $sdf$-IBM-2 calculation is 
quite different in nature,  
since in the present framework the intruder excitations and 
the subsequent configuration mixing between the normal and intruder 
states are not taken into account. 
However, the mapped $sdf$-IBM-2 
calculation generally provides a reasonable description 
of the observed negative-parity levels.

The electromagnetic transition properties of the low-lying states 
in $^{96}$Zr computed by the mapped $sdf$-IBM-2 Hamiltonian are shown 
in Table~\ref{tab:zr96}, together with the experimental data 
\cite{data,kremer2016}. 
Some disagreements between the calculated and experimental 
$E2$ transition properties for $^{96}$Zr, especially those related to the 
positive-parity states, are mainly due to the lack of the intruder 
configuration in the model. 
The predicted $B(E3)$ values for the transitions 
from the odd-spin negative-parity states are noticeably large, 
particularly, for the $\Delta I=1$ $E3$ transitions.

\begin{table}
\caption{\label{tab:zr96}
$B(E2)$, $B(M1)$, $B(E1)$, and $B(E3)$ transition rates (in W.u.), and the 
electric quadrupole $Q(I)$ (in $e$b), and magnetic 
dipole $\mu(I)$ (in $\mu_N$) moments, predicted by the mapped $sdf$-IBM-2 
for $^{96}$Zr in comparison with the available experimental data 
\cite{data,kremer2016}. 
}
 \begin{center}
 \begin{ruledtabular}
  \begin{tabular}{lcc}
 & $sdf$-IBM-2 & Experiment \\
\hline
$B(E2;0^{+}_{3}\to2^{+}_{2})$ & 11 & $34\pm9$ \\
$B(E2;2^{+}_{1}\to0^{+}_{1})$ & 21 & $2.3\pm0.3$ \\
$B(E2;2^{+}_{2}\to2^{+}_{1})$ & 26 & $>0.16$ \\
$B(E2;2^{+}_{2}\to0^{+}_{2})$ & 4.5 & $36\pm11$ \\
$B(E2;2^{+}_{2}\to0^{+}_{1})$ & 0.31 & $0.26\pm0.08$ \\
$B(E2;2^{+}_{3}\to2^{+}_{1})$ & 0.079 & $(5.0\pm0.7)E+1$ \\
$B(E2;3^{+}_{1}\to2^{+}_{1})$ & 0.43 & $0.1^{+0.3}_{-0.1}$ \\
$B(E2;4^{+}_{1}\to2^{+}_{2})$ & 0.0016 & $56^{+20}_{-44}$ \\
$B(E2;4^{+}_{1}\to2^{+}_{1})$ & 29 & $16^{+5}_{-13}$ \\
$B(E2;4^{+}_{2}\to2^{+}_{1})$ & 0.068 & ${}$ \\
$B(E2;4^{+}_{2}\to2^{+}_{2})$ & 17 & $<1.6$ \\
$B(E2;4^{+}_{3}\to2^{+}_{1})$ & 0.00042 & $0.4^{+0.4}_{-0.4}$ \\
$B(E2;3^{-}_{2}\to3^{-}_{1})$ & 1.7 & $<4.2$ \\
$B(E2;5^{-}_{1}\to3^{-}_{1})$ & 10 & $14^{+5}_{-14}$ \\
\hline
$B(M1;2^{+}_{2}\to2^{+}_{1})$ & 0.00082 & $0.14\pm0.05$ \\
$B(M1;2^{+}_{3}\to2^{+}_{1})$ & 0.022 & $0.04\pm0.06$ \\
$B(M1;3^{+}_{1}\to2^{+}_{1})$ & $9.0\times10^{-5}$ & $0.18^{+0.05}_{-0.09}$ \\
$B(M1;3^{-}_{2}\to3^{-}_{1})$ & 0.011 & $<0.0027$ \\
\hline
$B(E1;1^{-}_{1}\to0^{+}_{1})$ & $1.1\times10^{-4}$ & ${}$ \\
$B(E1;2^{+}_{2}\to3^{-}_{1})$ & $7.1\times10^{-4}$ & $(2.8\pm0.9)\times10^{-3}$ \\
$B(E1;2^{+}_{3}\to3^{-}_{1})$ & $1.4\times10^{-5}$ & $(0.0007^{+0.0004}_{-0.0007})$ \\
$B(E1;4^{+}_{1}\to3^{-}_{1})$ & $5.1\times10^{-7}$ & $(7.0^{+0.3}_{-0.6})E-5$ \\
$B(E1;4^{+}_{3}\to3^{-}_{1})$ & $8.3\times10^{-7}$ & $<0.00010$ \\
$B(E1;3^{-}_{1}\to2^{+}_{1})$ & $2.1\times10^{-4}$ & $0.00123\pm0.00010$ \\
$B(E1;3^{-}_{2}\to3^{+}_{1})$ & $1.8\times10^{-4}$ & $<0.00100$ \\
$B(E1;5^{-}_{1}\to4^{+}_{1})$ & $5.3\times10^{-4}$ & ${}$ \\
$B(E1;7^{-}_{1}\to6^{+}_{1})$ & $1.0\times10^{-3}$ & ${}$ \\
\hline
$B(E3;3^{-}_{1}\to0^{+}_{1})$ & 21 & $57\pm4$ \\
$B(E3;3^{-}_{1}\to0^{+}_{2})$ & 0.069 & ${}$ \\
$B(E3;3^{-}_{1}\to2^{+}_{1})$ & 38 & ${}$ \\
$B(E3;5^{-}_{1}\to2^{+}_{1})$ & 17 & ${}$ \\
$B(E3;5^{-}_{1}\to4^{+}_{1})$ & 32 & ${}$ \\
$B(E3;7^{-}_{1}\to4^{+}_{1})$ & 15 & ${}$ \\
$B(E3;7^{-}_{1}\to6^{+}_{1})$ & 28 & ${}$ \\
\hline
$Q(2^{+}_{1})$ & 0.17 & ${}$ \\
$Q(3^{-}_{1})$ & $-0.42$ & $+2.9\pm0.5$ \\
$\mu(2^{+}_{1})$ & 0.4 & $+0.06\pm0.14$ \\
  \end{tabular}
 \end{ruledtabular}
 \end{center}
\end{table}

The predicted low-energy spectrum for $^{98}$Mo, 
shown in Fig.~\ref{fig:n56}(d), is basically similar to the 
one for $^{96}$Zr. Comparing with the experimental data 
\cite{data,lalkovski2007,thomas2013}, the predicted 
$K=0^+_1$ band suggests a slightly stronger 
quadrupole collectivity with the ratio $R_{4/2}=2.58$. 
The calculated $0^+_2$ excitation energy  $E_x(0^+_2)=2178$ keV is much 
higher than the experimental counterpart $E_x(0^+_2)=735$ keV. 
Experimentally, the $0^+_2$ state is also the first excited 
state of $^{98}$Mo. 
As already remarked, the $0^+_2$ level could be lowered 
by the inclusion of the 
configuration mixing between the normal and intruder states. 
A previous mapped IBM-2 configuration-mixing
calculation, which is based on a Skyrme-type EDF, 
showed \cite{thomas2013} that the coexistence of a nearly 
spherical prolate and $\gamma$-soft prolate SCMF minima 
is present in $^{98}$Mo and that the $0^+_2$ state is characterized 
by the strong mixing between the nearly spherical normal 
configuration and the deformed intruder configuration 
arising from the proton $2p-2h$ excitation across the $Z=40$ 
subshell closure. 
On the other hand, the observed negative-parity states 
up to the $4^-_1$ state are reasonably reproduced. 
Some relevant electromagnetic properties for $^{98}$Mo 
resulting from the mapped $sdf$-IBM-2 calculation are shown 
in Table~\ref{tab:mo98} in comparison to the available data 
\cite{data,thomas2013}. 
The agreement with the experimental data is fair. 
The calculation suggests strong $\Delta I=3$ $E3$ 
transitions of the odd-spin negative-parity to the 
even-spin positive-parity yrast states.

\begin{table}
\caption{
\label{tab:mo98}
Same as Table~\ref{tab:zr96}, but for $^{98}$Mo.
The experimental data are taken from 
Refs.~\cite{data,lalkovski2007,thomas2013}. 
}
 \begin{center}
 \begin{ruledtabular}
  \begin{tabular}{lcc}
 & $sdf$-IBM-2 & Experiment \\
\hline
$B(E2;2^{+}_{1}\to0^{+}_{1})$ & 18 & $20.1\pm0.4$ \\
$B(E2;2^{+}_{1}\to0^{+}_{2})$ & 0.35 & $9.7^{+1.0}_{-2.5}$ \\
$B(E2;2^{+}_{2}\to0^{+}_{1})$ & 0.36 & $1.02^{+0.15}_{-0.12}$ \\
$B(E2;2^{+}_{2}\to0^{+}_{2})$ & 4.3 & $2.3^{+0.5}_{-0.4}$ \\
$B(E2;2^{+}_{2}\to2^{+}_{1})$ & 17 & $48^{+9}_{-8}$ \\
$B(E2;2^{+}_{3}\to4^{+}_{1})$ & 0.037 & $14\pm4$ \\
$B(E2;2^{+}_{3}\to2^{+}_{2})$ & 0.11 & $<22$ \\
$B(E2;2^{+}_{3}\to2^{+}_{1})$ & 0.19 & $3.0\pm0.7$ \\
$B(E2;2^{+}_{3}\to0^{+}_{2})$ & 0.36 & $7.5^{+0.6}_{-0.5}$ \\
$B(E2;2^{+}_{3}\to0^{+}_{1})$ & 0.0033 & $0.032^{+0.007}_{-0.006}$ \\
$B(E2;2^{+}_{4}\to2^{+}_{1})$ & 0.00031 & $>0.49$ \\
$B(E2;4^{+}_{1}\to2^{+}_{1})$ & 25 & $42.3^{+0.9}_{-0.8}$ \\
$B(E2;4^{+}_{1}\to2^{+}_{2})$ & 0.023 & $15.2^{+3.3}_{-3.0}$ \\
$B(E2;6^{+}_{1}\to4^{+}_{1})$ & 26 & $10.1\pm0.4$ \\
$B(E2;3^{-}_{2}\to3^{-}_{1})$ & 0.5 & ${}$ \\
\hline
$B(M1;2^{+}_{2}\to2^{+}_{1})$ & 0.0021 & $0.0073^{+0.0023}_{-0.0017}$ \\
$B(M1;2^{+}_{3}\to2^{+}_{2})$ & 0.00013 & $0.0157^{+0.0027}_{-0.0034}$ \\
$B(M1;2^{+}_{3}\to2^{+}_{1})$ & 0.023 & $0.0032^{+0.0008}_{-0.0007}$ \\
$B(M1;2^{+}_{4}\to2^{+}_{1})$ & 0.0016 & $>0.019$ \\
$B(M1;2^{+}_{5}\to2^{+}_{2})$ & 0.0019 & $>0.054$ \\
$B(M1;3^{-}_{2}\to3^{-}_{1})$ & 0.013 & ${}$ \\
\hline
$B(E1;1^{-}_{1}\to0^{+}_{1})$ & $2.5\times10^{-4}$ & ${}$ \\
$B(E1;3^{-}_{1}\to2^{+}_{3})$ & $1.9\times10^{-5}$ & $(4.9^{+0.9}_{-0.7})\times10^{-5}$ \\
$B(E1;3^{-}_{1}\to4^{+}_{1})$ & $4.0\times10^{-6}$ & $(1.02^{+0.31}_{-0.24})\times10^{-6}$ \\
$B(E1;3^{-}_{1}\to2^{+}_{2})$ & $1.8\times10^{-4}$ & $<5.7\times10^{-8}$ \\
$B(E1;3^{-}_{1}\to2^{+}_{1})$ & $6.8\times10^{-4}$ & $(1.76^{+0.28}_{-0.22})\times10^{-6}$ \\
$B(E1;3^{-}_{2}\to2^{+}_{1})$ & $3.8\times10^{-5}$ & ${}$ \\
$B(E1;3^{-}_{2}\to2^{+}_{2})$ & $1.5\times10^{-4}$ & ${}$ \\
$B(E1;5^{-}_{1}\to4^{+}_{1})$ & $1.3\times10^{-3}$ & ${}$ \\
$B(E1;7^{-}_{1}\to6^{+}_{1})$ & $1.9\times10^{-3}$ & ${}$ \\
\hline
$B(E3;3^{-}_{1}\to0^{+}_{2})$ & 0.038 & $<58$ \\
$B(E3;3^{-}_{1}\to0^{+}_{1})$ & 21 & $30^{+7}_{-5}$ \\
$B(E3;3^{-}_{1}\to2^{+}_{1})$ & 25 & ${}$ \\
$B(E3;3^{-}_{1}\to2^{+}_{2})$ & 5.2 & ${}$ \\
$B(E3;3^{-}_{2}\to2^{+}_{1})$ & 0.17 & ${}$ \\
$B(E3;3^{-}_{2}\to2^{+}_{2})$ & 1.8 & ${}$ \\
$B(E3;5^{-}_{1}\to2^{+}_{1})$ & 26 & ${}$ \\
$B(E3;5^{-}_{1}\to4^{+}_{1})$ & 19 & ${}$ \\
$B(E3;7^{-}_{1}\to4^{+}_{1})$ & 25 & ${}$ \\
$B(E3;7^{-}_{1}\to6^{+}_{1})$ & 18 & ${}$ \\
\hline
$Q(2^{+}_{1})$ & $-0.25$ & $-0.26\pm0.09$ \\
$\mu(2^{+}_{1})$ & 0.31 & $+0.97\pm0.06$ \\
  \end{tabular}
 \end{ruledtabular}
 \end{center}
\end{table}

\section{Concluding remarks\label{sec:summary}}

Octupole correlations and the related spectroscopic properties of 
the neutron-rich $A\approx100$ nuclei near the empirical 
octupole magic numbers $Z=34$ and $N=56$ have been investigated 
based on the nuclear density functional theory and the mapped 
$sdf$-IBM-2 framework. 
The axially-symmetric quadrupole and octupole constrained 
SCMF calculations using the relativistic DD-PC1 EDF and the 
separable pairing force of finite range have been carried out 
to provide the potential energy surfaces for the considered 
even-even nuclei $^{86-94}$Se, $^{88-96}$Kr, $^{90-98}$Sr, 
$^{92-100}$Zr, and $^{94-102}$Mo as functions of the $\beta_2$ and 
$\beta_3$ deformations. At the SCMF level, no nonzero $\beta_3$ 
minimum was obtained on the $\beta_2-\beta_3$ energy surfaces, 
whereas the potential becomes soft in the $\beta_3$ deformation 
around the neutron number $N=56$ for 
each of the considered isotopic chains. 
The excitation spectra for the low-lying states with both parities 
and the electromagnetic transition rates 
have been computed by the diagonalization of the 
$sdf$-IBM-2 Hamiltonian with the strength parameters 
determined by mapping the SCMF energy surface onto the 
expectation value of the Hamiltonian in the boson condensate state. 

Evolution of the predicted positive-parity yrast spectra has 
indicated a smooth onset of quadrupole collectivity in the Se, Kr, 
and Mo isotopes, and an abrupt nuclear 
structural change at $N\approx60$ in the Sr and Zr chains, as 
empirically suggested. 
The present theoretical analysis put much emphasis on the 
description of the negative-parity states. 
The predicted negative-parity odd-spin states show a 
parabolic dependence on $N$ centered around $N=56$ at which 
the corresponding $\beta_2-\beta_3$ energy surface becomes 
softest along the $\beta_3$ direction. 
The lowest negative-parity state $3^-_1$ has been predicted to be 
typically at the excitation energy $E_x\approx1.5-2.5$ MeV, 
in a reasonable agreement with experiment. 
The detailed comparisons have been made 
between the calculated and experimental 
low-lying energy levels and transition properties for the $N=56$ isotones 
and some of the neutron-rich Kr nuclei that are of interest for 
recent measurements. The present calculation has produced 
finite $E3$ transitions from the odd-spin negative-parity states 
to the ground-state band. These findings indicate the relevance of 
the octupole degrees of freedom in the description of the low-energy 
negative-parity states in the neutron-rich nuclei near the 
octupole magic numbers $N=56$ and $Z=34$ within the framework 
of the $sdf$-IBM-2. 

For a more complete description of the 
low-energy states with both parities and their spectroscopy in this 
mass region, some extensions of the model will be required. 
In particular, the $sdf$-IBM-2 Hamiltonian has been here determined 
by the constrained SCMF calculations based only on the 
axially-symmetric shape degrees of freedom. On the other hand, 
the studied neutron-rich nuclei should have a more complex low-lying 
structure characterized, e.g., by the nonaxial deformation and 
shape coexistence. Indeed, it has been revealed that 
the mapped $sdf$-IBM-2 employed 
in this paper is not able to describe the low-lying $0^+_2$ state 
and the band built on it for the $N>56$ nuclei. 
The model would need to be extended in such a way that it simultaneously 
handles the octupole boson degrees of freedom, the triaxial deformation, 
and the intruder states and configuration mixing. 
Work along this direction is in progress and will be reported elsewhere.

\acknowledgments
This work is financed within the Tenure Track Pilot Programme of 
the Croatian Science Foundation and the \'Ecole Polytechnique 
F\'ed\'erale de Lausanne, and the Project TTP-2018-07-3554 Exotic 
Nuclear Structure and Dynamics, with funds of the Croatian-Swiss 
Research Programme.

\bibliography{refs}

\begin{thebibliography}{96}%
\makeatletter
\providecommand \@ifxundefined [1]{%
 \@ifx{#1\undefined}
}%
\providecommand \@ifnum [1]{%
 \ifnum #1\expandafter \@firstoftwo
 \else \expandafter \@secondoftwo
 \fi
}%
\providecommand \@ifx [1]{%
 \ifx #1\expandafter \@firstoftwo
 \else \expandafter \@secondoftwo
 \fi
}%
\providecommand \natexlab [1]{#1}%
\providecommand \enquote  [1]{``#1''}%
\providecommand \bibnamefont  [1]{#1}%
\providecommand \bibfnamefont [1]{#1}%
\providecommand \citenamefont [1]{#1}%
\providecommand \href@noop [0]{\@secondoftwo}%
\providecommand \href [0]{\begingroup \@sanitize@url \@href}%
\providecommand \@href[1]{\@@startlink{#1}\@@href}%
\providecommand \@@href[1]{\endgroup#1\@@endlink}%
\providecommand \@sanitize@url [0]{\catcode `\\12\catcode `\$12\catcode
  `\&12\catcode `\#12\catcode `\^12\catcode `\_12\catcode `\%12\relax}%
\providecommand \@@startlink[1]{}%
\providecommand \@@endlink[0]{}%
\providecommand \url  [0]{\begingroup\@sanitize@url \@url }%
\providecommand \@url [1]{\endgroup\@href {#1}{\urlprefix }}%
\providecommand \urlprefix  [0]{URL }%
\providecommand \Eprint [0]{\href }%
\providecommand \doibase [0]{https://doi.org/}%
\providecommand \selectlanguage [0]{\@gobble}%
\providecommand \bibinfo  [0]{\@secondoftwo}%
\providecommand \bibfield  [0]{\@secondoftwo}%
\providecommand \translation [1]{[#1]}%
\providecommand \BibitemOpen [0]{}%
\providecommand \bibitemStop [0]{}%
\providecommand \bibitemNoStop [0]{.\EOS\space}%
\providecommand \EOS [0]{\spacefactor3000\relax}%
\providecommand \BibitemShut  [1]{\csname bibitem#1\endcsname}%
\let\auto@bib@innerbib\@empty
\bibitem [{\citenamefont {Heyde}\ and\ \citenamefont {Wood}(2011)}]{heyde2011}%
  \BibitemOpen
  \bibfield  {author} {\bibinfo {author} {\bibfnamefont {K.}~\bibnamefont
  {Heyde}}\ and\ \bibinfo {author} {\bibfnamefont {J.~L.}\ \bibnamefont
  {Wood}},\ }\href {https://doi.org/10.1103/RevModPhys.83.1467} {\bibfield
  {journal} {\bibinfo  {journal} {Rev. Mod. Phys.}\ }\textbf {\bibinfo {volume}
  {83}},\ \bibinfo {pages} {1467} (\bibinfo {year} {2011})}\BibitemShut
  {NoStop}%
\bibitem [{\citenamefont {Cejnar}\ \emph {et~al.}(2010)\citenamefont {Cejnar},
  \citenamefont {Jolie},\ and\ \citenamefont {Casten}}]{cejnar2010}%
  \BibitemOpen
  \bibfield  {author} {\bibinfo {author} {\bibfnamefont {P.}~\bibnamefont
  {Cejnar}}, \bibinfo {author} {\bibfnamefont {J.}~\bibnamefont {Jolie}},\ and\
  \bibinfo {author} {\bibfnamefont {R.~F.}\ \bibnamefont {Casten}},\ }\href
  {https://doi.org/10.1103/RevModPhys.82.2155} {\bibfield  {journal} {\bibinfo
  {journal} {Rev. Mod. Phys.}\ }\textbf {\bibinfo {volume} {82}},\ \bibinfo
  {pages} {2155} (\bibinfo {year} {2010})}\BibitemShut {NoStop}%
\bibitem [{\citenamefont {Butler}\ and\ \citenamefont
  {Nazarewicz}(1996)}]{butler1996}%
  \BibitemOpen
  \bibfield  {author} {\bibinfo {author} {\bibfnamefont {P.~A.}\ \bibnamefont
  {Butler}}\ and\ \bibinfo {author} {\bibfnamefont {W.}~\bibnamefont
  {Nazarewicz}},\ }\href {https://doi.org/10.1103/RevModPhys.68.349} {\bibfield
   {journal} {\bibinfo  {journal} {Rev. Mod. Phys.}\ }\textbf {\bibinfo
  {volume} {68}},\ \bibinfo {pages} {349} (\bibinfo {year} {1996})}\BibitemShut
  {NoStop}%
\bibitem [{\citenamefont {Butler}(2016)}]{butler2016}%
  \BibitemOpen
  \bibfield  {author} {\bibinfo {author} {\bibfnamefont {P.~A.}\ \bibnamefont
  {Butler}},\ }\href {https://doi.org/10.1088/0954-3899/43/7/073002} {\bibfield
   {journal} {\bibinfo  {journal} {J. Phys. G: Nucl. Part. Phys.}\ }\textbf
  {\bibinfo {volume} {43}},\ \bibinfo {pages} {073002} (\bibinfo {year}
  {2016})}\BibitemShut {NoStop}%
\bibitem [{\citenamefont {Gaffney}\ \emph {et~al.}(2013)\citenamefont
  {Gaffney}, \citenamefont {Butler}, \citenamefont {Scheck}, \citenamefont
  {Hayes}, \citenamefont {Wenander}, \citenamefont {Albers}, \citenamefont
  {Bastin}, \citenamefont {Bauer}, \citenamefont {Blazhev}, \citenamefont
  {B\"onig}, \citenamefont {Bree}, \citenamefont {Cederk\"all}, \citenamefont
  {Chupp}, \citenamefont {Cline}, \citenamefont {Cocolios}, \citenamefont
  {Davinson}, \citenamefont {Witte}, \citenamefont {Diriken}, \citenamefont
  {Grahn}, \citenamefont {Herzan}, \citenamefont {Huyse}, \citenamefont
  {Jenkins}, \citenamefont {Joss}, \citenamefont {Kesteloot}, \citenamefont
  {Konki}, \citenamefont {Kowalczyk}, \citenamefont {Kröll}, \citenamefont
  {Kwan}, \citenamefont {Lutter}, \citenamefont {Moschner}, \citenamefont
  {Napiorkowski}, \citenamefont {Pakarinen}, \citenamefont {Pfeiffer},
  \citenamefont {Radeck}, \citenamefont {Reiter}, \citenamefont {Reynders},
  \citenamefont {Rigby}, \citenamefont {Robledo}, \citenamefont {Rudigier},
  \citenamefont {Sambi}, \citenamefont {Seidlitz}, \citenamefont {Siebeck},
  \citenamefont {Stora}, \citenamefont {Thoele}, \citenamefont {Duppen},
  \citenamefont {Vermeulen}, \citenamefont {von Schmid}, \citenamefont
  {Voulot}, \citenamefont {Warr}, \citenamefont {Wimmer}, \citenamefont
  {Wrzosek-Lipska}, \citenamefont {Wu},\ and\ \citenamefont
  {Zielinska}}]{gaffney2013}%
  \BibitemOpen
  \bibfield  {author} {\bibinfo {author} {\bibfnamefont {L.~P.}\ \bibnamefont
  {Gaffney}}, \bibinfo {author} {\bibfnamefont {P.~A.}\ \bibnamefont {Butler}},
  \bibinfo {author} {\bibfnamefont {M.}~\bibnamefont {Scheck}}, \bibinfo
  {author} {\bibfnamefont {A.~B.}\ \bibnamefont {Hayes}}, \bibinfo {author}
  {\bibfnamefont {F.}~\bibnamefont {Wenander}}, \bibinfo {author}
  {\bibfnamefont {M.}~\bibnamefont {Albers}}, \bibinfo {author} {\bibfnamefont
  {B.}~\bibnamefont {Bastin}}, \bibinfo {author} {\bibfnamefont
  {C.}~\bibnamefont {Bauer}}, \bibinfo {author} {\bibfnamefont
  {A.}~\bibnamefont {Blazhev}}, \bibinfo {author} {\bibfnamefont
  {S.}~\bibnamefont {B\"onig}}, \bibinfo {author} {\bibfnamefont
  {N.}~\bibnamefont {Bree}}, \bibinfo {author} {\bibfnamefont {J.}~\bibnamefont
  {Cederk\"all}}, \bibinfo {author} {\bibfnamefont {T.}~\bibnamefont {Chupp}},
  \bibinfo {author} {\bibfnamefont {D.}~\bibnamefont {Cline}}, \bibinfo
  {author} {\bibfnamefont {T.~E.}\ \bibnamefont {Cocolios}}, \bibinfo {author}
  {\bibfnamefont {T.}~\bibnamefont {Davinson}}, \bibinfo {author}
  {\bibfnamefont {H.~D.}\ \bibnamefont {Witte}}, \bibinfo {author}
  {\bibfnamefont {J.}~\bibnamefont {Diriken}}, \bibinfo {author} {\bibfnamefont
  {T.}~\bibnamefont {Grahn}}, \bibinfo {author} {\bibfnamefont
  {A.}~\bibnamefont {Herzan}}, \bibinfo {author} {\bibfnamefont
  {M.}~\bibnamefont {Huyse}}, \bibinfo {author} {\bibfnamefont {D.~G.}\
  \bibnamefont {Jenkins}}, \bibinfo {author} {\bibfnamefont {D.~T.}\
  \bibnamefont {Joss}}, \bibinfo {author} {\bibfnamefont {N.}~\bibnamefont
  {Kesteloot}}, \bibinfo {author} {\bibfnamefont {J.}~\bibnamefont {Konki}},
  \bibinfo {author} {\bibfnamefont {M.}~\bibnamefont {Kowalczyk}}, \bibinfo
  {author} {\bibfnamefont {T.}~\bibnamefont {Kröll}}, \bibinfo {author}
  {\bibfnamefont {E.}~\bibnamefont {Kwan}}, \bibinfo {author} {\bibfnamefont
  {R.}~\bibnamefont {Lutter}}, \bibinfo {author} {\bibfnamefont
  {K.}~\bibnamefont {Moschner}}, \bibinfo {author} {\bibfnamefont
  {P.}~\bibnamefont {Napiorkowski}}, \bibinfo {author} {\bibfnamefont
  {J.}~\bibnamefont {Pakarinen}}, \bibinfo {author} {\bibfnamefont
  {M.}~\bibnamefont {Pfeiffer}}, \bibinfo {author} {\bibfnamefont
  {D.}~\bibnamefont {Radeck}}, \bibinfo {author} {\bibfnamefont
  {P.}~\bibnamefont {Reiter}}, \bibinfo {author} {\bibfnamefont
  {K.}~\bibnamefont {Reynders}}, \bibinfo {author} {\bibfnamefont {S.~V.}\
  \bibnamefont {Rigby}}, \bibinfo {author} {\bibfnamefont {L.~M.}\ \bibnamefont
  {Robledo}}, \bibinfo {author} {\bibfnamefont {M.}~\bibnamefont {Rudigier}},
  \bibinfo {author} {\bibfnamefont {S.}~\bibnamefont {Sambi}}, \bibinfo
  {author} {\bibfnamefont {M.}~\bibnamefont {Seidlitz}}, \bibinfo {author}
  {\bibfnamefont {B.}~\bibnamefont {Siebeck}}, \bibinfo {author} {\bibfnamefont
  {T.}~\bibnamefont {Stora}}, \bibinfo {author} {\bibfnamefont
  {P.}~\bibnamefont {Thoele}}, \bibinfo {author} {\bibfnamefont {P.~V.}\
  \bibnamefont {Duppen}}, \bibinfo {author} {\bibfnamefont {M.~J.}\
  \bibnamefont {Vermeulen}}, \bibinfo {author} {\bibfnamefont {M.}~\bibnamefont
  {von Schmid}}, \bibinfo {author} {\bibfnamefont {D.}~\bibnamefont {Voulot}},
  \bibinfo {author} {\bibfnamefont {N.}~\bibnamefont {Warr}}, \bibinfo {author}
  {\bibfnamefont {K.}~\bibnamefont {Wimmer}}, \bibinfo {author} {\bibfnamefont
  {K.}~\bibnamefont {Wrzosek-Lipska}}, \bibinfo {author} {\bibfnamefont
  {C.~Y.}\ \bibnamefont {Wu}},\ and\ \bibinfo {author} {\bibfnamefont
  {M.}~\bibnamefont {Zielinska}},\ }\href {https://doi.org/10.1038/nature12073}
  {\bibfield  {journal} {\bibinfo  {journal} {Nature (London)}\ }\textbf
  {\bibinfo {volume} {497}},\ \bibinfo {pages} {199} (\bibinfo {year}
  {2013})}\BibitemShut {NoStop}%
\bibitem [{\citenamefont {Chishti}\ \emph {et~al.}(2020)\citenamefont
  {Chishti}, \citenamefont {O'Donnell}, \citenamefont {Battaglia},
  \citenamefont {Bowry}, \citenamefont {Jaroszynski}, \citenamefont {Singh},
  \citenamefont {Scheck}, \citenamefont {Spagnoletti},\ and\ \citenamefont
  {Smith}}]{chishti2020}%
  \BibitemOpen
  \bibfield  {author} {\bibinfo {author} {\bibfnamefont {M.~M.~R.}\
  \bibnamefont {Chishti}}, \bibinfo {author} {\bibfnamefont {D.}~\bibnamefont
  {O'Donnell}}, \bibinfo {author} {\bibfnamefont {G.}~\bibnamefont
  {Battaglia}}, \bibinfo {author} {\bibfnamefont {M.}~\bibnamefont {Bowry}},
  \bibinfo {author} {\bibfnamefont {D.~A.}\ \bibnamefont {Jaroszynski}},
  \bibinfo {author} {\bibfnamefont {B.~S.~N.}\ \bibnamefont {Singh}}, \bibinfo
  {author} {\bibfnamefont {M.}~\bibnamefont {Scheck}}, \bibinfo {author}
  {\bibfnamefont {P.}~\bibnamefont {Spagnoletti}},\ and\ \bibinfo {author}
  {\bibfnamefont {J.~F.}\ \bibnamefont {Smith}},\ }\href
  {https://doi.org/10.1038/s41567-020-0899-4} {\bibfield  {journal} {\bibinfo
  {journal} {Nat. Phys.}\ }\textbf {\bibinfo {volume} {16}},\ \bibinfo {pages}
  {853} (\bibinfo {year} {2020})}\BibitemShut {NoStop}%
\bibitem [{\citenamefont {Bucher}\ \emph {et~al.}(2016)\citenamefont {Bucher},
  \citenamefont {Zhu}, \citenamefont {Wu}, \citenamefont {Janssens},
  \citenamefont {Cline}, \citenamefont {Hayes}, \citenamefont {Albers},
  \citenamefont {Ayangeakaa}, \citenamefont {Butler}, \citenamefont {Campbell},
  \citenamefont {Carpenter}, \citenamefont {Chiara}, \citenamefont {Clark},
  \citenamefont {Crawford}, \citenamefont {Cromaz}, \citenamefont {David},
  \citenamefont {Dickerson}, \citenamefont {Gregor}, \citenamefont {Harker},
  \citenamefont {Hoffman}, \citenamefont {Kay}, \citenamefont {Kondev},
  \citenamefont {Korichi}, \citenamefont {Lauritsen}, \citenamefont
  {Macchiavelli}, \citenamefont {Pardo}, \citenamefont {Richard}, \citenamefont
  {Riley}, \citenamefont {Savard}, \citenamefont {Scheck}, \citenamefont
  {Seweryniak}, \citenamefont {Smith}, \citenamefont {Vondrasek},\ and\
  \citenamefont {Wiens}}]{bucher2016}%
  \BibitemOpen
  \bibfield  {author} {\bibinfo {author} {\bibfnamefont {B.}~\bibnamefont
  {Bucher}}, \bibinfo {author} {\bibfnamefont {S.}~\bibnamefont {Zhu}},
  \bibinfo {author} {\bibfnamefont {C.~Y.}\ \bibnamefont {Wu}}, \bibinfo
  {author} {\bibfnamefont {R.~V.~F.}\ \bibnamefont {Janssens}}, \bibinfo
  {author} {\bibfnamefont {D.}~\bibnamefont {Cline}}, \bibinfo {author}
  {\bibfnamefont {A.~B.}\ \bibnamefont {Hayes}}, \bibinfo {author}
  {\bibfnamefont {M.}~\bibnamefont {Albers}}, \bibinfo {author} {\bibfnamefont
  {A.~D.}\ \bibnamefont {Ayangeakaa}}, \bibinfo {author} {\bibfnamefont
  {P.~A.}\ \bibnamefont {Butler}}, \bibinfo {author} {\bibfnamefont {C.~M.}\
  \bibnamefont {Campbell}}, \bibinfo {author} {\bibfnamefont {M.~P.}\
  \bibnamefont {Carpenter}}, \bibinfo {author} {\bibfnamefont {C.~J.}\
  \bibnamefont {Chiara}}, \bibinfo {author} {\bibfnamefont {J.~A.}\
  \bibnamefont {Clark}}, \bibinfo {author} {\bibfnamefont {H.~L.}\ \bibnamefont
  {Crawford}}, \bibinfo {author} {\bibfnamefont {M.}~\bibnamefont {Cromaz}},
  \bibinfo {author} {\bibfnamefont {H.~M.}\ \bibnamefont {David}}, \bibinfo
  {author} {\bibfnamefont {C.}~\bibnamefont {Dickerson}}, \bibinfo {author}
  {\bibfnamefont {E.~T.}\ \bibnamefont {Gregor}}, \bibinfo {author}
  {\bibfnamefont {J.}~\bibnamefont {Harker}}, \bibinfo {author} {\bibfnamefont
  {C.~R.}\ \bibnamefont {Hoffman}}, \bibinfo {author} {\bibfnamefont {B.~P.}\
  \bibnamefont {Kay}}, \bibinfo {author} {\bibfnamefont {F.~G.}\ \bibnamefont
  {Kondev}}, \bibinfo {author} {\bibfnamefont {A.}~\bibnamefont {Korichi}},
  \bibinfo {author} {\bibfnamefont {T.}~\bibnamefont {Lauritsen}}, \bibinfo
  {author} {\bibfnamefont {A.~O.}\ \bibnamefont {Macchiavelli}}, \bibinfo
  {author} {\bibfnamefont {R.~C.}\ \bibnamefont {Pardo}}, \bibinfo {author}
  {\bibfnamefont {A.}~\bibnamefont {Richard}}, \bibinfo {author} {\bibfnamefont
  {M.~A.}\ \bibnamefont {Riley}}, \bibinfo {author} {\bibfnamefont
  {G.}~\bibnamefont {Savard}}, \bibinfo {author} {\bibfnamefont
  {M.}~\bibnamefont {Scheck}}, \bibinfo {author} {\bibfnamefont
  {D.}~\bibnamefont {Seweryniak}}, \bibinfo {author} {\bibfnamefont {M.~K.}\
  \bibnamefont {Smith}}, \bibinfo {author} {\bibfnamefont {R.}~\bibnamefont
  {Vondrasek}},\ and\ \bibinfo {author} {\bibfnamefont {A.}~\bibnamefont
  {Wiens}},\ }\href {https://doi.org/10.1103/PhysRevLett.116.112503} {\bibfield
   {journal} {\bibinfo  {journal} {Phys. Rev. Lett.}\ }\textbf {\bibinfo
  {volume} {116}},\ \bibinfo {pages} {112503} (\bibinfo {year}
  {2016})}\BibitemShut {NoStop}%
\bibitem [{\citenamefont {Bucher}\ \emph {et~al.}(2017)\citenamefont {Bucher},
  \citenamefont {Zhu}, \citenamefont {Wu}, \citenamefont {Janssens},
  \citenamefont {Bernard}, \citenamefont {Robledo}, \citenamefont
  {Rodr\'{\i}guez}, \citenamefont {Cline}, \citenamefont {Hayes}, \citenamefont
  {Ayangeakaa}, \citenamefont {Buckner}, \citenamefont {Campbell},
  \citenamefont {Carpenter}, \citenamefont {Clark}, \citenamefont {Crawford},
  \citenamefont {David}, \citenamefont {Dickerson}, \citenamefont {Harker},
  \citenamefont {Hoffman}, \citenamefont {Kay}, \citenamefont {Kondev},
  \citenamefont {Lauritsen}, \citenamefont {Macchiavelli}, \citenamefont
  {Pardo}, \citenamefont {Savard}, \citenamefont {Seweryniak},\ and\
  \citenamefont {Vondrasek}}]{bucher2017}%
  \BibitemOpen
  \bibfield  {author} {\bibinfo {author} {\bibfnamefont {B.}~\bibnamefont
  {Bucher}}, \bibinfo {author} {\bibfnamefont {S.}~\bibnamefont {Zhu}},
  \bibinfo {author} {\bibfnamefont {C.~Y.}\ \bibnamefont {Wu}}, \bibinfo
  {author} {\bibfnamefont {R.~V.~F.}\ \bibnamefont {Janssens}}, \bibinfo
  {author} {\bibfnamefont {R.~N.}\ \bibnamefont {Bernard}}, \bibinfo {author}
  {\bibfnamefont {L.~M.}\ \bibnamefont {Robledo}}, \bibinfo {author}
  {\bibfnamefont {T.~R.}\ \bibnamefont {Rodr\'{\i}guez}}, \bibinfo {author}
  {\bibfnamefont {D.}~\bibnamefont {Cline}}, \bibinfo {author} {\bibfnamefont
  {A.~B.}\ \bibnamefont {Hayes}}, \bibinfo {author} {\bibfnamefont {A.~D.}\
  \bibnamefont {Ayangeakaa}}, \bibinfo {author} {\bibfnamefont {M.~Q.}\
  \bibnamefont {Buckner}}, \bibinfo {author} {\bibfnamefont {C.~M.}\
  \bibnamefont {Campbell}}, \bibinfo {author} {\bibfnamefont {M.~P.}\
  \bibnamefont {Carpenter}}, \bibinfo {author} {\bibfnamefont {J.~A.}\
  \bibnamefont {Clark}}, \bibinfo {author} {\bibfnamefont {H.~L.}\ \bibnamefont
  {Crawford}}, \bibinfo {author} {\bibfnamefont {H.~M.}\ \bibnamefont {David}},
  \bibinfo {author} {\bibfnamefont {C.}~\bibnamefont {Dickerson}}, \bibinfo
  {author} {\bibfnamefont {J.}~\bibnamefont {Harker}}, \bibinfo {author}
  {\bibfnamefont {C.~R.}\ \bibnamefont {Hoffman}}, \bibinfo {author}
  {\bibfnamefont {B.~P.}\ \bibnamefont {Kay}}, \bibinfo {author} {\bibfnamefont
  {F.~G.}\ \bibnamefont {Kondev}}, \bibinfo {author} {\bibfnamefont
  {T.}~\bibnamefont {Lauritsen}}, \bibinfo {author} {\bibfnamefont {A.~O.}\
  \bibnamefont {Macchiavelli}}, \bibinfo {author} {\bibfnamefont {R.~C.}\
  \bibnamefont {Pardo}}, \bibinfo {author} {\bibfnamefont {G.}~\bibnamefont
  {Savard}}, \bibinfo {author} {\bibfnamefont {D.}~\bibnamefont {Seweryniak}},\
  and\ \bibinfo {author} {\bibfnamefont {R.}~\bibnamefont {Vondrasek}},\ }\href
  {https://doi.org/10.1103/PhysRevLett.118.152504} {\bibfield  {journal}
  {\bibinfo  {journal} {Phys. Rev. Lett.}\ }\textbf {\bibinfo {volume} {118}},\
  \bibinfo {pages} {152504} (\bibinfo {year} {2017})}\BibitemShut {NoStop}%
\bibitem [{\citenamefont {Rzaca-Urban}\ \emph {et~al.}(2000)\citenamefont
  {Rzaca-Urban}, \citenamefont {Urban}, \citenamefont {Kaczor}, \citenamefont
  {Durell}, \citenamefont {Leddy}, \citenamefont {Jones}, \citenamefont
  {Phillips}, \citenamefont {Smith}, \citenamefont {Varley}, \citenamefont
  {Ahmad}, \citenamefont {Morss}, \citenamefont {Bentaleb}, \citenamefont
  {Lubkiewicz},\ and\ \citenamefont {Schulz}}]{rzacaurban2000}%
  \BibitemOpen
  \bibfield  {author} {\bibinfo {author} {\bibfnamefont {T.}~\bibnamefont
  {Rzaca-Urban}}, \bibinfo {author} {\bibfnamefont {W.}~\bibnamefont {Urban}},
  \bibinfo {author} {\bibfnamefont {A.}~\bibnamefont {Kaczor}}, \bibinfo
  {author} {\bibfnamefont {J.~L.}\ \bibnamefont {Durell}}, \bibinfo {author}
  {\bibfnamefont {M.~J.}\ \bibnamefont {Leddy}}, \bibinfo {author}
  {\bibfnamefont {M.~A.}\ \bibnamefont {Jones}}, \bibinfo {author}
  {\bibfnamefont {W.~R.}\ \bibnamefont {Phillips}}, \bibinfo {author}
  {\bibfnamefont {A.~G.}\ \bibnamefont {Smith}}, \bibinfo {author}
  {\bibfnamefont {B.~J.}\ \bibnamefont {Varley}}, \bibinfo {author}
  {\bibfnamefont {I.}~\bibnamefont {Ahmad}}, \bibinfo {author} {\bibfnamefont
  {L.~R.}\ \bibnamefont {Morss}}, \bibinfo {author} {\bibfnamefont
  {M.}~\bibnamefont {Bentaleb}}, \bibinfo {author} {\bibfnamefont
  {E.}~\bibnamefont {Lubkiewicz}},\ and\ \bibinfo {author} {\bibfnamefont
  {N.}~\bibnamefont {Schulz}},\ }\href {https://doi.org/10.1007/s100500070033}
  {\bibfield  {journal} {\bibinfo  {journal} {Eur. Phys. J. A}\ }\textbf
  {\bibinfo {volume} {9}},\ \bibinfo {pages} {165} (\bibinfo {year}
  {2000})}\BibitemShut {NoStop}%
\bibitem [{\citenamefont {Lalkovski}\ \emph {et~al.}(2007)\citenamefont
  {Lalkovski}, \citenamefont {Ilieva}, \citenamefont {Minkova}, \citenamefont
  {Minkov}, \citenamefont {Kutsarova}, \citenamefont {Lopez-Martens},
  \citenamefont {Korichi}, \citenamefont {H\"ubel}, \citenamefont {G\"orgen},
  \citenamefont {Jansen}, \citenamefont {Sch\"onwasser}, \citenamefont
  {Herskind}, \citenamefont {Bergstr\"om},\ and\ \citenamefont
  {Podoly\'ak}}]{lalkovski2007}%
  \BibitemOpen
  \bibfield  {author} {\bibinfo {author} {\bibfnamefont {S.}~\bibnamefont
  {Lalkovski}}, \bibinfo {author} {\bibfnamefont {S.}~\bibnamefont {Ilieva}},
  \bibinfo {author} {\bibfnamefont {A.}~\bibnamefont {Minkova}}, \bibinfo
  {author} {\bibfnamefont {N.}~\bibnamefont {Minkov}}, \bibinfo {author}
  {\bibfnamefont {T.}~\bibnamefont {Kutsarova}}, \bibinfo {author}
  {\bibfnamefont {A.}~\bibnamefont {Lopez-Martens}}, \bibinfo {author}
  {\bibfnamefont {A.}~\bibnamefont {Korichi}}, \bibinfo {author} {\bibfnamefont
  {H.}~\bibnamefont {H\"ubel}}, \bibinfo {author} {\bibfnamefont
  {A.}~\bibnamefont {G\"orgen}}, \bibinfo {author} {\bibfnamefont
  {A.}~\bibnamefont {Jansen}}, \bibinfo {author} {\bibfnamefont
  {G.}~\bibnamefont {Sch\"onwasser}}, \bibinfo {author} {\bibfnamefont
  {B.}~\bibnamefont {Herskind}}, \bibinfo {author} {\bibfnamefont
  {M.}~\bibnamefont {Bergstr\"om}},\ and\ \bibinfo {author} {\bibfnamefont
  {Z.}~\bibnamefont {Podoly\'ak}},\ }\href
  {https://doi.org/10.1103/PhysRevC.75.014314} {\bibfield  {journal} {\bibinfo
  {journal} {Phys. Rev. C}\ }\textbf {\bibinfo {volume} {75}},\ \bibinfo
  {pages} {014314} (\bibinfo {year} {2007})}\BibitemShut {NoStop}%
\bibitem [{\citenamefont {Scheck}\ \emph {et~al.}(2010)\citenamefont {Scheck},
  \citenamefont {Butler}, \citenamefont {Fransen}, \citenamefont {Werner},\
  and\ \citenamefont {Yates}}]{scheck2010}%
  \BibitemOpen
  \bibfield  {author} {\bibinfo {author} {\bibfnamefont {M.}~\bibnamefont
  {Scheck}}, \bibinfo {author} {\bibfnamefont {P.~A.}\ \bibnamefont {Butler}},
  \bibinfo {author} {\bibfnamefont {C.}~\bibnamefont {Fransen}}, \bibinfo
  {author} {\bibfnamefont {V.}~\bibnamefont {Werner}},\ and\ \bibinfo {author}
  {\bibfnamefont {S.~W.}\ \bibnamefont {Yates}},\ }\href
  {https://doi.org/10.1103/PhysRevC.81.064305} {\bibfield  {journal} {\bibinfo
  {journal} {Phys. Rev. C}\ }\textbf {\bibinfo {volume} {81}},\ \bibinfo
  {pages} {064305} (\bibinfo {year} {2010})}\BibitemShut {NoStop}%
\bibitem [{\citenamefont {LI}\ \emph {et~al.}(2011)\citenamefont {LI},
  \citenamefont {HAMILTON}, \citenamefont {RAMAYYA}, \citenamefont {LIU},
  \citenamefont {ZHANG}, \citenamefont {BREWER}, \citenamefont {HWANG},
  \citenamefont {GOODIN}, \citenamefont {ZHU}, \citenamefont {LUO},
  \citenamefont {RASMUSSEN}, \citenamefont {LEE}, \citenamefont {WU},
  \citenamefont {DONANGELO}, \citenamefont {DANIEL}, \citenamefont
  {TER-AKOPIAN}, \citenamefont {OGANESSIAN}, \citenamefont {UNZHAKOVA},
  \citenamefont {COLE}, \citenamefont {MA},\ and\ \citenamefont
  {STOYER}}]{li2011}%
  \BibitemOpen
  \bibfield  {author} {\bibinfo {author} {\bibfnamefont {K.}~\bibnamefont
  {LI}}, \bibinfo {author} {\bibfnamefont {J.~H.}\ \bibnamefont {HAMILTON}},
  \bibinfo {author} {\bibfnamefont {A.~V.}\ \bibnamefont {RAMAYYA}}, \bibinfo
  {author} {\bibfnamefont {S.~H.}\ \bibnamefont {LIU}}, \bibinfo {author}
  {\bibfnamefont {X.~Q.}\ \bibnamefont {ZHANG}}, \bibinfo {author}
  {\bibfnamefont {N.~T.}\ \bibnamefont {BREWER}}, \bibinfo {author}
  {\bibfnamefont {J.~K.}\ \bibnamefont {HWANG}}, \bibinfo {author}
  {\bibfnamefont {C.}~\bibnamefont {GOODIN}}, \bibinfo {author} {\bibfnamefont
  {S.~J.}\ \bibnamefont {ZHU}}, \bibinfo {author} {\bibfnamefont {Y.~X.}\
  \bibnamefont {LUO}}, \bibinfo {author} {\bibfnamefont {J.~O.}\ \bibnamefont
  {RASMUSSEN}}, \bibinfo {author} {\bibfnamefont {I.~Y.}\ \bibnamefont {LEE}},
  \bibinfo {author} {\bibfnamefont {S.~C.}\ \bibnamefont {WU}}, \bibinfo
  {author} {\bibfnamefont {R.}~\bibnamefont {DONANGELO}}, \bibinfo {author}
  {\bibfnamefont {A.~V.}\ \bibnamefont {DANIEL}}, \bibinfo {author}
  {\bibfnamefont {G.~M.}\ \bibnamefont {TER-AKOPIAN}}, \bibinfo {author}
  {\bibfnamefont {Y.~T.}\ \bibnamefont {OGANESSIAN}}, \bibinfo {author}
  {\bibfnamefont {A.}~\bibnamefont {UNZHAKOVA}}, \bibinfo {author}
  {\bibfnamefont {J.~D.}\ \bibnamefont {COLE}}, \bibinfo {author}
  {\bibfnamefont {W.~C.}\ \bibnamefont {MA}},\ and\ \bibinfo {author}
  {\bibfnamefont {M.~A.}\ \bibnamefont {STOYER}},\ }\href
  {https://doi.org/10.1142/S0218301311019635} {\bibfield  {journal} {\bibinfo
  {journal} {Int. J. Mod. Phys. E}\ }\textbf {\bibinfo {volume} {20}},\
  \bibinfo {pages} {1825} (\bibinfo {year} {2011})},\ \Eprint
  {https://arxiv.org/abs/https://doi.org/10.1142/S0218301311019635}
  {https://doi.org/10.1142/S0218301311019635} \BibitemShut {NoStop}%
\bibitem [{\citenamefont {Gregor}\ \emph {et~al.}(2017)\citenamefont {Gregor},
  \citenamefont {Scheck}, \citenamefont {Chapman}, \citenamefont {Gaffney},
  \citenamefont {Keatings}, \citenamefont {Mashtakov}, \citenamefont
  {O'Donnell}, \citenamefont {Smith}, \citenamefont {Spagnoletti},
  \citenamefont {Th{\"u}rauf}, \citenamefont {Werner},\ and\ \citenamefont
  {Wiseman}}]{gregor2017}%
  \BibitemOpen
  \bibfield  {author} {\bibinfo {author} {\bibfnamefont {E.~T.}\ \bibnamefont
  {Gregor}}, \bibinfo {author} {\bibfnamefont {M.}~\bibnamefont {Scheck}},
  \bibinfo {author} {\bibfnamefont {R.}~\bibnamefont {Chapman}}, \bibinfo
  {author} {\bibfnamefont {L.~P.}\ \bibnamefont {Gaffney}}, \bibinfo {author}
  {\bibfnamefont {J.}~\bibnamefont {Keatings}}, \bibinfo {author}
  {\bibfnamefont {K.~R.}\ \bibnamefont {Mashtakov}}, \bibinfo {author}
  {\bibfnamefont {D.}~\bibnamefont {O'Donnell}}, \bibinfo {author}
  {\bibfnamefont {J.~F.}\ \bibnamefont {Smith}}, \bibinfo {author}
  {\bibfnamefont {P.}~\bibnamefont {Spagnoletti}}, \bibinfo {author}
  {\bibfnamefont {M.}~\bibnamefont {Th{\"u}rauf}}, \bibinfo {author}
  {\bibfnamefont {V.}~\bibnamefont {Werner}},\ and\ \bibinfo {author}
  {\bibfnamefont {C.}~\bibnamefont {Wiseman}},\ }\href
  {https://doi.org/10.1140/epja/i2017-12224-7} {\bibfield  {journal} {\bibinfo
  {journal} {Eur. Phys. J. A}\ }\textbf {\bibinfo {volume} {53}},\ \bibinfo
  {pages} {50} (\bibinfo {year} {2017})}\BibitemShut {NoStop}%
\bibitem [{\citenamefont {Dudouet}\ \emph {et~al.}(2017)\citenamefont
  {Dudouet}, \citenamefont {Lemasson}, \citenamefont {Duch\^ene}, \citenamefont
  {Rejmund}, \citenamefont {Cl\'ement}, \citenamefont {Michelagnoli},
  \citenamefont {Didierjean}, \citenamefont {Korichi}, \citenamefont {Maquart},
  \citenamefont {Stezowski}, \citenamefont {Lizarazo}, \citenamefont
  {P\'erez-Vidal}, \citenamefont {Andreoiu}, \citenamefont {de~Angelis},
  \citenamefont {Astier}, \citenamefont {Delafosse}, \citenamefont {Deloncle},
  \citenamefont {Dombradi}, \citenamefont {de~France}, \citenamefont {Gadea},
  \citenamefont {Gottardo}, \citenamefont {Jacquot}, \citenamefont {Jones},
  \citenamefont {Konstantinopoulos}, \citenamefont {Kuti}, \citenamefont
  {Le~Blanc}, \citenamefont {Lenzi}, \citenamefont {Li}, \citenamefont
  {Lozeva}, \citenamefont {Million}, \citenamefont {Napoli}, \citenamefont
  {Navin}, \citenamefont {Petrache}, \citenamefont {Pietralla}, \citenamefont
  {Ralet}, \citenamefont {Ramdhane}, \citenamefont {Redon}, \citenamefont
  {Schmitt}, \citenamefont {Sohler}, \citenamefont {Verney}, \citenamefont
  {Barrientos}, \citenamefont {Birkenbach}, \citenamefont {Burrows},
  \citenamefont {Charles}, \citenamefont {Collado}, \citenamefont {Cullen},
  \citenamefont {D\'esesquelles}, \citenamefont {Domingo~Pardo}, \citenamefont
  {Gonz\'alez}, \citenamefont {Harkness-Brennan}, \citenamefont {Hess},
  \citenamefont {Judson}, \citenamefont {Karolak}, \citenamefont {Korten},
  \citenamefont {Labiche}, \citenamefont {Ljungvall}, \citenamefont
  {Menegazzo}, \citenamefont {Mengoni}, \citenamefont {Pullia}, \citenamefont
  {Recchia}, \citenamefont {Reiter}, \citenamefont {Salsac}, \citenamefont
  {Sanchis}, \citenamefont {Theisen}, \citenamefont {Valiente-Dob\'on},\ and\
  \citenamefont {Zieli\ifmmode~\acute{n}\else \'{n}\fi{}ska}}]{dudouet2017}%
  \BibitemOpen
  \bibfield  {author} {\bibinfo {author} {\bibfnamefont {J.}~\bibnamefont
  {Dudouet}}, \bibinfo {author} {\bibfnamefont {A.}~\bibnamefont {Lemasson}},
  \bibinfo {author} {\bibfnamefont {G.}~\bibnamefont {Duch\^ene}}, \bibinfo
  {author} {\bibfnamefont {M.}~\bibnamefont {Rejmund}}, \bibinfo {author}
  {\bibfnamefont {E.}~\bibnamefont {Cl\'ement}}, \bibinfo {author}
  {\bibfnamefont {C.}~\bibnamefont {Michelagnoli}}, \bibinfo {author}
  {\bibfnamefont {F.}~\bibnamefont {Didierjean}}, \bibinfo {author}
  {\bibfnamefont {A.}~\bibnamefont {Korichi}}, \bibinfo {author} {\bibfnamefont
  {G.}~\bibnamefont {Maquart}}, \bibinfo {author} {\bibfnamefont
  {O.}~\bibnamefont {Stezowski}}, \bibinfo {author} {\bibfnamefont
  {C.}~\bibnamefont {Lizarazo}}, \bibinfo {author} {\bibfnamefont {R.~M.}\
  \bibnamefont {P\'erez-Vidal}}, \bibinfo {author} {\bibfnamefont
  {C.}~\bibnamefont {Andreoiu}}, \bibinfo {author} {\bibfnamefont
  {G.}~\bibnamefont {de~Angelis}}, \bibinfo {author} {\bibfnamefont
  {A.}~\bibnamefont {Astier}}, \bibinfo {author} {\bibfnamefont
  {C.}~\bibnamefont {Delafosse}}, \bibinfo {author} {\bibfnamefont
  {I.}~\bibnamefont {Deloncle}}, \bibinfo {author} {\bibfnamefont
  {Z.}~\bibnamefont {Dombradi}}, \bibinfo {author} {\bibfnamefont
  {G.}~\bibnamefont {de~France}}, \bibinfo {author} {\bibfnamefont
  {A.}~\bibnamefont {Gadea}}, \bibinfo {author} {\bibfnamefont
  {A.}~\bibnamefont {Gottardo}}, \bibinfo {author} {\bibfnamefont
  {B.}~\bibnamefont {Jacquot}}, \bibinfo {author} {\bibfnamefont
  {P.}~\bibnamefont {Jones}}, \bibinfo {author} {\bibfnamefont
  {T.}~\bibnamefont {Konstantinopoulos}}, \bibinfo {author} {\bibfnamefont
  {I.}~\bibnamefont {Kuti}}, \bibinfo {author} {\bibfnamefont {F.}~\bibnamefont
  {Le~Blanc}}, \bibinfo {author} {\bibfnamefont {S.~M.}\ \bibnamefont {Lenzi}},
  \bibinfo {author} {\bibfnamefont {G.}~\bibnamefont {Li}}, \bibinfo {author}
  {\bibfnamefont {R.}~\bibnamefont {Lozeva}}, \bibinfo {author} {\bibfnamefont
  {B.}~\bibnamefont {Million}}, \bibinfo {author} {\bibfnamefont {D.~R.}\
  \bibnamefont {Napoli}}, \bibinfo {author} {\bibfnamefont {A.}~\bibnamefont
  {Navin}}, \bibinfo {author} {\bibfnamefont {C.~M.}\ \bibnamefont {Petrache}},
  \bibinfo {author} {\bibfnamefont {N.}~\bibnamefont {Pietralla}}, \bibinfo
  {author} {\bibfnamefont {D.}~\bibnamefont {Ralet}}, \bibinfo {author}
  {\bibfnamefont {M.}~\bibnamefont {Ramdhane}}, \bibinfo {author}
  {\bibfnamefont {N.}~\bibnamefont {Redon}}, \bibinfo {author} {\bibfnamefont
  {C.}~\bibnamefont {Schmitt}}, \bibinfo {author} {\bibfnamefont
  {D.}~\bibnamefont {Sohler}}, \bibinfo {author} {\bibfnamefont
  {D.}~\bibnamefont {Verney}}, \bibinfo {author} {\bibfnamefont
  {D.}~\bibnamefont {Barrientos}}, \bibinfo {author} {\bibfnamefont
  {B.}~\bibnamefont {Birkenbach}}, \bibinfo {author} {\bibfnamefont
  {I.}~\bibnamefont {Burrows}}, \bibinfo {author} {\bibfnamefont
  {L.}~\bibnamefont {Charles}}, \bibinfo {author} {\bibfnamefont
  {J.}~\bibnamefont {Collado}}, \bibinfo {author} {\bibfnamefont {D.~M.}\
  \bibnamefont {Cullen}}, \bibinfo {author} {\bibfnamefont {P.}~\bibnamefont
  {D\'esesquelles}}, \bibinfo {author} {\bibfnamefont {C.}~\bibnamefont
  {Domingo~Pardo}}, \bibinfo {author} {\bibfnamefont {V.}~\bibnamefont
  {Gonz\'alez}}, \bibinfo {author} {\bibfnamefont {L.}~\bibnamefont
  {Harkness-Brennan}}, \bibinfo {author} {\bibfnamefont {H.}~\bibnamefont
  {Hess}}, \bibinfo {author} {\bibfnamefont {D.~S.}\ \bibnamefont {Judson}},
  \bibinfo {author} {\bibfnamefont {M.}~\bibnamefont {Karolak}}, \bibinfo
  {author} {\bibfnamefont {W.}~\bibnamefont {Korten}}, \bibinfo {author}
  {\bibfnamefont {M.}~\bibnamefont {Labiche}}, \bibinfo {author} {\bibfnamefont
  {J.}~\bibnamefont {Ljungvall}}, \bibinfo {author} {\bibfnamefont
  {R.}~\bibnamefont {Menegazzo}}, \bibinfo {author} {\bibfnamefont
  {D.}~\bibnamefont {Mengoni}}, \bibinfo {author} {\bibfnamefont
  {A.}~\bibnamefont {Pullia}}, \bibinfo {author} {\bibfnamefont
  {F.}~\bibnamefont {Recchia}}, \bibinfo {author} {\bibfnamefont
  {P.}~\bibnamefont {Reiter}}, \bibinfo {author} {\bibfnamefont {M.~D.}\
  \bibnamefont {Salsac}}, \bibinfo {author} {\bibfnamefont {E.}~\bibnamefont
  {Sanchis}}, \bibinfo {author} {\bibfnamefont {C.}~\bibnamefont {Theisen}},
  \bibinfo {author} {\bibfnamefont {J.~J.}\ \bibnamefont {Valiente-Dob\'on}},\
  and\ \bibinfo {author} {\bibfnamefont {M.}~\bibnamefont
  {Zieli\ifmmode~\acute{n}\else \'{n}\fi{}ska}},\ }\href
  {https://doi.org/10.1103/PhysRevLett.118.162501} {\bibfield  {journal}
  {\bibinfo  {journal} {Phys. Rev. Lett.}\ }\textbf {\bibinfo {volume} {118}},\
  \bibinfo {pages} {162501} (\bibinfo {year} {2017})}\BibitemShut {NoStop}%
\bibitem [{\citenamefont {Gregor}\ \emph {et~al.}(2019)\citenamefont {Gregor},
  \citenamefont {Arsenyev}, \citenamefont {Scheck}, \citenamefont {Shneidman},
  \citenamefont {Thürauf}, \citenamefont {Bernards}, \citenamefont {Blanc},
  \citenamefont {Chapman}, \citenamefont {Drouet}, \citenamefont {Dzhioev},
  \citenamefont {de~France}, \citenamefont {Jentschel}, \citenamefont {Jolie},
  \citenamefont {Keatings}, \citenamefont {Kröll}, \citenamefont {Köster},
  \citenamefont {Leguillon}, \citenamefont {Mashtakov}, \citenamefont {Mutti},
  \citenamefont {O'Donnell}, \citenamefont {Petrache}, \citenamefont {Simpson},
  \citenamefont {Sinclair}, \citenamefont {Smith}, \citenamefont {Soldner},
  \citenamefont {Spagnoletti}, \citenamefont {Sushkov}, \citenamefont {Urban},
  \citenamefont {Vancraeyenest}, \citenamefont {Vanhoy}, \citenamefont
  {Werner}, \citenamefont {Zell},\ and\ \citenamefont
  {Zieli{\'{n}}ska}}]{gregor2019}%
  \BibitemOpen
  \bibfield  {author} {\bibinfo {author} {\bibfnamefont {E.~T.}\ \bibnamefont
  {Gregor}}, \bibinfo {author} {\bibfnamefont {N.~N.}\ \bibnamefont
  {Arsenyev}}, \bibinfo {author} {\bibfnamefont {M.}~\bibnamefont {Scheck}},
  \bibinfo {author} {\bibfnamefont {T.~M.}\ \bibnamefont {Shneidman}}, \bibinfo
  {author} {\bibfnamefont {M.}~\bibnamefont {Thürauf}}, \bibinfo {author}
  {\bibfnamefont {C.}~\bibnamefont {Bernards}}, \bibinfo {author}
  {\bibfnamefont {A.}~\bibnamefont {Blanc}}, \bibinfo {author} {\bibfnamefont
  {R.}~\bibnamefont {Chapman}}, \bibinfo {author} {\bibfnamefont
  {F.}~\bibnamefont {Drouet}}, \bibinfo {author} {\bibfnamefont {A.~A.}\
  \bibnamefont {Dzhioev}}, \bibinfo {author} {\bibfnamefont {G.}~\bibnamefont
  {de~France}}, \bibinfo {author} {\bibfnamefont {M.}~\bibnamefont
  {Jentschel}}, \bibinfo {author} {\bibfnamefont {J.}~\bibnamefont {Jolie}},
  \bibinfo {author} {\bibfnamefont {J.~M.}\ \bibnamefont {Keatings}}, \bibinfo
  {author} {\bibfnamefont {T.}~\bibnamefont {Kröll}}, \bibinfo {author}
  {\bibfnamefont {U.}~\bibnamefont {Köster}}, \bibinfo {author} {\bibfnamefont
  {R.}~\bibnamefont {Leguillon}}, \bibinfo {author} {\bibfnamefont {K.~R.}\
  \bibnamefont {Mashtakov}}, \bibinfo {author} {\bibfnamefont {P.}~\bibnamefont
  {Mutti}}, \bibinfo {author} {\bibfnamefont {D.}~\bibnamefont {O'Donnell}},
  \bibinfo {author} {\bibfnamefont {C.~M.}\ \bibnamefont {Petrache}}, \bibinfo
  {author} {\bibfnamefont {G.~S.}\ \bibnamefont {Simpson}}, \bibinfo {author}
  {\bibfnamefont {J.}~\bibnamefont {Sinclair}}, \bibinfo {author}
  {\bibfnamefont {J.~F.}\ \bibnamefont {Smith}}, \bibinfo {author}
  {\bibfnamefont {T.}~\bibnamefont {Soldner}}, \bibinfo {author} {\bibfnamefont
  {P.}~\bibnamefont {Spagnoletti}}, \bibinfo {author} {\bibfnamefont {A.~V.}\
  \bibnamefont {Sushkov}}, \bibinfo {author} {\bibfnamefont {W.}~\bibnamefont
  {Urban}}, \bibinfo {author} {\bibfnamefont {A.}~\bibnamefont
  {Vancraeyenest}}, \bibinfo {author} {\bibfnamefont {J.~R.}\ \bibnamefont
  {Vanhoy}}, \bibinfo {author} {\bibfnamefont {V.}~\bibnamefont {Werner}},
  \bibinfo {author} {\bibfnamefont {K.~O.}\ \bibnamefont {Zell}},\ and\
  \bibinfo {author} {\bibfnamefont {M.}~\bibnamefont {Zieli{\'{n}}ska}},\
  }\href {https://doi.org/10.1088/1361-6471/ab0b5e} {\bibfield  {journal}
  {\bibinfo  {journal} {J. Phys. G: Nucl. Part. Phys.}\ }\textbf {\bibinfo
  {volume} {46}},\ \bibinfo {pages} {075101} (\bibinfo {year}
  {2019})}\BibitemShut {NoStop}%
\bibitem [{\citenamefont {Gerst}\ \emph {et~al.}(2022)\citenamefont {Gerst},
  \citenamefont {Blazhev}, \citenamefont {Moschner}, \citenamefont
  {Doornenbal}, \citenamefont {Obertelli}, \citenamefont {Nomura},
  \citenamefont {Ebran}, \citenamefont {Hilaire}, \citenamefont {Libert},
  \citenamefont {Authelet}, \citenamefont {Baba}, \citenamefont {Calvet},
  \citenamefont {Ch\^ateau}, \citenamefont {Chen}, \citenamefont {Corsi},
  \citenamefont {Delbart}, \citenamefont {Gheller}, \citenamefont {Giganon},
  \citenamefont {Gillibert}, \citenamefont {Lapoux}, \citenamefont
  {Motobayashi}, \citenamefont {Niikura}, \citenamefont {Paul}, \citenamefont
  {Rouss\'e}, \citenamefont {Sakurai}, \citenamefont {Santamaria},
  \citenamefont {Steppenbeck}, \citenamefont {Taniuchi}, \citenamefont
  {Uesaka}, \citenamefont {Ando}, \citenamefont {Arici}, \citenamefont
  {Browne}, \citenamefont {Bruce}, \citenamefont {Caroll}, \citenamefont
  {Chung}, \citenamefont {Cort\'es}, \citenamefont {Dewald}, \citenamefont
  {Ding}, \citenamefont {Flavigny}, \citenamefont {Franchoo}, \citenamefont
  {G\'orska}, \citenamefont {Gottardo}, \citenamefont {Jolie}, \citenamefont
  {Jungclaus}, \citenamefont {Lee}, \citenamefont {Lettmann}, \citenamefont
  {Linh}, \citenamefont {Liu}, \citenamefont {Liu}, \citenamefont {Lizarazo},
  \citenamefont {Momiyama}, \citenamefont {Nagamine}, \citenamefont
  {Nakatsuka}, \citenamefont {Nita}, \citenamefont {Nobs}, \citenamefont
  {Olivier}, \citenamefont {Orlandi}, \citenamefont {Patel}, \citenamefont
  {Podoly\'ak}, \citenamefont {Rudigier}, \citenamefont {Saito}, \citenamefont
  {Shand}, \citenamefont {S\"oderstr\"om}, \citenamefont {Stefan},
  \citenamefont {Vaquero}, \citenamefont {Werner}, \citenamefont {Wimmer},\
  and\ \citenamefont {Xu}}]{gerst2022}%
  \BibitemOpen
  \bibfield  {author} {\bibinfo {author} {\bibfnamefont {R.-B.}\ \bibnamefont
  {Gerst}}, \bibinfo {author} {\bibfnamefont {A.}~\bibnamefont {Blazhev}},
  \bibinfo {author} {\bibfnamefont {K.}~\bibnamefont {Moschner}}, \bibinfo
  {author} {\bibfnamefont {P.}~\bibnamefont {Doornenbal}}, \bibinfo {author}
  {\bibfnamefont {A.}~\bibnamefont {Obertelli}}, \bibinfo {author}
  {\bibfnamefont {K.}~\bibnamefont {Nomura}}, \bibinfo {author} {\bibfnamefont
  {J.-P.}\ \bibnamefont {Ebran}}, \bibinfo {author} {\bibfnamefont
  {S.}~\bibnamefont {Hilaire}}, \bibinfo {author} {\bibfnamefont
  {J.}~\bibnamefont {Libert}}, \bibinfo {author} {\bibfnamefont
  {G.}~\bibnamefont {Authelet}}, \bibinfo {author} {\bibfnamefont
  {H.}~\bibnamefont {Baba}}, \bibinfo {author} {\bibfnamefont {D.}~\bibnamefont
  {Calvet}}, \bibinfo {author} {\bibfnamefont {F.}~\bibnamefont {Ch\^ateau}},
  \bibinfo {author} {\bibfnamefont {S.}~\bibnamefont {Chen}}, \bibinfo {author}
  {\bibfnamefont {A.}~\bibnamefont {Corsi}}, \bibinfo {author} {\bibfnamefont
  {A.}~\bibnamefont {Delbart}}, \bibinfo {author} {\bibfnamefont {J.-M.}\
  \bibnamefont {Gheller}}, \bibinfo {author} {\bibfnamefont {A.}~\bibnamefont
  {Giganon}}, \bibinfo {author} {\bibfnamefont {A.}~\bibnamefont {Gillibert}},
  \bibinfo {author} {\bibfnamefont {V.}~\bibnamefont {Lapoux}}, \bibinfo
  {author} {\bibfnamefont {T.}~\bibnamefont {Motobayashi}}, \bibinfo {author}
  {\bibfnamefont {M.}~\bibnamefont {Niikura}}, \bibinfo {author} {\bibfnamefont
  {N.}~\bibnamefont {Paul}}, \bibinfo {author} {\bibfnamefont {J.-Y.}\
  \bibnamefont {Rouss\'e}}, \bibinfo {author} {\bibfnamefont {H.}~\bibnamefont
  {Sakurai}}, \bibinfo {author} {\bibfnamefont {C.}~\bibnamefont {Santamaria}},
  \bibinfo {author} {\bibfnamefont {D.}~\bibnamefont {Steppenbeck}}, \bibinfo
  {author} {\bibfnamefont {R.}~\bibnamefont {Taniuchi}}, \bibinfo {author}
  {\bibfnamefont {T.}~\bibnamefont {Uesaka}}, \bibinfo {author} {\bibfnamefont
  {T.}~\bibnamefont {Ando}}, \bibinfo {author} {\bibfnamefont {T.}~\bibnamefont
  {Arici}}, \bibinfo {author} {\bibfnamefont {F.}~\bibnamefont {Browne}},
  \bibinfo {author} {\bibfnamefont {A.~M.}\ \bibnamefont {Bruce}}, \bibinfo
  {author} {\bibfnamefont {R.}~\bibnamefont {Caroll}}, \bibinfo {author}
  {\bibfnamefont {L.~X.}\ \bibnamefont {Chung}}, \bibinfo {author}
  {\bibfnamefont {M.~L.}\ \bibnamefont {Cort\'es}}, \bibinfo {author}
  {\bibfnamefont {M.}~\bibnamefont {Dewald}}, \bibinfo {author} {\bibfnamefont
  {B.}~\bibnamefont {Ding}}, \bibinfo {author} {\bibfnamefont {F.}~\bibnamefont
  {Flavigny}}, \bibinfo {author} {\bibfnamefont {S.}~\bibnamefont {Franchoo}},
  \bibinfo {author} {\bibfnamefont {M.}~\bibnamefont {G\'orska}}, \bibinfo
  {author} {\bibfnamefont {A.}~\bibnamefont {Gottardo}}, \bibinfo {author}
  {\bibfnamefont {J.}~\bibnamefont {Jolie}}, \bibinfo {author} {\bibfnamefont
  {A.}~\bibnamefont {Jungclaus}}, \bibinfo {author} {\bibfnamefont
  {J.}~\bibnamefont {Lee}}, \bibinfo {author} {\bibfnamefont {M.}~\bibnamefont
  {Lettmann}}, \bibinfo {author} {\bibfnamefont {B.~D.}\ \bibnamefont {Linh}},
  \bibinfo {author} {\bibfnamefont {J.}~\bibnamefont {Liu}}, \bibinfo {author}
  {\bibfnamefont {Z.}~\bibnamefont {Liu}}, \bibinfo {author} {\bibfnamefont
  {C.}~\bibnamefont {Lizarazo}}, \bibinfo {author} {\bibfnamefont
  {S.}~\bibnamefont {Momiyama}}, \bibinfo {author} {\bibfnamefont
  {S.}~\bibnamefont {Nagamine}}, \bibinfo {author} {\bibfnamefont
  {N.}~\bibnamefont {Nakatsuka}}, \bibinfo {author} {\bibfnamefont {C.~R.}\
  \bibnamefont {Nita}}, \bibinfo {author} {\bibfnamefont {C.}~\bibnamefont
  {Nobs}}, \bibinfo {author} {\bibfnamefont {L.}~\bibnamefont {Olivier}},
  \bibinfo {author} {\bibfnamefont {R.}~\bibnamefont {Orlandi}}, \bibinfo
  {author} {\bibfnamefont {Z.}~\bibnamefont {Patel}}, \bibinfo {author}
  {\bibfnamefont {Z.}~\bibnamefont {Podoly\'ak}}, \bibinfo {author}
  {\bibfnamefont {M.}~\bibnamefont {Rudigier}}, \bibinfo {author}
  {\bibfnamefont {T.}~\bibnamefont {Saito}}, \bibinfo {author} {\bibfnamefont
  {C.}~\bibnamefont {Shand}}, \bibinfo {author} {\bibfnamefont {P.-A.}\
  \bibnamefont {S\"oderstr\"om}}, \bibinfo {author} {\bibfnamefont
  {I.}~\bibnamefont {Stefan}}, \bibinfo {author} {\bibfnamefont
  {V.}~\bibnamefont {Vaquero}}, \bibinfo {author} {\bibfnamefont
  {V.}~\bibnamefont {Werner}}, \bibinfo {author} {\bibfnamefont
  {K.}~\bibnamefont {Wimmer}},\ and\ \bibinfo {author} {\bibfnamefont
  {Z.}~\bibnamefont {Xu}},\ }\href
  {https://doi.org/10.1103/PhysRevC.105.024302} {\bibfield  {journal} {\bibinfo
   {journal} {Phys. Rev. C}\ }\textbf {\bibinfo {volume} {105}},\ \bibinfo
  {pages} {024302} (\bibinfo {year} {2022})}\BibitemShut {NoStop}%
\bibitem [{\citenamefont {Nazarewicz}\ \emph {et~al.}(1984)\citenamefont
  {Nazarewicz}, \citenamefont {Olanders}, \citenamefont {Ragnarsson},
  \citenamefont {Dudek}, \citenamefont {Leander}, \citenamefont {M\"oller},\
  and\ \citenamefont {Ruchowsa}}]{naza1984b}%
  \BibitemOpen
  \bibfield  {author} {\bibinfo {author} {\bibfnamefont {W.}~\bibnamefont
  {Nazarewicz}}, \bibinfo {author} {\bibfnamefont {P.}~\bibnamefont
  {Olanders}}, \bibinfo {author} {\bibfnamefont {I.}~\bibnamefont
  {Ragnarsson}}, \bibinfo {author} {\bibfnamefont {J.}~\bibnamefont {Dudek}},
  \bibinfo {author} {\bibfnamefont {G.~A.}\ \bibnamefont {Leander}}, \bibinfo
  {author} {\bibfnamefont {P.}~\bibnamefont {M\"oller}},\ and\ \bibinfo
  {author} {\bibfnamefont {E.}~\bibnamefont {Ruchowsa}},\ }\href
  {https://doi.org/10.1016/0375-9474(84)90208-2} {\bibfield  {journal}
  {\bibinfo  {journal} {Nucl. Phys. A}\ }\textbf {\bibinfo {volume} {429}},\
  \bibinfo {pages} {269 } (\bibinfo {year} {1984})}\BibitemShut {NoStop}%
\bibitem [{\citenamefont {Leander}\ \emph {et~al.}(1985)\citenamefont
  {Leander}, \citenamefont {Nazarewicz}, \citenamefont {Olanders},
  \citenamefont {Ragnarsson},\ and\ \citenamefont {Dudek}}]{leander1985}%
  \BibitemOpen
  \bibfield  {author} {\bibinfo {author} {\bibfnamefont {G.}~\bibnamefont
  {Leander}}, \bibinfo {author} {\bibfnamefont {W.}~\bibnamefont {Nazarewicz}},
  \bibinfo {author} {\bibfnamefont {P.}~\bibnamefont {Olanders}}, \bibinfo
  {author} {\bibfnamefont {I.}~\bibnamefont {Ragnarsson}},\ and\ \bibinfo
  {author} {\bibfnamefont {J.}~\bibnamefont {Dudek}},\ }\href
  {https://doi.org/https://doi.org/10.1016/0370-2693(85)90496-4} {\bibfield
  {journal} {\bibinfo  {journal} {Phys. Lett. B}\ }\textbf {\bibinfo {volume}
  {152}},\ \bibinfo {pages} {284 } (\bibinfo {year} {1985})}\BibitemShut
  {NoStop}%
\bibitem [{\citenamefont {Möller}\ \emph {et~al.}(2008)\citenamefont
  {Möller}, \citenamefont {Bengtsson}, \citenamefont {Carlsson}, \citenamefont
  {Olivius}, \citenamefont {Ichikawa}, \citenamefont {Sagawa},\ and\
  \citenamefont {Iwamoto}}]{moeller2008}%
  \BibitemOpen
  \bibfield  {author} {\bibinfo {author} {\bibfnamefont {P.}~\bibnamefont
  {Möller}}, \bibinfo {author} {\bibfnamefont {R.}~\bibnamefont {Bengtsson}},
  \bibinfo {author} {\bibfnamefont {B.}~\bibnamefont {Carlsson}}, \bibinfo
  {author} {\bibfnamefont {P.}~\bibnamefont {Olivius}}, \bibinfo {author}
  {\bibfnamefont {T.}~\bibnamefont {Ichikawa}}, \bibinfo {author}
  {\bibfnamefont {H.}~\bibnamefont {Sagawa}},\ and\ \bibinfo {author}
  {\bibfnamefont {A.}~\bibnamefont {Iwamoto}},\ }\href
  {https://doi.org/https://doi.org/10.1016/j.adt.2008.05.002} {\bibfield
  {journal} {\bibinfo  {journal} {At. Dat. Nucl. Dat. Tab.}\ }\textbf {\bibinfo
  {volume} {94}},\ \bibinfo {pages} {758 } (\bibinfo {year}
  {2008})}\BibitemShut {NoStop}%
\bibitem [{\citenamefont {Bonche}\ \emph {et~al.}(1986)\citenamefont {Bonche},
  \citenamefont {Heenen}, \citenamefont {Flocard},\ and\ \citenamefont
  {Vautherin}}]{bonche1986}%
  \BibitemOpen
  \bibfield  {author} {\bibinfo {author} {\bibfnamefont {P.}~\bibnamefont
  {Bonche}}, \bibinfo {author} {\bibfnamefont {P.-H.}\ \bibnamefont {Heenen}},
  \bibinfo {author} {\bibfnamefont {H.}~\bibnamefont {Flocard}},\ and\ \bibinfo
  {author} {\bibfnamefont {D.}~\bibnamefont {Vautherin}},\ }\href
  {https://doi.org/10.1016/0370-2693(86)90609-X} {\bibfield  {journal}
  {\bibinfo  {journal} {Phys. Lett. B}\ }\textbf {\bibinfo {volume} {175}},\
  \bibinfo {pages} {387 } (\bibinfo {year} {1986})}\BibitemShut {NoStop}%
\bibitem [{\citenamefont {Heenen}\ \emph {et~al.}(1994)\citenamefont {Heenen},
  \citenamefont {Skalski}, \citenamefont {Bonche},\ and\ \citenamefont
  {Flocard}}]{heenen1994}%
  \BibitemOpen
  \bibfield  {author} {\bibinfo {author} {\bibfnamefont {P.-H.}\ \bibnamefont
  {Heenen}}, \bibinfo {author} {\bibfnamefont {J.}~\bibnamefont {Skalski}},
  \bibinfo {author} {\bibfnamefont {P.}~\bibnamefont {Bonche}},\ and\ \bibinfo
  {author} {\bibfnamefont {H.}~\bibnamefont {Flocard}},\ }\href
  {https://doi.org/10.1103/PhysRevC.50.802} {\bibfield  {journal} {\bibinfo
  {journal} {Phys. Rev. C}\ }\textbf {\bibinfo {volume} {50}},\ \bibinfo
  {pages} {802} (\bibinfo {year} {1994})}\BibitemShut {NoStop}%
\bibitem [{\citenamefont {Robledo}\ \emph {et~al.}(1987)\citenamefont
  {Robledo}, \citenamefont {Egido}, \citenamefont {Berger},\ and\ \citenamefont
  {Girod}}]{robledo1987}%
  \BibitemOpen
  \bibfield  {author} {\bibinfo {author} {\bibfnamefont {L.}~\bibnamefont
  {Robledo}}, \bibinfo {author} {\bibfnamefont {J.}~\bibnamefont {Egido}},
  \bibinfo {author} {\bibfnamefont {J.}~\bibnamefont {Berger}},\ and\ \bibinfo
  {author} {\bibfnamefont {M.}~\bibnamefont {Girod}},\ }\href
  {https://doi.org/https://doi.org/10.1016/0370-2693(87)91085-9} {\bibfield
  {journal} {\bibinfo  {journal} {Phys. Lett. B}\ }\textbf {\bibinfo {volume}
  {187}},\ \bibinfo {pages} {223 } (\bibinfo {year} {1987})}\BibitemShut
  {NoStop}%
\bibitem [{\citenamefont {Egido}\ and\ \citenamefont
  {Robledo}(1992)}]{egido1992}%
  \BibitemOpen
  \bibfield  {author} {\bibinfo {author} {\bibfnamefont {J.}~\bibnamefont
  {Egido}}\ and\ \bibinfo {author} {\bibfnamefont {L.}~\bibnamefont
  {Robledo}},\ }\href
  {https://doi.org/https://doi.org/10.1016/0375-9474(92)90294-T} {\bibfield
  {journal} {\bibinfo  {journal} {Nucl. Phys. A}\ }\textbf {\bibinfo {volume}
  {545}},\ \bibinfo {pages} {589 } (\bibinfo {year} {1992})}\BibitemShut
  {NoStop}%
\bibitem [{\citenamefont {Robledo}\ \emph {et~al.}(2010)\citenamefont
  {Robledo}, \citenamefont {Baldo}, \citenamefont {Schuck},\ and\ \citenamefont
  {Vi\~nas}}]{robledo2010}%
  \BibitemOpen
  \bibfield  {author} {\bibinfo {author} {\bibfnamefont {L.~M.}\ \bibnamefont
  {Robledo}}, \bibinfo {author} {\bibfnamefont {M.}~\bibnamefont {Baldo}},
  \bibinfo {author} {\bibfnamefont {P.}~\bibnamefont {Schuck}},\ and\ \bibinfo
  {author} {\bibfnamefont {X.}~\bibnamefont {Vi\~nas}},\ }\href
  {https://doi.org/10.1103/PhysRevC.81.034315} {\bibfield  {journal} {\bibinfo
  {journal} {Phys. Rev. C}\ }\textbf {\bibinfo {volume} {81}},\ \bibinfo
  {pages} {034315} (\bibinfo {year} {2010})}\BibitemShut {NoStop}%
\bibitem [{\citenamefont {Robledo}\ and\ \citenamefont
  {Bertsch}(2011)}]{robledo2011}%
  \BibitemOpen
  \bibfield  {author} {\bibinfo {author} {\bibfnamefont {L.~M.}\ \bibnamefont
  {Robledo}}\ and\ \bibinfo {author} {\bibfnamefont {G.~F.}\ \bibnamefont
  {Bertsch}},\ }\href {https://doi.org/10.1103/PhysRevC.84.054302} {\bibfield
  {journal} {\bibinfo  {journal} {Phys. Rev. C}\ }\textbf {\bibinfo {volume}
  {84}},\ \bibinfo {pages} {054302} (\bibinfo {year} {2011})}\BibitemShut
  {NoStop}%
\bibitem [{\citenamefont {Li}\ \emph {et~al.}(2013)\citenamefont {Li},
  \citenamefont {Song}, \citenamefont {Yao}, \citenamefont {Vretenar},\ and\
  \citenamefont {Meng}}]{li2013}%
  \BibitemOpen
  \bibfield  {author} {\bibinfo {author} {\bibfnamefont {Z.~P.}\ \bibnamefont
  {Li}}, \bibinfo {author} {\bibfnamefont {B.~Y.}\ \bibnamefont {Song}},
  \bibinfo {author} {\bibfnamefont {J.~M.}\ \bibnamefont {Yao}}, \bibinfo
  {author} {\bibfnamefont {D.}~\bibnamefont {Vretenar}},\ and\ \bibinfo
  {author} {\bibfnamefont {J.}~\bibnamefont {Meng}},\ }\href
  {https://doi.org/https://doi.org/10.1016/j.physletb.2013.09.035} {\bibfield
  {journal} {\bibinfo  {journal} {Phys. Lett. B}\ }\textbf {\bibinfo {volume}
  {726}},\ \bibinfo {pages} {866 } (\bibinfo {year} {2013})}\BibitemShut
  {NoStop}%
\bibitem [{\citenamefont {Robledo}\ and\ \citenamefont
  {Butler}(2013)}]{robledo2013}%
  \BibitemOpen
  \bibfield  {author} {\bibinfo {author} {\bibfnamefont {L.~M.}\ \bibnamefont
  {Robledo}}\ and\ \bibinfo {author} {\bibfnamefont {P.~A.}\ \bibnamefont
  {Butler}},\ }\href {https://doi.org/10.1103/PhysRevC.88.051302} {\bibfield
  {journal} {\bibinfo  {journal} {Phys. Rev. C}\ }\textbf {\bibinfo {volume}
  {88}},\ \bibinfo {pages} {051302} (\bibinfo {year} {2013})}\BibitemShut
  {NoStop}%
\bibitem [{\citenamefont {Yao}\ \emph {et~al.}(2015)\citenamefont {Yao},
  \citenamefont {Zhou},\ and\ \citenamefont {Li}}]{yao2015}%
  \BibitemOpen
  \bibfield  {author} {\bibinfo {author} {\bibfnamefont {J.~M.}\ \bibnamefont
  {Yao}}, \bibinfo {author} {\bibfnamefont {E.~F.}\ \bibnamefont {Zhou}},\ and\
  \bibinfo {author} {\bibfnamefont {Z.~P.}\ \bibnamefont {Li}},\ }\href
  {https://doi.org/10.1103/PhysRevC.92.041304} {\bibfield  {journal} {\bibinfo
  {journal} {Phys. Rev. C}\ }\textbf {\bibinfo {volume} {92}},\ \bibinfo
  {pages} {041304} (\bibinfo {year} {2015})}\BibitemShut {NoStop}%
\bibitem [{\citenamefont {Bernard}\ \emph {et~al.}(2016)\citenamefont
  {Bernard}, \citenamefont {Robledo},\ and\ \citenamefont
  {Rodr\'{\i}guez}}]{bernard2016}%
  \BibitemOpen
  \bibfield  {author} {\bibinfo {author} {\bibfnamefont {R.~N.}\ \bibnamefont
  {Bernard}}, \bibinfo {author} {\bibfnamefont {L.~M.}\ \bibnamefont
  {Robledo}},\ and\ \bibinfo {author} {\bibfnamefont {T.~R.}\ \bibnamefont
  {Rodr\'{\i}guez}},\ }\href {https://doi.org/10.1103/PhysRevC.93.061302}
  {\bibfield  {journal} {\bibinfo  {journal} {Phys. Rev. C}\ }\textbf {\bibinfo
  {volume} {93}},\ \bibinfo {pages} {061302} (\bibinfo {year}
  {2016})}\BibitemShut {NoStop}%
\bibitem [{\citenamefont {Xia}\ \emph {et~al.}(2017)\citenamefont {Xia},
  \citenamefont {Tao}, \citenamefont {Lu}, \citenamefont {Li}, \citenamefont
  {Nik\ifmmode \check{s}\else \v{s}\fi{}i\ifmmode~\acute{c}\else \'{c}\fi{}},\
  and\ \citenamefont {Vretenar}}]{xia2017}%
  \BibitemOpen
  \bibfield  {author} {\bibinfo {author} {\bibfnamefont {S.~Y.}\ \bibnamefont
  {Xia}}, \bibinfo {author} {\bibfnamefont {H.}~\bibnamefont {Tao}}, \bibinfo
  {author} {\bibfnamefont {Y.}~\bibnamefont {Lu}}, \bibinfo {author}
  {\bibfnamefont {Z.~P.}\ \bibnamefont {Li}}, \bibinfo {author} {\bibfnamefont
  {T.}~\bibnamefont {Nik\ifmmode \check{s}\else
  \v{s}\fi{}i\ifmmode~\acute{c}\else \'{c}\fi{}}},\ and\ \bibinfo {author}
  {\bibfnamefont {D.}~\bibnamefont {Vretenar}},\ }\href
  {https://doi.org/10.1103/PhysRevC.96.054303} {\bibfield  {journal} {\bibinfo
  {journal} {Phys. Rev. C}\ }\textbf {\bibinfo {volume} {96}},\ \bibinfo
  {pages} {054303} (\bibinfo {year} {2017})}\BibitemShut {NoStop}%
\bibitem [{\citenamefont {Agbemava}\ and\ \citenamefont
  {Afanasjev}(2017)}]{agbemava2017}%
  \BibitemOpen
  \bibfield  {author} {\bibinfo {author} {\bibfnamefont {S.~E.}\ \bibnamefont
  {Agbemava}}\ and\ \bibinfo {author} {\bibfnamefont {A.~V.}\ \bibnamefont
  {Afanasjev}},\ }\href {https://doi.org/10.1103/PhysRevC.96.024301} {\bibfield
   {journal} {\bibinfo  {journal} {Phys. Rev. C}\ }\textbf {\bibinfo {volume}
  {96}},\ \bibinfo {pages} {024301} (\bibinfo {year} {2017})}\BibitemShut
  {NoStop}%
\bibitem [{\citenamefont {Ebata}\ and\ \citenamefont
  {Nakatsukasa}(2017)}]{ebata2017}%
  \BibitemOpen
  \bibfield  {author} {\bibinfo {author} {\bibfnamefont {S.}~\bibnamefont
  {Ebata}}\ and\ \bibinfo {author} {\bibfnamefont {T.}~\bibnamefont
  {Nakatsukasa}},\ }\href {https://doi.org/10.1088/1402-4896/aa6c4c} {\bibfield
   {journal} {\bibinfo  {journal} {Phys. Scr.}\ }\textbf {\bibinfo {volume}
  {92}},\ \bibinfo {pages} {064005} (\bibinfo {year} {2017})}\BibitemShut
  {NoStop}%
\bibitem [{\citenamefont {Marevi\ifmmode~\acute{c}\else \'{c}\fi{}}\ \emph
  {et~al.}(2018)\citenamefont {Marevi\ifmmode~\acute{c}\else \'{c}\fi{}},
  \citenamefont {Ebran}, \citenamefont {Khan}, \citenamefont {Nik\ifmmode
  \check{s}\else \v{s}\fi{}i\ifmmode~\acute{c}\else \'{c}\fi{}},\ and\
  \citenamefont {Vretenar}}]{marevic2018}%
  \BibitemOpen
  \bibfield  {author} {\bibinfo {author} {\bibfnamefont {P.}~\bibnamefont
  {Marevi\ifmmode~\acute{c}\else \'{c}\fi{}}}, \bibinfo {author} {\bibfnamefont
  {J.-P.}\ \bibnamefont {Ebran}}, \bibinfo {author} {\bibfnamefont
  {E.}~\bibnamefont {Khan}}, \bibinfo {author} {\bibfnamefont {T.}~\bibnamefont
  {Nik\ifmmode \check{s}\else \v{s}\fi{}i\ifmmode~\acute{c}\else \'{c}\fi{}}},\
  and\ \bibinfo {author} {\bibfnamefont {D.}~\bibnamefont {Vretenar}},\ }\href
  {https://doi.org/10.1103/PhysRevC.97.024334} {\bibfield  {journal} {\bibinfo
  {journal} {Phys. Rev. C}\ }\textbf {\bibinfo {volume} {97}},\ \bibinfo
  {pages} {024334} (\bibinfo {year} {2018})}\BibitemShut {NoStop}%
\bibitem [{\citenamefont {Robledo}\ \emph {et~al.}(2019)\citenamefont
  {Robledo}, \citenamefont {Rodríguez},\ and\ \citenamefont
  {Rodr{\'{i}}guez-Guzm{\'{a}}n}}]{robledo2019}%
  \BibitemOpen
  \bibfield  {author} {\bibinfo {author} {\bibfnamefont {L.~M.}\ \bibnamefont
  {Robledo}}, \bibinfo {author} {\bibfnamefont {T.~R.}\ \bibnamefont
  {Rodríguez}},\ and\ \bibinfo {author} {\bibfnamefont {R.~R.}\ \bibnamefont
  {Rodr{\'{i}}guez-Guzm{\'{a}}n}},\ }\href
  {http://stacks.iop.org/0954-3899/46/i=1/a=013001} {\bibfield  {journal}
  {\bibinfo  {journal} {J. Phys. G: Nucl. Part. Phys.}\ }\textbf {\bibinfo
  {volume} {46}},\ \bibinfo {pages} {013001} (\bibinfo {year}
  {2019})}\BibitemShut {NoStop}%
\bibitem [{\citenamefont {Cao}\ \emph {et~al.}(2020)\citenamefont {Cao},
  \citenamefont {Agbemava}, \citenamefont {Afanasjev}, \citenamefont
  {Nazarewicz},\ and\ \citenamefont {Olsen}}]{cao2020}%
  \BibitemOpen
  \bibfield  {author} {\bibinfo {author} {\bibfnamefont {Y.}~\bibnamefont
  {Cao}}, \bibinfo {author} {\bibfnamefont {S.~E.}\ \bibnamefont {Agbemava}},
  \bibinfo {author} {\bibfnamefont {A.~V.}\ \bibnamefont {Afanasjev}}, \bibinfo
  {author} {\bibfnamefont {W.}~\bibnamefont {Nazarewicz}},\ and\ \bibinfo
  {author} {\bibfnamefont {E.}~\bibnamefont {Olsen}},\ }\href
  {https://doi.org/10.1103/PhysRevC.102.024311} {\bibfield  {journal} {\bibinfo
   {journal} {Phys. Rev. C}\ }\textbf {\bibinfo {volume} {102}},\ \bibinfo
  {pages} {024311} (\bibinfo {year} {2020})}\BibitemShut {NoStop}%
\bibitem [{\citenamefont {Rodr{\'{\i}}guez-Guzm{\'{a}}n}\ \emph
  {et~al.}(2020)\citenamefont {Rodr{\'{\i}}guez-Guzm{\'{a}}n}, \citenamefont
  {Humadi},\ and\ \citenamefont {Robledo}}]{rayner2020oct}%
  \BibitemOpen
  \bibfield  {author} {\bibinfo {author} {\bibfnamefont {R.}~\bibnamefont
  {Rodr{\'{\i}}guez-Guzm{\'{a}}n}}, \bibinfo {author} {\bibfnamefont {Y.~M.}\
  \bibnamefont {Humadi}},\ and\ \bibinfo {author} {\bibfnamefont {L.~M.}\
  \bibnamefont {Robledo}},\ }\href {https://doi.org/10.1088/1361-6471/abb000}
  {\bibfield  {journal} {\bibinfo  {journal} {J. Phys. G: Nucl. Part. Phys.}\
  }\textbf {\bibinfo {volume} {48}},\ \bibinfo {pages} {015103} (\bibinfo
  {year} {2020})}\BibitemShut {NoStop}%
\bibitem [{\citenamefont {Nomura}\ \emph
  {et~al.}(2021{\natexlab{a}})\citenamefont {Nomura}, \citenamefont {Lotina},
  \citenamefont {Nik\ifmmode \check{s}\else \v{s}\fi{}i\ifmmode~\acute{c}\else
  \'{c}\fi{}},\ and\ \citenamefont {Vretenar}}]{nomura2021qoch}%
  \BibitemOpen
  \bibfield  {author} {\bibinfo {author} {\bibfnamefont {K.}~\bibnamefont
  {Nomura}}, \bibinfo {author} {\bibfnamefont {L.}~\bibnamefont {Lotina}},
  \bibinfo {author} {\bibfnamefont {T.}~\bibnamefont {Nik\ifmmode
  \check{s}\else \v{s}\fi{}i\ifmmode~\acute{c}\else \'{c}\fi{}}},\ and\
  \bibinfo {author} {\bibfnamefont {D.}~\bibnamefont {Vretenar}},\ }\href
  {https://doi.org/10.1103/PhysRevC.103.054301} {\bibfield  {journal} {\bibinfo
   {journal} {Phys. Rev. C}\ }\textbf {\bibinfo {volume} {103}},\ \bibinfo
  {pages} {054301} (\bibinfo {year} {2021}{\natexlab{a}})}\BibitemShut
  {NoStop}%
\bibitem [{\citenamefont {Engel}\ and\ \citenamefont
  {Iachello}(1985)}]{engel1985}%
  \BibitemOpen
  \bibfield  {author} {\bibinfo {author} {\bibfnamefont {J.}~\bibnamefont
  {Engel}}\ and\ \bibinfo {author} {\bibfnamefont {F.}~\bibnamefont
  {Iachello}},\ }\href {https://doi.org/10.1103/PhysRevLett.54.1126} {\bibfield
   {journal} {\bibinfo  {journal} {Phys. Rev. Lett.}\ }\textbf {\bibinfo
  {volume} {54}},\ \bibinfo {pages} {1126} (\bibinfo {year}
  {1985})}\BibitemShut {NoStop}%
\bibitem [{\citenamefont {Engel}\ and\ \citenamefont
  {Iachello}(1987)}]{engel1987}%
  \BibitemOpen
  \bibfield  {author} {\bibinfo {author} {\bibfnamefont {J.}~\bibnamefont
  {Engel}}\ and\ \bibinfo {author} {\bibfnamefont {F.}~\bibnamefont
  {Iachello}},\ }\href {https://doi.org/10.1016/0375-9474(87)90220-X}
  {\bibfield  {journal} {\bibinfo  {journal} {Nucl. Phys. A}\ }\textbf
  {\bibinfo {volume} {472}},\ \bibinfo {pages} {61 } (\bibinfo {year}
  {1987})}\BibitemShut {NoStop}%
\bibitem [{\citenamefont {Otsuka}(1986)}]{otsuka1986}%
  \BibitemOpen
  \bibfield  {author} {\bibinfo {author} {\bibfnamefont {T.}~\bibnamefont
  {Otsuka}},\ }\href {https://doi.org/10.1016/0370-2693(86)90085-7} {\bibfield
  {journal} {\bibinfo  {journal} {Phys. Lett. B}\ }\textbf {\bibinfo {volume}
  {182}},\ \bibinfo {pages} {256 } (\bibinfo {year} {1986})}\BibitemShut
  {NoStop}%
\bibitem [{\citenamefont {Otsuka}\ and\ \citenamefont
  {Sugita}(1988)}]{otsuka1988}%
  \BibitemOpen
  \bibfield  {author} {\bibinfo {author} {\bibfnamefont {T.}~\bibnamefont
  {Otsuka}}\ and\ \bibinfo {author} {\bibfnamefont {M.}~\bibnamefont
  {Sugita}},\ }\href {https://doi.org/10.1016/0370-2693(88)90920-3} {\bibfield
  {journal} {\bibinfo  {journal} {Phys. Lett. B}\ }\textbf {\bibinfo {volume}
  {209}},\ \bibinfo {pages} {140 } (\bibinfo {year} {1988})}\BibitemShut
  {NoStop}%
\bibitem [{\citenamefont {Sugita}\ \emph {et~al.}(1996)\citenamefont {Sugita},
  \citenamefont {Otsuka},\ and\ \citenamefont {von Brentano}}]{sugita1996}%
  \BibitemOpen
  \bibfield  {author} {\bibinfo {author} {\bibfnamefont {M.}~\bibnamefont
  {Sugita}}, \bibinfo {author} {\bibfnamefont {T.}~\bibnamefont {Otsuka}},\
  and\ \bibinfo {author} {\bibfnamefont {P.}~\bibnamefont {von Brentano}},\
  }\href {https://doi.org/http://dx.doi.org/10.1016/S0370-2693(96)80003-7}
  {\bibfield  {journal} {\bibinfo  {journal} {Phys. Lett. B}\ }\textbf
  {\bibinfo {volume} {389}},\ \bibinfo {pages} {642 } (\bibinfo {year}
  {1996})}\BibitemShut {NoStop}%
\bibitem [{\citenamefont {Kusnezov}\ and\ \citenamefont
  {Iachello}(1988)}]{kusnezov1988}%
  \BibitemOpen
  \bibfield  {author} {\bibinfo {author} {\bibfnamefont {D.}~\bibnamefont
  {Kusnezov}}\ and\ \bibinfo {author} {\bibfnamefont {F.}~\bibnamefont
  {Iachello}},\ }\href
  {https://doi.org/https://doi.org/10.1016/0370-2693(88)91166-5} {\bibfield
  {journal} {\bibinfo  {journal} {Physics Letters B}\ }\textbf {\bibinfo
  {volume} {209}},\ \bibinfo {pages} {420 } (\bibinfo {year}
  {1988})}\BibitemShut {NoStop}%
\bibitem [{\citenamefont {Yoshinaga}\ \emph {et~al.}(1993)\citenamefont
  {Yoshinaga}, \citenamefont {Mizusaki},\ and\ \citenamefont
  {Otsuka}}]{yoshinaga1993}%
  \BibitemOpen
  \bibfield  {author} {\bibinfo {author} {\bibfnamefont {N.}~\bibnamefont
  {Yoshinaga}}, \bibinfo {author} {\bibfnamefont {T.}~\bibnamefont
  {Mizusaki}},\ and\ \bibinfo {author} {\bibfnamefont {T.}~\bibnamefont
  {Otsuka}},\ }\href
  {https://doi.org/https://doi.org/10.1016/0375-9474(93)90186-2} {\bibfield
  {journal} {\bibinfo  {journal} {Nucl. Phys. A}\ }\textbf {\bibinfo {volume}
  {559}},\ \bibinfo {pages} {193} (\bibinfo {year} {1993})}\BibitemShut
  {NoStop}%
\bibitem [{\citenamefont {Zamfir}\ and\ \citenamefont
  {Kusnezov}(2001)}]{zamfir2001}%
  \BibitemOpen
  \bibfield  {author} {\bibinfo {author} {\bibfnamefont {N.~V.}\ \bibnamefont
  {Zamfir}}\ and\ \bibinfo {author} {\bibfnamefont {D.}~\bibnamefont
  {Kusnezov}},\ }\href {https://doi.org/10.1103/PhysRevC.63.054306} {\bibfield
  {journal} {\bibinfo  {journal} {Phys. Rev. C}\ }\textbf {\bibinfo {volume}
  {63}},\ \bibinfo {pages} {054306} (\bibinfo {year} {2001})}\BibitemShut
  {NoStop}%
\bibitem [{\citenamefont {Smirnova}\ \emph {et~al.}(2000)\citenamefont
  {Smirnova}, \citenamefont {Pietralla}, \citenamefont {Mizusaki},\ and\
  \citenamefont {{Van Isacker}}}]{SMIRNOVA2000}%
  \BibitemOpen
  \bibfield  {author} {\bibinfo {author} {\bibfnamefont {N.~A.}\ \bibnamefont
  {Smirnova}}, \bibinfo {author} {\bibfnamefont {N.}~\bibnamefont {Pietralla}},
  \bibinfo {author} {\bibfnamefont {T.}~\bibnamefont {Mizusaki}},\ and\
  \bibinfo {author} {\bibfnamefont {P.}~\bibnamefont {{Van Isacker}}},\ }\href
  {https://doi.org/https://doi.org/10.1016/S0375-9474(00)00331-6} {\bibfield
  {journal} {\bibinfo  {journal} {Nuclear Physics A}\ }\textbf {\bibinfo
  {volume} {678}},\ \bibinfo {pages} {235} (\bibinfo {year}
  {2000})}\BibitemShut {NoStop}%
\bibitem [{\citenamefont {Pietralla}\ \emph {et~al.}(2003)\citenamefont
  {Pietralla}, \citenamefont {Fransen}, \citenamefont {Gade}, \citenamefont
  {Smirnova}, \citenamefont {von Brentano}, \citenamefont {Werner},\ and\
  \citenamefont {Yates}}]{pietralla2003}%
  \BibitemOpen
  \bibfield  {author} {\bibinfo {author} {\bibfnamefont {N.}~\bibnamefont
  {Pietralla}}, \bibinfo {author} {\bibfnamefont {C.}~\bibnamefont {Fransen}},
  \bibinfo {author} {\bibfnamefont {A.}~\bibnamefont {Gade}}, \bibinfo {author}
  {\bibfnamefont {N.~A.}\ \bibnamefont {Smirnova}}, \bibinfo {author}
  {\bibfnamefont {P.}~\bibnamefont {von Brentano}}, \bibinfo {author}
  {\bibfnamefont {V.}~\bibnamefont {Werner}},\ and\ \bibinfo {author}
  {\bibfnamefont {S.~W.}\ \bibnamefont {Yates}},\ }\href
  {https://doi.org/10.1103/PhysRevC.68.031305} {\bibfield  {journal} {\bibinfo
  {journal} {Phys. Rev. C}\ }\textbf {\bibinfo {volume} {68}},\ \bibinfo
  {pages} {031305} (\bibinfo {year} {2003})}\BibitemShut {NoStop}%
\bibitem [{\citenamefont {Nomura}\ \emph {et~al.}(2013)\citenamefont {Nomura},
  \citenamefont {Vretenar},\ and\ \citenamefont {Lu}}]{nomura2013oct}%
  \BibitemOpen
  \bibfield  {author} {\bibinfo {author} {\bibfnamefont {K.}~\bibnamefont
  {Nomura}}, \bibinfo {author} {\bibfnamefont {D.}~\bibnamefont {Vretenar}},\
  and\ \bibinfo {author} {\bibfnamefont {B.-N.}\ \bibnamefont {Lu}},\ }\href
  {https://doi.org/10.1103/PhysRevC.88.021303} {\bibfield  {journal} {\bibinfo
  {journal} {Phys. Rev. C}\ }\textbf {\bibinfo {volume} {88}},\ \bibinfo
  {pages} {021303} (\bibinfo {year} {2013})}\BibitemShut {NoStop}%
\bibitem [{\citenamefont {Nomura}\ \emph {et~al.}(2014)\citenamefont {Nomura},
  \citenamefont {Vretenar}, \citenamefont {Nik\ifmmode \check{s}\else
  \v{s}\fi{}i\ifmmode~\acute{c}\else \'{c}\fi{}},\ and\ \citenamefont
  {Lu}}]{nomura2014}%
  \BibitemOpen
  \bibfield  {author} {\bibinfo {author} {\bibfnamefont {K.}~\bibnamefont
  {Nomura}}, \bibinfo {author} {\bibfnamefont {D.}~\bibnamefont {Vretenar}},
  \bibinfo {author} {\bibfnamefont {T.}~\bibnamefont {Nik\ifmmode
  \check{s}\else \v{s}\fi{}i\ifmmode~\acute{c}\else \'{c}\fi{}}},\ and\
  \bibinfo {author} {\bibfnamefont {B.-N.}\ \bibnamefont {Lu}},\ }\href
  {https://doi.org/10.1103/PhysRevC.89.024312} {\bibfield  {journal} {\bibinfo
  {journal} {Phys. Rev. C}\ }\textbf {\bibinfo {volume} {89}},\ \bibinfo
  {pages} {024312} (\bibinfo {year} {2014})}\BibitemShut {NoStop}%
\bibitem [{\citenamefont {Nomura}\ \emph {et~al.}(2015)\citenamefont {Nomura},
  \citenamefont {Rodr\'{\i}guez-Guzm\'an},\ and\ \citenamefont
  {Robledo}}]{nomura2015}%
  \BibitemOpen
  \bibfield  {author} {\bibinfo {author} {\bibfnamefont {K.}~\bibnamefont
  {Nomura}}, \bibinfo {author} {\bibfnamefont {R.}~\bibnamefont
  {Rodr\'{\i}guez-Guzm\'an}},\ and\ \bibinfo {author} {\bibfnamefont {L.~M.}\
  \bibnamefont {Robledo}},\ }\href {https://doi.org/10.1103/PhysRevC.92.014312}
  {\bibfield  {journal} {\bibinfo  {journal} {Phys. Rev. C}\ }\textbf {\bibinfo
  {volume} {92}},\ \bibinfo {pages} {014312} (\bibinfo {year}
  {2015})}\BibitemShut {NoStop}%
\bibitem [{\citenamefont {Nomura}\ \emph {et~al.}(2020)\citenamefont {Nomura},
  \citenamefont {Rodr\'{\i}guez-Guzm\'an}, \citenamefont {Humadi},
  \citenamefont {Robledo},\ and\ \citenamefont
  {Garc\'{\i}a-Ramos}}]{nomura2020oct}%
  \BibitemOpen
  \bibfield  {author} {\bibinfo {author} {\bibfnamefont {K.}~\bibnamefont
  {Nomura}}, \bibinfo {author} {\bibfnamefont {R.}~\bibnamefont
  {Rodr\'{\i}guez-Guzm\'an}}, \bibinfo {author} {\bibfnamefont {Y.~M.}\
  \bibnamefont {Humadi}}, \bibinfo {author} {\bibfnamefont {L.~M.}\
  \bibnamefont {Robledo}},\ and\ \bibinfo {author} {\bibfnamefont {J.~E.}\
  \bibnamefont {Garc\'{\i}a-Ramos}},\ }\href
  {https://doi.org/10.1103/PhysRevC.102.064326} {\bibfield  {journal} {\bibinfo
   {journal} {Phys. Rev. C}\ }\textbf {\bibinfo {volume} {102}},\ \bibinfo
  {pages} {064326} (\bibinfo {year} {2020})}\BibitemShut {NoStop}%
\bibitem [{\citenamefont {Nomura}\ \emph
  {et~al.}(2021{\natexlab{b}})\citenamefont {Nomura}, \citenamefont
  {Rodr\'{\i}guez-Guzm\'an}, \citenamefont {Robledo},\ and\ \citenamefont
  {Garc\'{\i}a-Ramos}}]{nomura2021oct-u}%
  \BibitemOpen
  \bibfield  {author} {\bibinfo {author} {\bibfnamefont {K.}~\bibnamefont
  {Nomura}}, \bibinfo {author} {\bibfnamefont {R.}~\bibnamefont
  {Rodr\'{\i}guez-Guzm\'an}}, \bibinfo {author} {\bibfnamefont
  {L.}~\bibnamefont {Robledo}},\ and\ \bibinfo {author} {\bibfnamefont
  {J.}~\bibnamefont {Garc\'{\i}a-Ramos}},\ }\href
  {https://doi.org/10.1103/PhysRevC.103.044311} {\bibfield  {journal} {\bibinfo
   {journal} {Phys. Rev. C}\ }\textbf {\bibinfo {volume} {103}},\ \bibinfo
  {pages} {044311} (\bibinfo {year} {2021}{\natexlab{b}})}\BibitemShut
  {NoStop}%
\bibitem [{\citenamefont {Nomura}\ \emph
  {et~al.}(2021{\natexlab{c}})\citenamefont {Nomura}, \citenamefont
  {Rodr\'{\i}guez-Guzm\'an}, \citenamefont {Robledo}, \citenamefont
  {Garc\'{\i}a-Ramos},\ and\ \citenamefont {Hern\'andez}}]{nomura2021oct-ba}%
  \BibitemOpen
  \bibfield  {author} {\bibinfo {author} {\bibfnamefont {K.}~\bibnamefont
  {Nomura}}, \bibinfo {author} {\bibfnamefont {R.}~\bibnamefont
  {Rodr\'{\i}guez-Guzm\'an}}, \bibinfo {author} {\bibfnamefont {L.~M.}\
  \bibnamefont {Robledo}}, \bibinfo {author} {\bibfnamefont {J.~E.}\
  \bibnamefont {Garc\'{\i}a-Ramos}},\ and\ \bibinfo {author} {\bibfnamefont
  {N.~C.}\ \bibnamefont {Hern\'andez}},\ }\href
  {https://doi.org/10.1103/PhysRevC.104.044324} {\bibfield  {journal} {\bibinfo
   {journal} {Phys. Rev. C}\ }\textbf {\bibinfo {volume} {104}},\ \bibinfo
  {pages} {044324} (\bibinfo {year} {2021}{\natexlab{c}})}\BibitemShut
  {NoStop}%
\bibitem [{\citenamefont {Nomura}\ \emph
  {et~al.}(2021{\natexlab{d}})\citenamefont {Nomura}, \citenamefont
  {Rodr\'{\i}guez-Guzm\'an},\ and\ \citenamefont {Robledo}}]{nomura2021oct-zn}%
  \BibitemOpen
  \bibfield  {author} {\bibinfo {author} {\bibfnamefont {K.}~\bibnamefont
  {Nomura}}, \bibinfo {author} {\bibfnamefont {R.}~\bibnamefont
  {Rodr\'{\i}guez-Guzm\'an}},\ and\ \bibinfo {author} {\bibfnamefont {L.~M.}\
  \bibnamefont {Robledo}},\ }\href
  {https://doi.org/10.1103/PhysRevC.104.054320} {\bibfield  {journal} {\bibinfo
   {journal} {Phys. Rev. C}\ }\textbf {\bibinfo {volume} {104}},\ \bibinfo
  {pages} {054320} (\bibinfo {year} {2021}{\natexlab{d}})}\BibitemShut
  {NoStop}%
\bibitem [{\citenamefont {Hennig}\ \emph {et~al.}(2014)\citenamefont {Hennig},
  \citenamefont {Spieker}, \citenamefont {Werner}, \citenamefont {Ahn},
  \citenamefont {Anagnostatou}, \citenamefont {Cooper}, \citenamefont {Derya},
  \citenamefont {Elvers}, \citenamefont {Endres}, \citenamefont {Goddard},
  \citenamefont {Heinz}, \citenamefont {Hughes}, \citenamefont {Ilie},
  \citenamefont {Mineva}, \citenamefont {Petkov}, \citenamefont {Pickstone},
  \citenamefont {Pietralla}, \citenamefont {Radeck}, \citenamefont {Ross},
  \citenamefont {Savran},\ and\ \citenamefont {Zilges}}]{hennig2014}%
  \BibitemOpen
  \bibfield  {author} {\bibinfo {author} {\bibfnamefont {A.}~\bibnamefont
  {Hennig}}, \bibinfo {author} {\bibfnamefont {M.}~\bibnamefont {Spieker}},
  \bibinfo {author} {\bibfnamefont {V.}~\bibnamefont {Werner}}, \bibinfo
  {author} {\bibfnamefont {T.}~\bibnamefont {Ahn}}, \bibinfo {author}
  {\bibfnamefont {V.}~\bibnamefont {Anagnostatou}}, \bibinfo {author}
  {\bibfnamefont {N.}~\bibnamefont {Cooper}}, \bibinfo {author} {\bibfnamefont
  {V.}~\bibnamefont {Derya}}, \bibinfo {author} {\bibfnamefont
  {M.}~\bibnamefont {Elvers}}, \bibinfo {author} {\bibfnamefont
  {J.}~\bibnamefont {Endres}}, \bibinfo {author} {\bibfnamefont
  {P.}~\bibnamefont {Goddard}}, \bibinfo {author} {\bibfnamefont
  {A.}~\bibnamefont {Heinz}}, \bibinfo {author} {\bibfnamefont {R.~O.}\
  \bibnamefont {Hughes}}, \bibinfo {author} {\bibfnamefont {G.}~\bibnamefont
  {Ilie}}, \bibinfo {author} {\bibfnamefont {M.~N.}\ \bibnamefont {Mineva}},
  \bibinfo {author} {\bibfnamefont {P.}~\bibnamefont {Petkov}}, \bibinfo
  {author} {\bibfnamefont {S.~G.}\ \bibnamefont {Pickstone}}, \bibinfo {author}
  {\bibfnamefont {N.}~\bibnamefont {Pietralla}}, \bibinfo {author}
  {\bibfnamefont {D.}~\bibnamefont {Radeck}}, \bibinfo {author} {\bibfnamefont
  {T.~J.}\ \bibnamefont {Ross}}, \bibinfo {author} {\bibfnamefont
  {D.}~\bibnamefont {Savran}},\ and\ \bibinfo {author} {\bibfnamefont
  {A.}~\bibnamefont {Zilges}},\ }\href
  {https://doi.org/10.1103/PhysRevC.90.051302} {\bibfield  {journal} {\bibinfo
  {journal} {Phys. Rev. C}\ }\textbf {\bibinfo {volume} {90}},\ \bibinfo
  {pages} {051302} (\bibinfo {year} {2014})}\BibitemShut {NoStop}%
\bibitem [{\citenamefont {Vallejos}\ and\ \citenamefont
  {Barea}(2021)}]{vallejos2021}%
  \BibitemOpen
  \bibfield  {author} {\bibinfo {author} {\bibfnamefont {O.}~\bibnamefont
  {Vallejos}}\ and\ \bibinfo {author} {\bibfnamefont {J.}~\bibnamefont
  {Barea}},\ }\href {https://doi.org/10.1103/PhysRevC.104.014308} {\bibfield
  {journal} {\bibinfo  {journal} {Phys. Rev. C}\ }\textbf {\bibinfo {volume}
  {104}},\ \bibinfo {pages} {014308} (\bibinfo {year} {2021})}\BibitemShut
  {NoStop}%
\bibitem [{\citenamefont {Bonatsos}\ \emph {et~al.}(2005)\citenamefont
  {Bonatsos}, \citenamefont {Lenis}, \citenamefont {Minkov}, \citenamefont
  {Petrellis},\ and\ \citenamefont {Yotov}}]{bonatsos2005}%
  \BibitemOpen
  \bibfield  {author} {\bibinfo {author} {\bibfnamefont {D.}~\bibnamefont
  {Bonatsos}}, \bibinfo {author} {\bibfnamefont {D.}~\bibnamefont {Lenis}},
  \bibinfo {author} {\bibfnamefont {N.}~\bibnamefont {Minkov}}, \bibinfo
  {author} {\bibfnamefont {D.}~\bibnamefont {Petrellis}},\ and\ \bibinfo
  {author} {\bibfnamefont {P.}~\bibnamefont {Yotov}},\ }\href
  {https://doi.org/10.1103/PhysRevC.71.064309} {\bibfield  {journal} {\bibinfo
  {journal} {Phys. Rev. C}\ }\textbf {\bibinfo {volume} {71}},\ \bibinfo
  {pages} {064309} (\bibinfo {year} {2005})}\BibitemShut {NoStop}%
\bibitem [{\citenamefont {Lenis}\ and\ \citenamefont
  {Bonatsos}(2006)}]{lenis2006}%
  \BibitemOpen
  \bibfield  {author} {\bibinfo {author} {\bibfnamefont {D.}~\bibnamefont
  {Lenis}}\ and\ \bibinfo {author} {\bibfnamefont {D.}~\bibnamefont
  {Bonatsos}},\ }\href {https://doi.org/10.1016/j.physletb.2005.12.016}
  {\bibfield  {journal} {\bibinfo  {journal} {Phys. Lett. B}\ }\textbf
  {\bibinfo {volume} {633}},\ \bibinfo {pages} {474} (\bibinfo {year}
  {2006})}\BibitemShut {NoStop}%
\bibitem [{\citenamefont {Bizzeti}\ and\ \citenamefont
  {Bizzeti-Sona}(2013)}]{bizzeti2013}%
  \BibitemOpen
  \bibfield  {author} {\bibinfo {author} {\bibfnamefont {P.~G.}\ \bibnamefont
  {Bizzeti}}\ and\ \bibinfo {author} {\bibfnamefont {A.~M.}\ \bibnamefont
  {Bizzeti-Sona}},\ }\href {https://doi.org/10.1103/PhysRevC.88.011305}
  {\bibfield  {journal} {\bibinfo  {journal} {Phys. Rev. C}\ }\textbf {\bibinfo
  {volume} {88}},\ \bibinfo {pages} {011305} (\bibinfo {year}
  {2013})}\BibitemShut {NoStop}%
\bibitem [{\citenamefont {Bonatsos}\ \emph {et~al.}(2015)\citenamefont
  {Bonatsos}, \citenamefont {Martinou}, \citenamefont {Minkov}, \citenamefont
  {Karampagia},\ and\ \citenamefont {Petrellis}}]{bonatsos2015}%
  \BibitemOpen
  \bibfield  {author} {\bibinfo {author} {\bibfnamefont {D.}~\bibnamefont
  {Bonatsos}}, \bibinfo {author} {\bibfnamefont {A.}~\bibnamefont {Martinou}},
  \bibinfo {author} {\bibfnamefont {N.}~\bibnamefont {Minkov}}, \bibinfo
  {author} {\bibfnamefont {S.}~\bibnamefont {Karampagia}},\ and\ \bibinfo
  {author} {\bibfnamefont {D.}~\bibnamefont {Petrellis}},\ }\href
  {https://doi.org/10.1103/PhysRevC.91.054315} {\bibfield  {journal} {\bibinfo
  {journal} {Phys. Rev. C}\ }\textbf {\bibinfo {volume} {91}},\ \bibinfo
  {pages} {054315} (\bibinfo {year} {2015})}\BibitemShut {NoStop}%
\bibitem [{\citenamefont {Shneidman}\ \emph {et~al.}(2002)\citenamefont
  {Shneidman}, \citenamefont {Adamian}, \citenamefont {Antonenko},
  \citenamefont {Jolos},\ and\ \citenamefont {Scheid}}]{shneidman2002}%
  \BibitemOpen
  \bibfield  {author} {\bibinfo {author} {\bibfnamefont {T.~M.}\ \bibnamefont
  {Shneidman}}, \bibinfo {author} {\bibfnamefont {G.~G.}\ \bibnamefont
  {Adamian}}, \bibinfo {author} {\bibfnamefont {N.~V.}\ \bibnamefont
  {Antonenko}}, \bibinfo {author} {\bibfnamefont {R.~V.}\ \bibnamefont
  {Jolos}},\ and\ \bibinfo {author} {\bibfnamefont {W.}~\bibnamefont
  {Scheid}},\ }\href
  {https://doi.org/http://dx.doi.org/10.1016/S0370-2693(01)01512-X} {\bibfield
  {journal} {\bibinfo  {journal} {Phys. Lett. B}\ }\textbf {\bibinfo {volume}
  {526}},\ \bibinfo {pages} {322 } (\bibinfo {year} {2002})}\BibitemShut
  {NoStop}%
\bibitem [{\citenamefont {Shneidman}\ \emph {et~al.}(2003)\citenamefont
  {Shneidman}, \citenamefont {Adamian}, \citenamefont {Antonenko},
  \citenamefont {Jolos},\ and\ \citenamefont {Scheid}}]{shneidman2003}%
  \BibitemOpen
  \bibfield  {author} {\bibinfo {author} {\bibfnamefont {T.~M.}\ \bibnamefont
  {Shneidman}}, \bibinfo {author} {\bibfnamefont {G.~G.}\ \bibnamefont
  {Adamian}}, \bibinfo {author} {\bibfnamefont {N.~V.}\ \bibnamefont
  {Antonenko}}, \bibinfo {author} {\bibfnamefont {R.~V.}\ \bibnamefont
  {Jolos}},\ and\ \bibinfo {author} {\bibfnamefont {W.}~\bibnamefont
  {Scheid}},\ }\href {https://doi.org/10.1103/PhysRevC.67.014313} {\bibfield
  {journal} {\bibinfo  {journal} {Phys. Rev. C}\ }\textbf {\bibinfo {volume}
  {67}},\ \bibinfo {pages} {014313} (\bibinfo {year} {2003})}\BibitemShut
  {NoStop}%
\bibitem [{\citenamefont {Jolos}\ \emph {et~al.}(2012)\citenamefont {Jolos},
  \citenamefont {von Brentano},\ and\ \citenamefont {Jolie}}]{jolos2012}%
  \BibitemOpen
  \bibfield  {author} {\bibinfo {author} {\bibfnamefont {R.~V.}\ \bibnamefont
  {Jolos}}, \bibinfo {author} {\bibfnamefont {P.}~\bibnamefont {von
  Brentano}},\ and\ \bibinfo {author} {\bibfnamefont {J.}~\bibnamefont
  {Jolie}},\ }\href {https://doi.org/10.1103/PhysRevC.86.024319} {\bibfield
  {journal} {\bibinfo  {journal} {Phys. Rev. C}\ }\textbf {\bibinfo {volume}
  {86}},\ \bibinfo {pages} {024319} (\bibinfo {year} {2012})}\BibitemShut
  {NoStop}%
\bibitem [{\citenamefont {Brown}(2000)}]{brown2000}%
  \BibitemOpen
  \bibfield  {author} {\bibinfo {author} {\bibfnamefont {B.~A.}\ \bibnamefont
  {Brown}},\ }\href {https://doi.org/10.1103/PhysRevLett.85.5300} {\bibfield
  {journal} {\bibinfo  {journal} {Phys. Rev. Lett.}\ }\textbf {\bibinfo
  {volume} {85}},\ \bibinfo {pages} {5300} (\bibinfo {year}
  {2000})}\BibitemShut {NoStop}%
\bibitem [{\citenamefont {Kaneko}\ \emph {et~al.}(2002)\citenamefont {Kaneko},
  \citenamefont {Hasegawa},\ and\ \citenamefont {Mizusaki}}]{kaneko2002}%
  \BibitemOpen
  \bibfield  {author} {\bibinfo {author} {\bibfnamefont {K.}~\bibnamefont
  {Kaneko}}, \bibinfo {author} {\bibfnamefont {M.}~\bibnamefont {Hasegawa}},\
  and\ \bibinfo {author} {\bibfnamefont {T.}~\bibnamefont {Mizusaki}},\ }\href
  {https://doi.org/10.1103/PhysRevC.66.051306} {\bibfield  {journal} {\bibinfo
  {journal} {Phys. Rev. C}\ }\textbf {\bibinfo {volume} {66}},\ \bibinfo
  {pages} {051306} (\bibinfo {year} {2002})}\BibitemShut {NoStop}%
\bibitem [{\citenamefont {Yoshinaga}\ \emph {et~al.}(2018)\citenamefont
  {Yoshinaga}, \citenamefont {Yanase}, \citenamefont {Higashiyama},\ and\
  \citenamefont {Teruya}}]{yoshinaga2018}%
  \BibitemOpen
  \bibfield  {author} {\bibinfo {author} {\bibfnamefont {N.}~\bibnamefont
  {Yoshinaga}}, \bibinfo {author} {\bibfnamefont {K.}~\bibnamefont {Yanase}},
  \bibinfo {author} {\bibfnamefont {K.}~\bibnamefont {Higashiyama}},\ and\
  \bibinfo {author} {\bibfnamefont {E.}~\bibnamefont {Teruya}},\ }\href
  {https://doi.org/10.1103/PhysRevC.98.044321} {\bibfield  {journal} {\bibinfo
  {journal} {Phys. Rev. C}\ }\textbf {\bibinfo {volume} {98}},\ \bibinfo
  {pages} {044321} (\bibinfo {year} {2018})}\BibitemShut {NoStop}%
\bibitem [{\citenamefont {Van~Isacker}(2020)}]{vanisacker2020}%
  \BibitemOpen
  \bibfield  {author} {\bibinfo {author} {\bibfnamefont {P.}~\bibnamefont
  {Van~Isacker}},\ }\href {https://doi.org/10.1140/epjst/e2020-000026-x}
  {\bibfield  {journal} {\bibinfo  {journal} {Eur. Phys. J. Special Topics}\
  }\textbf {\bibinfo {volume} {229}},\ \bibinfo {pages} {2443} (\bibinfo {year}
  {2020})}\BibitemShut {NoStop}%
\bibitem [{\citenamefont {Bender}\ \emph {et~al.}(2003)\citenamefont {Bender},
  \citenamefont {Heenen},\ and\ \citenamefont {Reinhard}}]{bender2003}%
  \BibitemOpen
  \bibfield  {author} {\bibinfo {author} {\bibfnamefont {M.}~\bibnamefont
  {Bender}}, \bibinfo {author} {\bibfnamefont {P.-H.}\ \bibnamefont {Heenen}},\
  and\ \bibinfo {author} {\bibfnamefont {P.-G.}\ \bibnamefont {Reinhard}},\
  }\href {https://doi.org/10.1103/RevModPhys.75.121} {\bibfield  {journal}
  {\bibinfo  {journal} {Rev. Mod. Phys.}\ }\textbf {\bibinfo {volume} {75}},\
  \bibinfo {pages} {121} (\bibinfo {year} {2003})}\BibitemShut {NoStop}%
\bibitem [{\citenamefont {Vretenar}\ \emph {et~al.}(2005)\citenamefont
  {Vretenar}, \citenamefont {Afanasjev}, \citenamefont {Lalazissis},\ and\
  \citenamefont {Ring}}]{vretenar2005}%
  \BibitemOpen
  \bibfield  {author} {\bibinfo {author} {\bibfnamefont {D.}~\bibnamefont
  {Vretenar}}, \bibinfo {author} {\bibfnamefont {A.~V.}\ \bibnamefont
  {Afanasjev}}, \bibinfo {author} {\bibfnamefont {G.~A.}\ \bibnamefont
  {Lalazissis}},\ and\ \bibinfo {author} {\bibfnamefont {P.}~\bibnamefont
  {Ring}},\ }\href {https://doi.org/10.1016/j.physrep.2004.10.001} {\bibfield
  {journal} {\bibinfo  {journal} {Phys. Rep.}\ }\textbf {\bibinfo {volume}
  {409}},\ \bibinfo {pages} {101 } (\bibinfo {year} {2005})}\BibitemShut
  {NoStop}%
\bibitem [{\citenamefont {Nik\ifmmode \check{s}\else
  \v{s}\fi{}i\ifmmode~\acute{c}\else \'{c}\fi{}}\ \emph
  {et~al.}(2011)\citenamefont {Nik\ifmmode \check{s}\else
  \v{s}\fi{}i\ifmmode~\acute{c}\else \'{c}\fi{}}, \citenamefont {Vretenar},\
  and\ \citenamefont {Ring}}]{niksic2011}%
  \BibitemOpen
  \bibfield  {author} {\bibinfo {author} {\bibfnamefont {T.}~\bibnamefont
  {Nik\ifmmode \check{s}\else \v{s}\fi{}i\ifmmode~\acute{c}\else \'{c}\fi{}}},
  \bibinfo {author} {\bibfnamefont {D.}~\bibnamefont {Vretenar}},\ and\
  \bibinfo {author} {\bibfnamefont {P.}~\bibnamefont {Ring}},\ }\href
  {https://doi.org/10.1016/j.ppnp.2011.01.055} {\bibfield  {journal} {\bibinfo
  {journal} {Prog. Part. Nucl. Phys.}\ }\textbf {\bibinfo {volume} {66}},\
  \bibinfo {pages} {519} (\bibinfo {year} {2011})}\BibitemShut {NoStop}%
\bibitem [{\citenamefont {Ring}\ and\ \citenamefont {Schuck}(1980)}]{RS}%
  \BibitemOpen
  \bibfield  {author} {\bibinfo {author} {\bibfnamefont {P.}~\bibnamefont
  {Ring}}\ and\ \bibinfo {author} {\bibfnamefont {P.}~\bibnamefont {Schuck}},\
  }\href@noop {} {\emph {\bibinfo {title} {The Nuclear Many-Body Problem}}}\
  (\bibinfo  {publisher} {Springer-Verlag, Berlin},\ \bibinfo {year}
  {1980})\BibitemShut {NoStop}%
\bibitem [{\citenamefont {Nik\ifmmode \check{s}\else
  \v{s}\fi{}i\ifmmode~\acute{c}\else \'{c}\fi{}}\ \emph
  {et~al.}(2008)\citenamefont {Nik\ifmmode \check{s}\else
  \v{s}\fi{}i\ifmmode~\acute{c}\else \'{c}\fi{}}, \citenamefont {Vretenar},\
  and\ \citenamefont {Ring}}]{DDPC1}%
  \BibitemOpen
  \bibfield  {author} {\bibinfo {author} {\bibfnamefont {T.}~\bibnamefont
  {Nik\ifmmode \check{s}\else \v{s}\fi{}i\ifmmode~\acute{c}\else \'{c}\fi{}}},
  \bibinfo {author} {\bibfnamefont {D.}~\bibnamefont {Vretenar}},\ and\
  \bibinfo {author} {\bibfnamefont {P.}~\bibnamefont {Ring}},\ }\href
  {https://doi.org/10.1103/PhysRevC.78.034318} {\bibfield  {journal} {\bibinfo
  {journal} {Phys. Rev. C}\ }\textbf {\bibinfo {volume} {78}},\ \bibinfo
  {pages} {034318} (\bibinfo {year} {2008})}\BibitemShut {NoStop}%
\bibitem [{\citenamefont {Nik\v{s}i\'c}\ \emph {et~al.}(2014)\citenamefont
  {Nik\v{s}i\'c}, \citenamefont {Paar}, \citenamefont {Vretenar},\ and\
  \citenamefont {Ring}}]{DIRHB}%
  \BibitemOpen
  \bibfield  {author} {\bibinfo {author} {\bibfnamefont {T.}~\bibnamefont
  {Nik\v{s}i\'c}}, \bibinfo {author} {\bibfnamefont {N.}~\bibnamefont {Paar}},
  \bibinfo {author} {\bibfnamefont {D.}~\bibnamefont {Vretenar}},\ and\
  \bibinfo {author} {\bibfnamefont {P.}~\bibnamefont {Ring}},\ }\href
  {https://doi.org/https://doi.org/10.1016/j.cpc.2014.02.027} {\bibfield
  {journal} {\bibinfo  {journal} {Comput. Phys. Commun.}\ }\textbf {\bibinfo
  {volume} {185}},\ \bibinfo {pages} {1808} (\bibinfo {year}
  {2014})}\BibitemShut {NoStop}%
\bibitem [{\citenamefont {Tian}\ \emph {et~al.}(2009)\citenamefont {Tian},
  \citenamefont {Ma},\ and\ \citenamefont {Ring}}]{tian2009}%
  \BibitemOpen
  \bibfield  {author} {\bibinfo {author} {\bibfnamefont {Y.}~\bibnamefont
  {Tian}}, \bibinfo {author} {\bibfnamefont {Z.~Y.}\ \bibnamefont {Ma}},\ and\
  \bibinfo {author} {\bibfnamefont {P.}~\bibnamefont {Ring}},\ }\href
  {https://doi.org/10.1016/j.physletb.2009.04.067} {\bibfield  {journal}
  {\bibinfo  {journal} {Phys. Lett. B}\ }\textbf {\bibinfo {volume} {676}},\
  \bibinfo {pages} {44 } (\bibinfo {year} {2009})}\BibitemShut {NoStop}%
\bibitem [{\citenamefont {Teeti}\ and\ \citenamefont
  {Afanasjev}(2021)}]{teeti2021}%
  \BibitemOpen
  \bibfield  {author} {\bibinfo {author} {\bibfnamefont {S.}~\bibnamefont
  {Teeti}}\ and\ \bibinfo {author} {\bibfnamefont {A.~V.}\ \bibnamefont
  {Afanasjev}},\ }\href {https://doi.org/10.1103/PhysRevC.103.034310}
  {\bibfield  {journal} {\bibinfo  {journal} {Phys. Rev. C}\ }\textbf {\bibinfo
  {volume} {103}},\ \bibinfo {pages} {034310} (\bibinfo {year}
  {2021})}\BibitemShut {NoStop}%
\bibitem [{\citenamefont {Otsuka}\ \emph
  {et~al.}(1978{\natexlab{a}})\citenamefont {Otsuka}, \citenamefont {Arima},
  \citenamefont {Iachello},\ and\ \citenamefont {Talmi}}]{OAIT}%
  \BibitemOpen
  \bibfield  {author} {\bibinfo {author} {\bibfnamefont {T.}~\bibnamefont
  {Otsuka}}, \bibinfo {author} {\bibfnamefont {A.}~\bibnamefont {Arima}},
  \bibinfo {author} {\bibfnamefont {F.}~\bibnamefont {Iachello}},\ and\
  \bibinfo {author} {\bibfnamefont {I.}~\bibnamefont {Talmi}},\ }\href
  {https://doi.org/10.1016/0370-2693(78)90260-5} {\bibfield  {journal}
  {\bibinfo  {journal} {Phys. Lett. B}\ }\textbf {\bibinfo {volume} {76}},\
  \bibinfo {pages} {139 } (\bibinfo {year} {1978}{\natexlab{a}})}\BibitemShut
  {NoStop}%
\bibitem [{\citenamefont {Otsuka}\ \emph
  {et~al.}(1978{\natexlab{b}})\citenamefont {Otsuka}, \citenamefont {Arima},\
  and\ \citenamefont {Iachello}}]{OAI}%
  \BibitemOpen
  \bibfield  {author} {\bibinfo {author} {\bibfnamefont {T.}~\bibnamefont
  {Otsuka}}, \bibinfo {author} {\bibfnamefont {A.}~\bibnamefont {Arima}},\ and\
  \bibinfo {author} {\bibfnamefont {F.}~\bibnamefont {Iachello}},\ }\href
  {https://doi.org/10.1016/0375-9474(78)90532-8} {\bibfield  {journal}
  {\bibinfo  {journal} {Nucl. Phys. A}\ }\textbf {\bibinfo {volume} {309}},\
  \bibinfo {pages} {1} (\bibinfo {year} {1978}{\natexlab{b}})}\BibitemShut
  {NoStop}%
\bibitem [{\citenamefont {Ginocchio}\ and\ \citenamefont
  {Kirson}(1980)}]{ginocchio1980}%
  \BibitemOpen
  \bibfield  {author} {\bibinfo {author} {\bibfnamefont {J.~N.}\ \bibnamefont
  {Ginocchio}}\ and\ \bibinfo {author} {\bibfnamefont {M.~W.}\ \bibnamefont
  {Kirson}},\ }\href {https://doi.org/10.1016/0375-9474(80)90387-5} {\bibfield
  {journal} {\bibinfo  {journal} {Nucl. Phys. A}\ }\textbf {\bibinfo {volume}
  {350}},\ \bibinfo {pages} {31} (\bibinfo {year} {1980})}\BibitemShut
  {NoStop}%
\bibitem [{\citenamefont {Nomura}\ \emph {et~al.}(2008)\citenamefont {Nomura},
  \citenamefont {Shimizu},\ and\ \citenamefont {Otsuka}}]{nomura2008}%
  \BibitemOpen
  \bibfield  {author} {\bibinfo {author} {\bibfnamefont {K.}~\bibnamefont
  {Nomura}}, \bibinfo {author} {\bibfnamefont {N.}~\bibnamefont {Shimizu}},\
  and\ \bibinfo {author} {\bibfnamefont {T.}~\bibnamefont {Otsuka}},\ }\href
  {https://doi.org/10.1103/PhysRevLett.101.142501} {\bibfield  {journal}
  {\bibinfo  {journal} {Phys. Rev. Lett.}\ }\textbf {\bibinfo {volume} {101}},\
  \bibinfo {pages} {142501} (\bibinfo {year} {2008})}\BibitemShut {NoStop}%
\bibitem [{\citenamefont {Nomura}\ \emph {et~al.}(2010)\citenamefont {Nomura},
  \citenamefont {Shimizu},\ and\ \citenamefont {Otsuka}}]{nomura2010}%
  \BibitemOpen
  \bibfield  {author} {\bibinfo {author} {\bibfnamefont {K.}~\bibnamefont
  {Nomura}}, \bibinfo {author} {\bibfnamefont {N.}~\bibnamefont {Shimizu}},\
  and\ \bibinfo {author} {\bibfnamefont {T.}~\bibnamefont {Otsuka}},\ }\href
  {https://doi.org/10.1103/PhysRevC.81.044307} {\bibfield  {journal} {\bibinfo
  {journal} {Phys. Rev. C}\ }\textbf {\bibinfo {volume} {81}},\ \bibinfo
  {pages} {044307} (\bibinfo {year} {2010})}\BibitemShut {NoStop}%
\bibitem [{\citenamefont {Heinze}()}]{arbmodel}%
  \BibitemOpen
  \bibfield  {author} {\bibinfo {author} {\bibfnamefont {S.}~\bibnamefont
  {Heinze}},\ }\href@noop {} {}\bibinfo {note} {{}computer program ARBMODEL,
  University of Cologne (2008)}\BibitemShut {NoStop}%
\bibitem [{\citenamefont {Lalazissis}\ \emph {et~al.}(2005)\citenamefont
  {Lalazissis}, \citenamefont {Nik\ifmmode \check{s}\else
  \v{s}\fi{}i\ifmmode~\acute{c}\else \'{c}\fi{}}, \citenamefont {Vretenar},\
  and\ \citenamefont {Ring}}]{DDME2}%
  \BibitemOpen
  \bibfield  {author} {\bibinfo {author} {\bibfnamefont {G.~A.}\ \bibnamefont
  {Lalazissis}}, \bibinfo {author} {\bibfnamefont {T.}~\bibnamefont
  {Nik\ifmmode \check{s}\else \v{s}\fi{}i\ifmmode~\acute{c}\else \'{c}\fi{}}},
  \bibinfo {author} {\bibfnamefont {D.}~\bibnamefont {Vretenar}},\ and\
  \bibinfo {author} {\bibfnamefont {P.}~\bibnamefont {Ring}},\ }\href
  {https://doi.org/10.1103/PhysRevC.71.024312} {\bibfield  {journal} {\bibinfo
  {journal} {Phys. Rev. C}\ }\textbf {\bibinfo {volume} {71}},\ \bibinfo
  {pages} {024312} (\bibinfo {year} {2005})}\BibitemShut {NoStop}%
\bibitem [{\citenamefont {{Brookhaven National Nuclear Data Center}}()}]{data}%
  \BibitemOpen
  \bibfield  {author} {\bibinfo {author} {\bibnamefont {{Brookhaven National
  Nuclear Data Center}}},\ }\href@noop {} {}\bibinfo {howpublished}
  {{\url{http://www.nndc.bnl.gov}}}\BibitemShut {NoStop}%
\bibitem [{\citenamefont {Rzaca-Urban}\ \emph {et~al.}(2009)\citenamefont
  {Rzaca-Urban}, \citenamefont {Sieja}, \citenamefont {Urban}, \citenamefont
  {Nowacki}, \citenamefont {Durell}, \citenamefont {Smith},\ and\ \citenamefont
  {Ahmad}}]{rzacaurban2009}%
  \BibitemOpen
  \bibfield  {author} {\bibinfo {author} {\bibfnamefont {T.}~\bibnamefont
  {Rzaca-Urban}}, \bibinfo {author} {\bibfnamefont {K.}~\bibnamefont {Sieja}},
  \bibinfo {author} {\bibfnamefont {W.}~\bibnamefont {Urban}}, \bibinfo
  {author} {\bibfnamefont {F.}~\bibnamefont {Nowacki}}, \bibinfo {author}
  {\bibfnamefont {J.~L.}\ \bibnamefont {Durell}}, \bibinfo {author}
  {\bibfnamefont {A.~G.}\ \bibnamefont {Smith}},\ and\ \bibinfo {author}
  {\bibfnamefont {I.}~\bibnamefont {Ahmad}},\ }\href
  {https://doi.org/10.1103/PhysRevC.79.024319} {\bibfield  {journal} {\bibinfo
  {journal} {Phys. Rev. C}\ }\textbf {\bibinfo {volume} {79}},\ \bibinfo
  {pages} {024319} (\bibinfo {year} {2009})}\BibitemShut {NoStop}%
\bibitem [{\citenamefont {Chen}\ \emph {et~al.}(2017)\citenamefont {Chen},
  \citenamefont {Doornenbal}, \citenamefont {Obertelli}, \citenamefont
  {Rodr\'{\i}guez}, \citenamefont {Authelet}, \citenamefont {Baba},
  \citenamefont {Calvet}, \citenamefont {Ch\^ateau}, \citenamefont {Corsi},
  \citenamefont {Delbart}, \citenamefont {Gheller}, \citenamefont {Giganon},
  \citenamefont {Gillibert}, \citenamefont {Lapoux}, \citenamefont
  {Motobayashi}, \citenamefont {Niikura}, \citenamefont {Paul}, \citenamefont
  {Rouss\'e}, \citenamefont {Sakurai}, \citenamefont {Santamaria},
  \citenamefont {Steppenbeck}, \citenamefont {Taniuchi}, \citenamefont
  {Uesaka}, \citenamefont {Ando}, \citenamefont {Arici}, \citenamefont
  {Blazhev}, \citenamefont {Browne}, \citenamefont {Bruce}, \citenamefont
  {Caroll}, \citenamefont {Chung}, \citenamefont {Cort\'es}, \citenamefont
  {Dewald}, \citenamefont {Ding}, \citenamefont {Flavigny}, \citenamefont
  {Franchoo}, \citenamefont {G\'orska}, \citenamefont {Gottardo}, \citenamefont
  {Jungclaus}, \citenamefont {Lee}, \citenamefont {Lettmann}, \citenamefont
  {Linh}, \citenamefont {Liu}, \citenamefont {Liu}, \citenamefont {Lizarazo},
  \citenamefont {Momiyama}, \citenamefont {Moschner}, \citenamefont {Nagamine},
  \citenamefont {Nakatsuka}, \citenamefont {Nita}, \citenamefont {Nobs},
  \citenamefont {Olivier}, \citenamefont {Orlandi}, \citenamefont {Patel},
  \citenamefont {Podolyak}, \citenamefont {Rudigier}, \citenamefont {Saito},
  \citenamefont {Shand}, \citenamefont {S\"oderstr\"om}, \citenamefont
  {Stefan}, \citenamefont {Vaquero}, \citenamefont {Werner}, \citenamefont
  {Wimmer},\ and\ \citenamefont {Xu}}]{chen2017}%
  \BibitemOpen
  \bibfield  {author} {\bibinfo {author} {\bibfnamefont {S.}~\bibnamefont
  {Chen}}, \bibinfo {author} {\bibfnamefont {P.}~\bibnamefont {Doornenbal}},
  \bibinfo {author} {\bibfnamefont {A.}~\bibnamefont {Obertelli}}, \bibinfo
  {author} {\bibfnamefont {T.~R.}\ \bibnamefont {Rodr\'{\i}guez}}, \bibinfo
  {author} {\bibfnamefont {G.}~\bibnamefont {Authelet}}, \bibinfo {author}
  {\bibfnamefont {H.}~\bibnamefont {Baba}}, \bibinfo {author} {\bibfnamefont
  {D.}~\bibnamefont {Calvet}}, \bibinfo {author} {\bibfnamefont
  {F.}~\bibnamefont {Ch\^ateau}}, \bibinfo {author} {\bibfnamefont
  {A.}~\bibnamefont {Corsi}}, \bibinfo {author} {\bibfnamefont
  {A.}~\bibnamefont {Delbart}}, \bibinfo {author} {\bibfnamefont {J.-M.}\
  \bibnamefont {Gheller}}, \bibinfo {author} {\bibfnamefont {A.}~\bibnamefont
  {Giganon}}, \bibinfo {author} {\bibfnamefont {A.}~\bibnamefont {Gillibert}},
  \bibinfo {author} {\bibfnamefont {V.}~\bibnamefont {Lapoux}}, \bibinfo
  {author} {\bibfnamefont {T.}~\bibnamefont {Motobayashi}}, \bibinfo {author}
  {\bibfnamefont {M.}~\bibnamefont {Niikura}}, \bibinfo {author} {\bibfnamefont
  {N.}~\bibnamefont {Paul}}, \bibinfo {author} {\bibfnamefont {J.-Y.}\
  \bibnamefont {Rouss\'e}}, \bibinfo {author} {\bibfnamefont {H.}~\bibnamefont
  {Sakurai}}, \bibinfo {author} {\bibfnamefont {C.}~\bibnamefont {Santamaria}},
  \bibinfo {author} {\bibfnamefont {D.}~\bibnamefont {Steppenbeck}}, \bibinfo
  {author} {\bibfnamefont {R.}~\bibnamefont {Taniuchi}}, \bibinfo {author}
  {\bibfnamefont {T.}~\bibnamefont {Uesaka}}, \bibinfo {author} {\bibfnamefont
  {T.}~\bibnamefont {Ando}}, \bibinfo {author} {\bibfnamefont {T.}~\bibnamefont
  {Arici}}, \bibinfo {author} {\bibfnamefont {A.}~\bibnamefont {Blazhev}},
  \bibinfo {author} {\bibfnamefont {F.}~\bibnamefont {Browne}}, \bibinfo
  {author} {\bibfnamefont {A.~M.}\ \bibnamefont {Bruce}}, \bibinfo {author}
  {\bibfnamefont {R.}~\bibnamefont {Caroll}}, \bibinfo {author} {\bibfnamefont
  {L.~X.}\ \bibnamefont {Chung}}, \bibinfo {author} {\bibfnamefont {M.~L.}\
  \bibnamefont {Cort\'es}}, \bibinfo {author} {\bibfnamefont {M.}~\bibnamefont
  {Dewald}}, \bibinfo {author} {\bibfnamefont {B.}~\bibnamefont {Ding}},
  \bibinfo {author} {\bibfnamefont {F.}~\bibnamefont {Flavigny}}, \bibinfo
  {author} {\bibfnamefont {S.}~\bibnamefont {Franchoo}}, \bibinfo {author}
  {\bibfnamefont {M.}~\bibnamefont {G\'orska}}, \bibinfo {author}
  {\bibfnamefont {A.}~\bibnamefont {Gottardo}}, \bibinfo {author}
  {\bibfnamefont {A.}~\bibnamefont {Jungclaus}}, \bibinfo {author}
  {\bibfnamefont {J.}~\bibnamefont {Lee}}, \bibinfo {author} {\bibfnamefont
  {M.}~\bibnamefont {Lettmann}}, \bibinfo {author} {\bibfnamefont {B.~D.}\
  \bibnamefont {Linh}}, \bibinfo {author} {\bibfnamefont {J.}~\bibnamefont
  {Liu}}, \bibinfo {author} {\bibfnamefont {Z.}~\bibnamefont {Liu}}, \bibinfo
  {author} {\bibfnamefont {C.}~\bibnamefont {Lizarazo}}, \bibinfo {author}
  {\bibfnamefont {S.}~\bibnamefont {Momiyama}}, \bibinfo {author}
  {\bibfnamefont {K.}~\bibnamefont {Moschner}}, \bibinfo {author}
  {\bibfnamefont {S.}~\bibnamefont {Nagamine}}, \bibinfo {author}
  {\bibfnamefont {N.}~\bibnamefont {Nakatsuka}}, \bibinfo {author}
  {\bibfnamefont {C.~R.}\ \bibnamefont {Nita}}, \bibinfo {author}
  {\bibfnamefont {C.}~\bibnamefont {Nobs}}, \bibinfo {author} {\bibfnamefont
  {L.}~\bibnamefont {Olivier}}, \bibinfo {author} {\bibfnamefont
  {R.}~\bibnamefont {Orlandi}}, \bibinfo {author} {\bibfnamefont
  {Z.}~\bibnamefont {Patel}}, \bibinfo {author} {\bibfnamefont
  {Z.}~\bibnamefont {Podolyak}}, \bibinfo {author} {\bibfnamefont
  {M.}~\bibnamefont {Rudigier}}, \bibinfo {author} {\bibfnamefont
  {T.}~\bibnamefont {Saito}}, \bibinfo {author} {\bibfnamefont
  {C.}~\bibnamefont {Shand}}, \bibinfo {author} {\bibfnamefont {P.-A.}\
  \bibnamefont {S\"oderstr\"om}}, \bibinfo {author} {\bibfnamefont
  {I.}~\bibnamefont {Stefan}}, \bibinfo {author} {\bibfnamefont
  {V.}~\bibnamefont {Vaquero}}, \bibinfo {author} {\bibfnamefont
  {V.}~\bibnamefont {Werner}}, \bibinfo {author} {\bibfnamefont
  {K.}~\bibnamefont {Wimmer}},\ and\ \bibinfo {author} {\bibfnamefont
  {Z.}~\bibnamefont {Xu}},\ }\href {https://doi.org/10.1103/PhysRevC.95.041302}
  {\bibfield  {journal} {\bibinfo  {journal} {Phys. Rev. C}\ }\textbf {\bibinfo
  {volume} {95}},\ \bibinfo {pages} {041302} (\bibinfo {year}
  {2017})}\BibitemShut {NoStop}%
\bibitem [{\citenamefont {Lizarazo}\ \emph {et~al.}(2020)\citenamefont
  {Lizarazo}, \citenamefont {S\"oderstr\"om}, \citenamefont {Werner},
  \citenamefont {Pietralla}, \citenamefont {Walker}, \citenamefont {Dong},
  \citenamefont {Xu}, \citenamefont {Rodr\'{\i}guez}, \citenamefont {Browne},
  \citenamefont {Doornenbal}, \citenamefont {Nishimura}, \citenamefont
  {Ni\ifmmode \mbox{\c{t}}\else \c{t}\fi{}\ifmmode~\u{a}\else \u{a}\fi{}},
  \citenamefont {Obertelli}, \citenamefont {Ando}, \citenamefont {Arici},
  \citenamefont {Authelet}, \citenamefont {Baba}, \citenamefont {Blazhev},
  \citenamefont {Bruce}, \citenamefont {Calvet}, \citenamefont {Caroll},
  \citenamefont {Ch\^ateau}, \citenamefont {Chen}, \citenamefont {Chung},
  \citenamefont {Corsi}, \citenamefont {Cort\'es}, \citenamefont {Delbart},
  \citenamefont {Dewald}, \citenamefont {Ding}, \citenamefont {Flavigny},
  \citenamefont {Franchoo}, \citenamefont {Gerl}, \citenamefont {Gheller},
  \citenamefont {Giganon}, \citenamefont {Gillibert}, \citenamefont {G\'orska},
  \citenamefont {Gottardo}, \citenamefont {Kojouharov}, \citenamefont {Kurz},
  \citenamefont {Lapoux}, \citenamefont {Lee}, \citenamefont {Lettmann},
  \citenamefont {Linh}, \citenamefont {Liu}, \citenamefont {Liu}, \citenamefont
  {Momiyama}, \citenamefont {Moschner}, \citenamefont {Motobayashi},
  \citenamefont {Nagamine}, \citenamefont {Nakatsuka}, \citenamefont {Niikura},
  \citenamefont {Nobs}, \citenamefont {Olivier}, \citenamefont {Patel},
  \citenamefont {Paul}, \citenamefont {Podoly\'ak}, \citenamefont {Rouss\'e},
  \citenamefont {Rudigier}, \citenamefont {Saito}, \citenamefont {Sakurai},
  \citenamefont {Santamaria}, \citenamefont {Schaffner}, \citenamefont {Shand},
  \citenamefont {Stefan}, \citenamefont {Steppenbeck}, \citenamefont
  {Taniuchi}, \citenamefont {Uesaka}, \citenamefont {Vaquero}, \citenamefont
  {Wimmer},\ and\ \citenamefont {Xu}}]{lizarazo2020}%
  \BibitemOpen
  \bibfield  {author} {\bibinfo {author} {\bibfnamefont {C.}~\bibnamefont
  {Lizarazo}}, \bibinfo {author} {\bibfnamefont {P.-A.}\ \bibnamefont
  {S\"oderstr\"om}}, \bibinfo {author} {\bibfnamefont {V.}~\bibnamefont
  {Werner}}, \bibinfo {author} {\bibfnamefont {N.}~\bibnamefont {Pietralla}},
  \bibinfo {author} {\bibfnamefont {P.~M.}\ \bibnamefont {Walker}}, \bibinfo
  {author} {\bibfnamefont {G.~X.}\ \bibnamefont {Dong}}, \bibinfo {author}
  {\bibfnamefont {F.~R.}\ \bibnamefont {Xu}}, \bibinfo {author} {\bibfnamefont
  {T.~R.}\ \bibnamefont {Rodr\'{\i}guez}}, \bibinfo {author} {\bibfnamefont
  {F.}~\bibnamefont {Browne}}, \bibinfo {author} {\bibfnamefont
  {P.}~\bibnamefont {Doornenbal}}, \bibinfo {author} {\bibfnamefont
  {S.}~\bibnamefont {Nishimura}}, \bibinfo {author} {\bibfnamefont {C.~R.}\
  \bibnamefont {Ni\ifmmode \mbox{\c{t}}\else \c{t}\fi{}\ifmmode~\u{a}\else
  \u{a}\fi{}}}, \bibinfo {author} {\bibfnamefont {A.}~\bibnamefont
  {Obertelli}}, \bibinfo {author} {\bibfnamefont {T.}~\bibnamefont {Ando}},
  \bibinfo {author} {\bibfnamefont {T.}~\bibnamefont {Arici}}, \bibinfo
  {author} {\bibfnamefont {G.}~\bibnamefont {Authelet}}, \bibinfo {author}
  {\bibfnamefont {H.}~\bibnamefont {Baba}}, \bibinfo {author} {\bibfnamefont
  {A.}~\bibnamefont {Blazhev}}, \bibinfo {author} {\bibfnamefont {A.~M.}\
  \bibnamefont {Bruce}}, \bibinfo {author} {\bibfnamefont {D.}~\bibnamefont
  {Calvet}}, \bibinfo {author} {\bibfnamefont {R.~J.}\ \bibnamefont {Caroll}},
  \bibinfo {author} {\bibfnamefont {F.}~\bibnamefont {Ch\^ateau}}, \bibinfo
  {author} {\bibfnamefont {S.}~\bibnamefont {Chen}}, \bibinfo {author}
  {\bibfnamefont {L.~X.}\ \bibnamefont {Chung}}, \bibinfo {author}
  {\bibfnamefont {A.}~\bibnamefont {Corsi}}, \bibinfo {author} {\bibfnamefont
  {M.~L.}\ \bibnamefont {Cort\'es}}, \bibinfo {author} {\bibfnamefont
  {A.}~\bibnamefont {Delbart}}, \bibinfo {author} {\bibfnamefont
  {M.}~\bibnamefont {Dewald}}, \bibinfo {author} {\bibfnamefont
  {B.}~\bibnamefont {Ding}}, \bibinfo {author} {\bibfnamefont {F.}~\bibnamefont
  {Flavigny}}, \bibinfo {author} {\bibfnamefont {S.}~\bibnamefont {Franchoo}},
  \bibinfo {author} {\bibfnamefont {J.}~\bibnamefont {Gerl}}, \bibinfo {author}
  {\bibfnamefont {J.-M.}\ \bibnamefont {Gheller}}, \bibinfo {author}
  {\bibfnamefont {A.}~\bibnamefont {Giganon}}, \bibinfo {author} {\bibfnamefont
  {A.}~\bibnamefont {Gillibert}}, \bibinfo {author} {\bibfnamefont
  {M.}~\bibnamefont {G\'orska}}, \bibinfo {author} {\bibfnamefont
  {A.}~\bibnamefont {Gottardo}}, \bibinfo {author} {\bibfnamefont
  {I.}~\bibnamefont {Kojouharov}}, \bibinfo {author} {\bibfnamefont
  {N.}~\bibnamefont {Kurz}}, \bibinfo {author} {\bibfnamefont {V.}~\bibnamefont
  {Lapoux}}, \bibinfo {author} {\bibfnamefont {J.}~\bibnamefont {Lee}},
  \bibinfo {author} {\bibfnamefont {M.}~\bibnamefont {Lettmann}}, \bibinfo
  {author} {\bibfnamefont {B.~D.}\ \bibnamefont {Linh}}, \bibinfo {author}
  {\bibfnamefont {J.~J.}\ \bibnamefont {Liu}}, \bibinfo {author} {\bibfnamefont
  {Z.}~\bibnamefont {Liu}}, \bibinfo {author} {\bibfnamefont {S.}~\bibnamefont
  {Momiyama}}, \bibinfo {author} {\bibfnamefont {K.}~\bibnamefont {Moschner}},
  \bibinfo {author} {\bibfnamefont {T.}~\bibnamefont {Motobayashi}}, \bibinfo
  {author} {\bibfnamefont {S.}~\bibnamefont {Nagamine}}, \bibinfo {author}
  {\bibfnamefont {N.}~\bibnamefont {Nakatsuka}}, \bibinfo {author}
  {\bibfnamefont {M.}~\bibnamefont {Niikura}}, \bibinfo {author} {\bibfnamefont
  {C.}~\bibnamefont {Nobs}}, \bibinfo {author} {\bibfnamefont {L.}~\bibnamefont
  {Olivier}}, \bibinfo {author} {\bibfnamefont {Z.}~\bibnamefont {Patel}},
  \bibinfo {author} {\bibfnamefont {N.}~\bibnamefont {Paul}}, \bibinfo {author}
  {\bibfnamefont {Z.}~\bibnamefont {Podoly\'ak}}, \bibinfo {author}
  {\bibfnamefont {J.-Y.}\ \bibnamefont {Rouss\'e}}, \bibinfo {author}
  {\bibfnamefont {M.}~\bibnamefont {Rudigier}}, \bibinfo {author}
  {\bibfnamefont {T.~Y.}\ \bibnamefont {Saito}}, \bibinfo {author}
  {\bibfnamefont {H.}~\bibnamefont {Sakurai}}, \bibinfo {author} {\bibfnamefont
  {C.}~\bibnamefont {Santamaria}}, \bibinfo {author} {\bibfnamefont
  {H.}~\bibnamefont {Schaffner}}, \bibinfo {author} {\bibfnamefont
  {C.}~\bibnamefont {Shand}}, \bibinfo {author} {\bibfnamefont
  {I.}~\bibnamefont {Stefan}}, \bibinfo {author} {\bibfnamefont
  {D.}~\bibnamefont {Steppenbeck}}, \bibinfo {author} {\bibfnamefont
  {R.}~\bibnamefont {Taniuchi}}, \bibinfo {author} {\bibfnamefont
  {T.}~\bibnamefont {Uesaka}}, \bibinfo {author} {\bibfnamefont
  {V.}~\bibnamefont {Vaquero}}, \bibinfo {author} {\bibfnamefont
  {K.}~\bibnamefont {Wimmer}},\ and\ \bibinfo {author} {\bibfnamefont
  {Z.}~\bibnamefont {Xu}},\ }\href
  {https://doi.org/10.1103/PhysRevLett.124.222501} {\bibfield  {journal}
  {\bibinfo  {journal} {Phys. Rev. Lett.}\ }\textbf {\bibinfo {volume} {124}},\
  \bibinfo {pages} {222501} (\bibinfo {year} {2020})}\BibitemShut {NoStop}%
\bibitem [{\citenamefont {Gerst}\ \emph {et~al.}(2020)\citenamefont {Gerst},
  \citenamefont {Blazhev}, \citenamefont {Warr}, \citenamefont {Wilson},
  \citenamefont {Lebois}, \citenamefont {Jovan\ifmmode \check{c}\else
  \v{c}\fi{}evi\ifmmode~\acute{c}\else \'{c}\fi{}}, \citenamefont {Thisse},
  \citenamefont {Canavan}, \citenamefont {Rudigier}, \citenamefont {\'Etasse},
  \citenamefont {Adamska}, \citenamefont {Adsley}, \citenamefont {Algora},
  \citenamefont {Babo}, \citenamefont {Belvedere}, \citenamefont {Benito},
  \citenamefont {Benzoni}, \citenamefont {Boso}, \citenamefont {Bottoni},
  \citenamefont {Bunce}, \citenamefont {Chakma}, \citenamefont
  {Cieplicka-Ory\ifmmode~\acute{n}\else \'{n}\fi{}czak}, \citenamefont
  {Courtin}, \citenamefont {Cort\'es}, \citenamefont {Davies}, \citenamefont
  {Delafosse}, \citenamefont {Fallot}, \citenamefont {Fornal}, \citenamefont
  {Fraile}, \citenamefont {Gjestvang}, \citenamefont {Gottardo}, \citenamefont
  {Guadilla}, \citenamefont {H\"afner}, \citenamefont {Hauschild},
  \citenamefont {Heine}, \citenamefont {Henrich}, \citenamefont {Homm},
  \citenamefont {Ibrahim}, \citenamefont {Iskra}, \citenamefont {Ivanov},
  \citenamefont {Jazrawi}, \citenamefont {Korgul}, \citenamefont {Koseoglou},
  \citenamefont {Kr\"oll}, \citenamefont {Kurtukian-Nieto}, \citenamefont
  {Le~Meur}, \citenamefont {Leoni}, \citenamefont {Ljungvall}, \citenamefont
  {Lopez-Martens}, \citenamefont {Lozeva}, \citenamefont {Matea}, \citenamefont
  {Miernik}, \citenamefont {Nemer}, \citenamefont {Oberstedt}, \citenamefont
  {Paulsen}, \citenamefont {Piersa}, \citenamefont {Popovitch}, \citenamefont
  {Porzio}, \citenamefont {Qi}, \citenamefont {Ralet}, \citenamefont {Regan},
  \citenamefont {Reygadas-Tello}, \citenamefont {Rezynkina}, \citenamefont
  {S\'anchez-Tembleque}, \citenamefont {Schmitt}, \citenamefont
  {S\"oderstr\"om}, \citenamefont {S\"urder}, \citenamefont {Tocabens},
  \citenamefont {Vedia}, \citenamefont {Verney}, \citenamefont {Wasilewska},
  \citenamefont {Wiederhold}, \citenamefont {Yavachova}, \citenamefont
  {Zeiser},\ and\ \citenamefont {Ziliani}}]{gerst2020}%
  \BibitemOpen
  \bibfield  {author} {\bibinfo {author} {\bibfnamefont {R.-B.}\ \bibnamefont
  {Gerst}}, \bibinfo {author} {\bibfnamefont {A.}~\bibnamefont {Blazhev}},
  \bibinfo {author} {\bibfnamefont {N.}~\bibnamefont {Warr}}, \bibinfo {author}
  {\bibfnamefont {J.~N.}\ \bibnamefont {Wilson}}, \bibinfo {author}
  {\bibfnamefont {M.}~\bibnamefont {Lebois}}, \bibinfo {author} {\bibfnamefont
  {N.}~\bibnamefont {Jovan\ifmmode \check{c}\else
  \v{c}\fi{}evi\ifmmode~\acute{c}\else \'{c}\fi{}}}, \bibinfo {author}
  {\bibfnamefont {D.}~\bibnamefont {Thisse}}, \bibinfo {author} {\bibfnamefont
  {R.}~\bibnamefont {Canavan}}, \bibinfo {author} {\bibfnamefont
  {M.}~\bibnamefont {Rudigier}}, \bibinfo {author} {\bibfnamefont
  {D.}~\bibnamefont {\'Etasse}}, \bibinfo {author} {\bibfnamefont
  {E.}~\bibnamefont {Adamska}}, \bibinfo {author} {\bibfnamefont
  {P.}~\bibnamefont {Adsley}}, \bibinfo {author} {\bibfnamefont
  {A.}~\bibnamefont {Algora}}, \bibinfo {author} {\bibfnamefont
  {M.}~\bibnamefont {Babo}}, \bibinfo {author} {\bibfnamefont {K.}~\bibnamefont
  {Belvedere}}, \bibinfo {author} {\bibfnamefont {J.}~\bibnamefont {Benito}},
  \bibinfo {author} {\bibfnamefont {G.}~\bibnamefont {Benzoni}}, \bibinfo
  {author} {\bibfnamefont {A.}~\bibnamefont {Boso}}, \bibinfo {author}
  {\bibfnamefont {S.}~\bibnamefont {Bottoni}}, \bibinfo {author} {\bibfnamefont
  {M.}~\bibnamefont {Bunce}}, \bibinfo {author} {\bibfnamefont
  {R.}~\bibnamefont {Chakma}}, \bibinfo {author} {\bibfnamefont
  {N.}~\bibnamefont {Cieplicka-Ory\ifmmode~\acute{n}\else \'{n}\fi{}czak}},
  \bibinfo {author} {\bibfnamefont {S.}~\bibnamefont {Courtin}}, \bibinfo
  {author} {\bibfnamefont {M.~L.}\ \bibnamefont {Cort\'es}}, \bibinfo {author}
  {\bibfnamefont {P.}~\bibnamefont {Davies}}, \bibinfo {author} {\bibfnamefont
  {C.}~\bibnamefont {Delafosse}}, \bibinfo {author} {\bibfnamefont
  {M.}~\bibnamefont {Fallot}}, \bibinfo {author} {\bibfnamefont
  {B.}~\bibnamefont {Fornal}}, \bibinfo {author} {\bibfnamefont {L.~M.}\
  \bibnamefont {Fraile}}, \bibinfo {author} {\bibfnamefont {D.}~\bibnamefont
  {Gjestvang}}, \bibinfo {author} {\bibfnamefont {A.}~\bibnamefont {Gottardo}},
  \bibinfo {author} {\bibfnamefont {V.}~\bibnamefont {Guadilla}}, \bibinfo
  {author} {\bibfnamefont {G.}~\bibnamefont {H\"afner}}, \bibinfo {author}
  {\bibfnamefont {K.}~\bibnamefont {Hauschild}}, \bibinfo {author}
  {\bibfnamefont {M.}~\bibnamefont {Heine}}, \bibinfo {author} {\bibfnamefont
  {C.}~\bibnamefont {Henrich}}, \bibinfo {author} {\bibfnamefont
  {I.}~\bibnamefont {Homm}}, \bibinfo {author} {\bibfnamefont {F.}~\bibnamefont
  {Ibrahim}}, \bibinfo {author} {\bibfnamefont {L.~W.}\ \bibnamefont {Iskra}},
  \bibinfo {author} {\bibfnamefont {P.}~\bibnamefont {Ivanov}}, \bibinfo
  {author} {\bibfnamefont {S.}~\bibnamefont {Jazrawi}}, \bibinfo {author}
  {\bibfnamefont {A.}~\bibnamefont {Korgul}}, \bibinfo {author} {\bibfnamefont
  {P.}~\bibnamefont {Koseoglou}}, \bibinfo {author} {\bibfnamefont
  {T.}~\bibnamefont {Kr\"oll}}, \bibinfo {author} {\bibfnamefont
  {T.}~\bibnamefont {Kurtukian-Nieto}}, \bibinfo {author} {\bibfnamefont
  {L.}~\bibnamefont {Le~Meur}}, \bibinfo {author} {\bibfnamefont
  {S.}~\bibnamefont {Leoni}}, \bibinfo {author} {\bibfnamefont
  {J.}~\bibnamefont {Ljungvall}}, \bibinfo {author} {\bibfnamefont
  {A.}~\bibnamefont {Lopez-Martens}}, \bibinfo {author} {\bibfnamefont
  {R.}~\bibnamefont {Lozeva}}, \bibinfo {author} {\bibfnamefont
  {I.}~\bibnamefont {Matea}}, \bibinfo {author} {\bibfnamefont
  {K.}~\bibnamefont {Miernik}}, \bibinfo {author} {\bibfnamefont
  {J.}~\bibnamefont {Nemer}}, \bibinfo {author} {\bibfnamefont
  {S.}~\bibnamefont {Oberstedt}}, \bibinfo {author} {\bibfnamefont
  {W.}~\bibnamefont {Paulsen}}, \bibinfo {author} {\bibfnamefont
  {M.}~\bibnamefont {Piersa}}, \bibinfo {author} {\bibfnamefont
  {Y.}~\bibnamefont {Popovitch}}, \bibinfo {author} {\bibfnamefont
  {C.}~\bibnamefont {Porzio}}, \bibinfo {author} {\bibfnamefont
  {L.}~\bibnamefont {Qi}}, \bibinfo {author} {\bibfnamefont {D.}~\bibnamefont
  {Ralet}}, \bibinfo {author} {\bibfnamefont {P.~H.}\ \bibnamefont {Regan}},
  \bibinfo {author} {\bibfnamefont {D.}~\bibnamefont {Reygadas-Tello}},
  \bibinfo {author} {\bibfnamefont {K.}~\bibnamefont {Rezynkina}}, \bibinfo
  {author} {\bibfnamefont {V.}~\bibnamefont {S\'anchez-Tembleque}}, \bibinfo
  {author} {\bibfnamefont {C.}~\bibnamefont {Schmitt}}, \bibinfo {author}
  {\bibfnamefont {P.-A.}\ \bibnamefont {S\"oderstr\"om}}, \bibinfo {author}
  {\bibfnamefont {C.}~\bibnamefont {S\"urder}}, \bibinfo {author}
  {\bibfnamefont {G.}~\bibnamefont {Tocabens}}, \bibinfo {author}
  {\bibfnamefont {V.}~\bibnamefont {Vedia}}, \bibinfo {author} {\bibfnamefont
  {D.}~\bibnamefont {Verney}}, \bibinfo {author} {\bibfnamefont
  {B.}~\bibnamefont {Wasilewska}}, \bibinfo {author} {\bibfnamefont
  {J.}~\bibnamefont {Wiederhold}}, \bibinfo {author} {\bibfnamefont
  {M.}~\bibnamefont {Yavachova}}, \bibinfo {author} {\bibfnamefont
  {F.}~\bibnamefont {Zeiser}},\ and\ \bibinfo {author} {\bibfnamefont
  {S.}~\bibnamefont {Ziliani}},\ }\href
  {https://doi.org/10.1103/PhysRevC.102.064323} {\bibfield  {journal} {\bibinfo
   {journal} {Phys. Rev. C}\ }\textbf {\bibinfo {volume} {102}},\ \bibinfo
  {pages} {064323} (\bibinfo {year} {2020})}\BibitemShut {NoStop}%
\bibitem [{\citenamefont {Albers}\ \emph {et~al.}(2012)\citenamefont {Albers},
  \citenamefont {Warr}, \citenamefont {Nomura}, \citenamefont {Blazhev},
  \citenamefont {Jolie}, \citenamefont {M\"ucher}, \citenamefont {Bastin},
  \citenamefont {Bauer}, \citenamefont {Bernards}, \citenamefont {Bettermann},
  \citenamefont {Bildstein}, \citenamefont {Butterworth}, \citenamefont
  {Cappellazzo}, \citenamefont {Cederk\"all}, \citenamefont {Cline},
  \citenamefont {Darby}, \citenamefont {Das~Gupta}, \citenamefont {Daugas},
  \citenamefont {Davinson}, \citenamefont {De~Witte}, \citenamefont {Diriken},
  \citenamefont {Filipescu}, \citenamefont {Fiori}, \citenamefont {Fransen},
  \citenamefont {Gaffney}, \citenamefont {Georgiev}, \citenamefont
  {Gernh\"auser}, \citenamefont {Hackstein}, \citenamefont {Heinze},
  \citenamefont {Hess}, \citenamefont {Huyse}, \citenamefont {Jenkins},
  \citenamefont {Konki}, \citenamefont {Kowalczyk}, \citenamefont {Kr\"oll},
  \citenamefont {Kr\"ucken}, \citenamefont {Litzinger}, \citenamefont {Lutter},
  \citenamefont {Marginean}, \citenamefont {Mihai}, \citenamefont {Moschner},
  \citenamefont {Napiorkowski}, \citenamefont {Nara~Singh}, \citenamefont
  {Nowak}, \citenamefont {Otsuka}, \citenamefont {Pakarinen}, \citenamefont
  {Pfeiffer}, \citenamefont {Radeck}, \citenamefont {Reiter}, \citenamefont
  {Rigby}, \citenamefont {Robledo}, \citenamefont {Rodr\'iguez-Guzm\'an},
  \citenamefont {Rudigier}, \citenamefont {Sarriguren}, \citenamefont {Scheck},
  \citenamefont {Seidlitz}, \citenamefont {Siebeck}, \citenamefont {Simpson},
  \citenamefont {Th\"ole}, \citenamefont {Thomas}, \citenamefont {Van~de
  Walle}, \citenamefont {Van~Duppen}, \citenamefont {Vermeulen}, \citenamefont
  {Voulot}, \citenamefont {Wadsworth}, \citenamefont {Wenander}, \citenamefont
  {Wimmer}, \citenamefont {Zell},\ and\ \citenamefont
  {Zielinska}}]{albers2012}%
  \BibitemOpen
  \bibfield  {author} {\bibinfo {author} {\bibfnamefont {M.}~\bibnamefont
  {Albers}}, \bibinfo {author} {\bibfnamefont {N.}~\bibnamefont {Warr}},
  \bibinfo {author} {\bibfnamefont {K.}~\bibnamefont {Nomura}}, \bibinfo
  {author} {\bibfnamefont {A.}~\bibnamefont {Blazhev}}, \bibinfo {author}
  {\bibfnamefont {J.}~\bibnamefont {Jolie}}, \bibinfo {author} {\bibfnamefont
  {D.}~\bibnamefont {M\"ucher}}, \bibinfo {author} {\bibfnamefont
  {B.}~\bibnamefont {Bastin}}, \bibinfo {author} {\bibfnamefont
  {C.}~\bibnamefont {Bauer}}, \bibinfo {author} {\bibfnamefont
  {C.}~\bibnamefont {Bernards}}, \bibinfo {author} {\bibfnamefont
  {L.}~\bibnamefont {Bettermann}}, \bibinfo {author} {\bibfnamefont
  {V.}~\bibnamefont {Bildstein}}, \bibinfo {author} {\bibfnamefont
  {J.}~\bibnamefont {Butterworth}}, \bibinfo {author} {\bibfnamefont
  {M.}~\bibnamefont {Cappellazzo}}, \bibinfo {author} {\bibfnamefont
  {J.}~\bibnamefont {Cederk\"all}}, \bibinfo {author} {\bibfnamefont
  {D.}~\bibnamefont {Cline}}, \bibinfo {author} {\bibfnamefont
  {I.}~\bibnamefont {Darby}}, \bibinfo {author} {\bibfnamefont
  {S.}~\bibnamefont {Das~Gupta}}, \bibinfo {author} {\bibfnamefont {J.~M.}\
  \bibnamefont {Daugas}}, \bibinfo {author} {\bibfnamefont {T.}~\bibnamefont
  {Davinson}}, \bibinfo {author} {\bibfnamefont {H.}~\bibnamefont {De~Witte}},
  \bibinfo {author} {\bibfnamefont {J.}~\bibnamefont {Diriken}}, \bibinfo
  {author} {\bibfnamefont {D.}~\bibnamefont {Filipescu}}, \bibinfo {author}
  {\bibfnamefont {E.}~\bibnamefont {Fiori}}, \bibinfo {author} {\bibfnamefont
  {C.}~\bibnamefont {Fransen}}, \bibinfo {author} {\bibfnamefont {L.~P.}\
  \bibnamefont {Gaffney}}, \bibinfo {author} {\bibfnamefont {G.}~\bibnamefont
  {Georgiev}}, \bibinfo {author} {\bibfnamefont {R.}~\bibnamefont
  {Gernh\"auser}}, \bibinfo {author} {\bibfnamefont {M.}~\bibnamefont
  {Hackstein}}, \bibinfo {author} {\bibfnamefont {S.}~\bibnamefont {Heinze}},
  \bibinfo {author} {\bibfnamefont {H.}~\bibnamefont {Hess}}, \bibinfo {author}
  {\bibfnamefont {M.}~\bibnamefont {Huyse}}, \bibinfo {author} {\bibfnamefont
  {D.}~\bibnamefont {Jenkins}}, \bibinfo {author} {\bibfnamefont
  {J.}~\bibnamefont {Konki}}, \bibinfo {author} {\bibfnamefont
  {M.}~\bibnamefont {Kowalczyk}}, \bibinfo {author} {\bibfnamefont
  {T.}~\bibnamefont {Kr\"oll}}, \bibinfo {author} {\bibfnamefont
  {R.}~\bibnamefont {Kr\"ucken}}, \bibinfo {author} {\bibfnamefont
  {J.}~\bibnamefont {Litzinger}}, \bibinfo {author} {\bibfnamefont
  {R.}~\bibnamefont {Lutter}}, \bibinfo {author} {\bibfnamefont
  {N.}~\bibnamefont {Marginean}}, \bibinfo {author} {\bibfnamefont
  {C.}~\bibnamefont {Mihai}}, \bibinfo {author} {\bibfnamefont
  {K.}~\bibnamefont {Moschner}}, \bibinfo {author} {\bibfnamefont
  {P.}~\bibnamefont {Napiorkowski}}, \bibinfo {author} {\bibfnamefont {B.~S.}\
  \bibnamefont {Nara~Singh}}, \bibinfo {author} {\bibfnamefont
  {K.}~\bibnamefont {Nowak}}, \bibinfo {author} {\bibfnamefont
  {T.}~\bibnamefont {Otsuka}}, \bibinfo {author} {\bibfnamefont
  {J.}~\bibnamefont {Pakarinen}}, \bibinfo {author} {\bibfnamefont
  {M.}~\bibnamefont {Pfeiffer}}, \bibinfo {author} {\bibfnamefont
  {D.}~\bibnamefont {Radeck}}, \bibinfo {author} {\bibfnamefont
  {P.}~\bibnamefont {Reiter}}, \bibinfo {author} {\bibfnamefont
  {S.}~\bibnamefont {Rigby}}, \bibinfo {author} {\bibfnamefont {L.~M.}\
  \bibnamefont {Robledo}}, \bibinfo {author} {\bibfnamefont {R.}~\bibnamefont
  {Rodr\'iguez-Guzm\'an}}, \bibinfo {author} {\bibfnamefont {M.}~\bibnamefont
  {Rudigier}}, \bibinfo {author} {\bibfnamefont {P.}~\bibnamefont
  {Sarriguren}}, \bibinfo {author} {\bibfnamefont {M.}~\bibnamefont {Scheck}},
  \bibinfo {author} {\bibfnamefont {M.}~\bibnamefont {Seidlitz}}, \bibinfo
  {author} {\bibfnamefont {B.}~\bibnamefont {Siebeck}}, \bibinfo {author}
  {\bibfnamefont {G.}~\bibnamefont {Simpson}}, \bibinfo {author} {\bibfnamefont
  {P.}~\bibnamefont {Th\"ole}}, \bibinfo {author} {\bibfnamefont
  {T.}~\bibnamefont {Thomas}}, \bibinfo {author} {\bibfnamefont
  {J.}~\bibnamefont {Van~de Walle}}, \bibinfo {author} {\bibfnamefont
  {P.}~\bibnamefont {Van~Duppen}}, \bibinfo {author} {\bibfnamefont
  {M.}~\bibnamefont {Vermeulen}}, \bibinfo {author} {\bibfnamefont
  {D.}~\bibnamefont {Voulot}}, \bibinfo {author} {\bibfnamefont
  {R.}~\bibnamefont {Wadsworth}}, \bibinfo {author} {\bibfnamefont
  {F.}~\bibnamefont {Wenander}}, \bibinfo {author} {\bibfnamefont
  {K.}~\bibnamefont {Wimmer}}, \bibinfo {author} {\bibfnamefont {K.~O.}\
  \bibnamefont {Zell}},\ and\ \bibinfo {author} {\bibfnamefont
  {M.}~\bibnamefont {Zielinska}},\ }\href
  {https://doi.org/10.1103/PhysRevLett.108.062701} {\bibfield  {journal}
  {\bibinfo  {journal} {Phys. Rev. Lett.}\ }\textbf {\bibinfo {volume} {108}},\
  \bibinfo {pages} {062701} (\bibinfo {year} {2012})}\BibitemShut {NoStop}%
\bibitem [{\citenamefont {Duval}\ and\ \citenamefont
  {Barrett}(1981)}]{duval1981}%
  \BibitemOpen
  \bibfield  {author} {\bibinfo {author} {\bibfnamefont {P.~D.}\ \bibnamefont
  {Duval}}\ and\ \bibinfo {author} {\bibfnamefont {B.~R.}\ \bibnamefont
  {Barrett}},\ }\href {https://doi.org/10.1016/0370-2693(81)90321-X} {\bibfield
   {journal} {\bibinfo  {journal} {Phys. Lett. B}\ }\textbf {\bibinfo {volume}
  {100}},\ \bibinfo {pages} {223} (\bibinfo {year} {1981})}\BibitemShut
  {NoStop}%
\bibitem [{\citenamefont {Kib\'edi}\ and\ \citenamefont
  {Spear}(2002)}]{kibedi2002}%
  \BibitemOpen
  \bibfield  {author} {\bibinfo {author} {\bibfnamefont {T.}~\bibnamefont
  {Kib\'edi}}\ and\ \bibinfo {author} {\bibfnamefont {R.~H.}\ \bibnamefont
  {Spear}},\ }\href {https://doi.org/http://dx.doi.org/10.1006/adnd.2001.0871}
  {\bibfield  {journal} {\bibinfo  {journal} {At. Data and Nucl. Data Tables}\
  }\textbf {\bibinfo {volume} {80}},\ \bibinfo {pages} {35 } (\bibinfo {year}
  {2002})}\BibitemShut {NoStop}%
\bibitem [{\citenamefont {Elhami}\ \emph {et~al.}(2008)\citenamefont {Elhami},
  \citenamefont {Orce}, \citenamefont {Scheck}, \citenamefont {Mukhopadhyay},
  \citenamefont {Choudry}, \citenamefont {McEllistrem}, \citenamefont {Yates},
  \citenamefont {Angell}, \citenamefont {Boswell}, \citenamefont {Fallin},
  \citenamefont {Howell}, \citenamefont {Hutcheson}, \citenamefont {Karwowski},
  \citenamefont {Kelley}, \citenamefont {Parpottas}, \citenamefont {Tonchev},\
  and\ \citenamefont {Tornow}}]{elhami2008}%
  \BibitemOpen
  \bibfield  {author} {\bibinfo {author} {\bibfnamefont {E.}~\bibnamefont
  {Elhami}}, \bibinfo {author} {\bibfnamefont {J.~N.}\ \bibnamefont {Orce}},
  \bibinfo {author} {\bibfnamefont {M.}~\bibnamefont {Scheck}}, \bibinfo
  {author} {\bibfnamefont {S.}~\bibnamefont {Mukhopadhyay}}, \bibinfo {author}
  {\bibfnamefont {S.~N.}\ \bibnamefont {Choudry}}, \bibinfo {author}
  {\bibfnamefont {M.~T.}\ \bibnamefont {McEllistrem}}, \bibinfo {author}
  {\bibfnamefont {S.~W.}\ \bibnamefont {Yates}}, \bibinfo {author}
  {\bibfnamefont {C.}~\bibnamefont {Angell}}, \bibinfo {author} {\bibfnamefont
  {M.}~\bibnamefont {Boswell}}, \bibinfo {author} {\bibfnamefont
  {B.}~\bibnamefont {Fallin}}, \bibinfo {author} {\bibfnamefont {C.~R.}\
  \bibnamefont {Howell}}, \bibinfo {author} {\bibfnamefont {A.}~\bibnamefont
  {Hutcheson}}, \bibinfo {author} {\bibfnamefont {H.~J.}\ \bibnamefont
  {Karwowski}}, \bibinfo {author} {\bibfnamefont {J.~H.}\ \bibnamefont
  {Kelley}}, \bibinfo {author} {\bibfnamefont {Y.}~\bibnamefont {Parpottas}},
  \bibinfo {author} {\bibfnamefont {A.~P.}\ \bibnamefont {Tonchev}},\ and\
  \bibinfo {author} {\bibfnamefont {W.}~\bibnamefont {Tornow}},\ }\href
  {https://doi.org/10.1103/PhysRevC.78.064303} {\bibfield  {journal} {\bibinfo
  {journal} {Phys. Rev. C}\ }\textbf {\bibinfo {volume} {78}},\ \bibinfo
  {pages} {064303} (\bibinfo {year} {2008})}\BibitemShut {NoStop}%
\bibitem [{\citenamefont {Albers}\ \emph {et~al.}(2013)\citenamefont {Albers},
  \citenamefont {Nomura}, \citenamefont {Warr}, \citenamefont {Blazhev},
  \citenamefont {Jolie}, \citenamefont {Mücher}, \citenamefont {Bastin},
  \citenamefont {Bauer}, \citenamefont {Bernards}, \citenamefont {Bettermann},
  \citenamefont {Bildstein}, \citenamefont {Butterworth}, \citenamefont
  {Cappellazzo}, \citenamefont {Cederkäll}, \citenamefont {Cline},
  \citenamefont {Darby}, \citenamefont {{Das Gupta}}, \citenamefont {Daugas},
  \citenamefont {Davinson}, \citenamefont {{De Witte}}, \citenamefont
  {Diriken}, \citenamefont {Filipescu}, \citenamefont {Fiori}, \citenamefont
  {Fransen}, \citenamefont {Gaffney}, \citenamefont {Georgiev}, \citenamefont
  {Gernhäuser}, \citenamefont {Hackstein}, \citenamefont {Heinze},
  \citenamefont {Hess}, \citenamefont {Huyse}, \citenamefont {Jenkins},
  \citenamefont {Konki}, \citenamefont {Kowalczyk}, \citenamefont {Kröll},
  \citenamefont {Krücken}, \citenamefont {Litzinger}, \citenamefont {Lutter},
  \citenamefont {Marginean}, \citenamefont {Mihai}, \citenamefont {Moschner},
  \citenamefont {Napiorkowski}, \citenamefont {{Nara Singh}}, \citenamefont
  {Nowak}, \citenamefont {Pakarinen}, \citenamefont {Pfeiffer}, \citenamefont
  {Radeck}, \citenamefont {Reiter}, \citenamefont {Rigby}, \citenamefont
  {Robledo}, \citenamefont {Rodríguez-Guzmán}, \citenamefont {Rudigier},
  \citenamefont {Scheck}, \citenamefont {Seidlitz}, \citenamefont {Siebeck},
  \citenamefont {Simpson}, \citenamefont {Thöle}, \citenamefont {Thomas},
  \citenamefont {{Van de Walle}}, \citenamefont {{Van Duppen}}, \citenamefont
  {Vermeulen}, \citenamefont {Voulot}, \citenamefont {Wadsworth}, \citenamefont
  {Wenander}, \citenamefont {Wimmer}, \citenamefont {Zell},\ and\ \citenamefont
  {Zielinska}}]{albers2013}%
  \BibitemOpen
  \bibfield  {author} {\bibinfo {author} {\bibfnamefont {M.}~\bibnamefont
  {Albers}}, \bibinfo {author} {\bibfnamefont {K.}~\bibnamefont {Nomura}},
  \bibinfo {author} {\bibfnamefont {N.}~\bibnamefont {Warr}}, \bibinfo {author}
  {\bibfnamefont {A.}~\bibnamefont {Blazhev}}, \bibinfo {author} {\bibfnamefont
  {J.}~\bibnamefont {Jolie}}, \bibinfo {author} {\bibfnamefont
  {D.}~\bibnamefont {Mücher}}, \bibinfo {author} {\bibfnamefont
  {B.}~\bibnamefont {Bastin}}, \bibinfo {author} {\bibfnamefont
  {C.}~\bibnamefont {Bauer}}, \bibinfo {author} {\bibfnamefont
  {C.}~\bibnamefont {Bernards}}, \bibinfo {author} {\bibfnamefont
  {L.}~\bibnamefont {Bettermann}}, \bibinfo {author} {\bibfnamefont
  {V.}~\bibnamefont {Bildstein}}, \bibinfo {author} {\bibfnamefont
  {J.}~\bibnamefont {Butterworth}}, \bibinfo {author} {\bibfnamefont
  {M.}~\bibnamefont {Cappellazzo}}, \bibinfo {author} {\bibfnamefont
  {J.}~\bibnamefont {Cederkäll}}, \bibinfo {author} {\bibfnamefont
  {D.}~\bibnamefont {Cline}}, \bibinfo {author} {\bibfnamefont
  {I.}~\bibnamefont {Darby}}, \bibinfo {author} {\bibfnamefont
  {S.}~\bibnamefont {{Das Gupta}}}, \bibinfo {author} {\bibfnamefont
  {J.}~\bibnamefont {Daugas}}, \bibinfo {author} {\bibfnamefont
  {T.}~\bibnamefont {Davinson}}, \bibinfo {author} {\bibfnamefont
  {H.}~\bibnamefont {{De Witte}}}, \bibinfo {author} {\bibfnamefont
  {J.}~\bibnamefont {Diriken}}, \bibinfo {author} {\bibfnamefont
  {D.}~\bibnamefont {Filipescu}}, \bibinfo {author} {\bibfnamefont
  {E.}~\bibnamefont {Fiori}}, \bibinfo {author} {\bibfnamefont
  {C.}~\bibnamefont {Fransen}}, \bibinfo {author} {\bibfnamefont
  {L.}~\bibnamefont {Gaffney}}, \bibinfo {author} {\bibfnamefont
  {G.}~\bibnamefont {Georgiev}}, \bibinfo {author} {\bibfnamefont
  {R.}~\bibnamefont {Gernhäuser}}, \bibinfo {author} {\bibfnamefont
  {M.}~\bibnamefont {Hackstein}}, \bibinfo {author} {\bibfnamefont
  {S.}~\bibnamefont {Heinze}}, \bibinfo {author} {\bibfnamefont
  {H.}~\bibnamefont {Hess}}, \bibinfo {author} {\bibfnamefont {M.}~\bibnamefont
  {Huyse}}, \bibinfo {author} {\bibfnamefont {D.}~\bibnamefont {Jenkins}},
  \bibinfo {author} {\bibfnamefont {J.}~\bibnamefont {Konki}}, \bibinfo
  {author} {\bibfnamefont {M.}~\bibnamefont {Kowalczyk}}, \bibinfo {author}
  {\bibfnamefont {T.}~\bibnamefont {Kröll}}, \bibinfo {author} {\bibfnamefont
  {R.}~\bibnamefont {Krücken}}, \bibinfo {author} {\bibfnamefont
  {J.}~\bibnamefont {Litzinger}}, \bibinfo {author} {\bibfnamefont
  {R.}~\bibnamefont {Lutter}}, \bibinfo {author} {\bibfnamefont
  {N.}~\bibnamefont {Marginean}}, \bibinfo {author} {\bibfnamefont
  {C.}~\bibnamefont {Mihai}}, \bibinfo {author} {\bibfnamefont
  {K.}~\bibnamefont {Moschner}}, \bibinfo {author} {\bibfnamefont
  {P.}~\bibnamefont {Napiorkowski}}, \bibinfo {author} {\bibfnamefont
  {B.}~\bibnamefont {{Nara Singh}}}, \bibinfo {author} {\bibfnamefont
  {K.}~\bibnamefont {Nowak}}, \bibinfo {author} {\bibfnamefont
  {J.}~\bibnamefont {Pakarinen}}, \bibinfo {author} {\bibfnamefont
  {M.}~\bibnamefont {Pfeiffer}}, \bibinfo {author} {\bibfnamefont
  {D.}~\bibnamefont {Radeck}}, \bibinfo {author} {\bibfnamefont
  {P.}~\bibnamefont {Reiter}}, \bibinfo {author} {\bibfnamefont
  {S.}~\bibnamefont {Rigby}}, \bibinfo {author} {\bibfnamefont
  {L.}~\bibnamefont {Robledo}}, \bibinfo {author} {\bibfnamefont
  {R.}~\bibnamefont {Rodríguez-Guzmán}}, \bibinfo {author} {\bibfnamefont
  {M.}~\bibnamefont {Rudigier}}, \bibinfo {author} {\bibfnamefont
  {M.}~\bibnamefont {Scheck}}, \bibinfo {author} {\bibfnamefont
  {M.}~\bibnamefont {Seidlitz}}, \bibinfo {author} {\bibfnamefont
  {B.}~\bibnamefont {Siebeck}}, \bibinfo {author} {\bibfnamefont
  {G.}~\bibnamefont {Simpson}}, \bibinfo {author} {\bibfnamefont
  {P.}~\bibnamefont {Thöle}}, \bibinfo {author} {\bibfnamefont
  {T.}~\bibnamefont {Thomas}}, \bibinfo {author} {\bibfnamefont
  {J.}~\bibnamefont {{Van de Walle}}}, \bibinfo {author} {\bibfnamefont
  {P.}~\bibnamefont {{Van Duppen}}}, \bibinfo {author} {\bibfnamefont
  {M.}~\bibnamefont {Vermeulen}}, \bibinfo {author} {\bibfnamefont
  {D.}~\bibnamefont {Voulot}}, \bibinfo {author} {\bibfnamefont
  {R.}~\bibnamefont {Wadsworth}}, \bibinfo {author} {\bibfnamefont
  {F.}~\bibnamefont {Wenander}}, \bibinfo {author} {\bibfnamefont
  {K.}~\bibnamefont {Wimmer}}, \bibinfo {author} {\bibfnamefont
  {K.}~\bibnamefont {Zell}},\ and\ \bibinfo {author} {\bibfnamefont
  {M.}~\bibnamefont {Zielinska}},\ }\href
  {https://doi.org/https://doi.org/10.1016/j.nuclphysa.2013.01.013} {\bibfield
  {journal} {\bibinfo  {journal} {Nucl. Phys. A}\ }\textbf {\bibinfo {volume}
  {899}},\ \bibinfo {pages} {1} (\bibinfo {year} {2013})}\BibitemShut {NoStop}%
\bibitem [{\citenamefont {Kremer}\ \emph {et~al.}(2016)\citenamefont {Kremer},
  \citenamefont {Aslanidou}, \citenamefont {Bassauer}, \citenamefont {Hilcker},
  \citenamefont {Krugmann}, \citenamefont {von Neumann-Cosel}, \citenamefont
  {Otsuka}, \citenamefont {Pietralla}, \citenamefont {Ponomarev}, \citenamefont
  {Shimizu}, \citenamefont {Singer}, \citenamefont {Steinhilber}, \citenamefont
  {Togashi}, \citenamefont {Tsunoda}, \citenamefont {Werner},\ and\
  \citenamefont {Zweidinger}}]{kremer2016}%
  \BibitemOpen
  \bibfield  {author} {\bibinfo {author} {\bibfnamefont {C.}~\bibnamefont
  {Kremer}}, \bibinfo {author} {\bibfnamefont {S.}~\bibnamefont {Aslanidou}},
  \bibinfo {author} {\bibfnamefont {S.}~\bibnamefont {Bassauer}}, \bibinfo
  {author} {\bibfnamefont {M.}~\bibnamefont {Hilcker}}, \bibinfo {author}
  {\bibfnamefont {A.}~\bibnamefont {Krugmann}}, \bibinfo {author}
  {\bibfnamefont {P.}~\bibnamefont {von Neumann-Cosel}}, \bibinfo {author}
  {\bibfnamefont {T.}~\bibnamefont {Otsuka}}, \bibinfo {author} {\bibfnamefont
  {N.}~\bibnamefont {Pietralla}}, \bibinfo {author} {\bibfnamefont {V.~Y.}\
  \bibnamefont {Ponomarev}}, \bibinfo {author} {\bibfnamefont {N.}~\bibnamefont
  {Shimizu}}, \bibinfo {author} {\bibfnamefont {M.}~\bibnamefont {Singer}},
  \bibinfo {author} {\bibfnamefont {G.}~\bibnamefont {Steinhilber}}, \bibinfo
  {author} {\bibfnamefont {T.}~\bibnamefont {Togashi}}, \bibinfo {author}
  {\bibfnamefont {Y.}~\bibnamefont {Tsunoda}}, \bibinfo {author} {\bibfnamefont
  {V.}~\bibnamefont {Werner}},\ and\ \bibinfo {author} {\bibfnamefont
  {M.}~\bibnamefont {Zweidinger}},\ }\href
  {https://doi.org/10.1103/PhysRevLett.117.172503} {\bibfield  {journal}
  {\bibinfo  {journal} {Phys. Rev. Lett.}\ }\textbf {\bibinfo {volume} {117}},\
  \bibinfo {pages} {172503} (\bibinfo {year} {2016})}\BibitemShut {NoStop}%
\bibitem [{\citenamefont {Iachello}\ and\ \citenamefont {Arima}(1987)}]{IBM}%
  \BibitemOpen
  \bibfield  {author} {\bibinfo {author} {\bibfnamefont {F.}~\bibnamefont
  {Iachello}}\ and\ \bibinfo {author} {\bibfnamefont {A.}~\bibnamefont
  {Arima}},\ }\href@noop {} {\emph {\bibinfo {title} {The interacting boson
  model}}}\ (\bibinfo  {publisher} {Cambridge University Press, Cambridge},\
  \bibinfo {year} {1987})\BibitemShut {NoStop}%
\bibitem [{\citenamefont {Thomas}\ \emph {et~al.}(2013)\citenamefont {Thomas},
  \citenamefont {Nomura}, \citenamefont {Werner}, \citenamefont {Ahn},
  \citenamefont {Cooper}, \citenamefont {Duckwitz}, \citenamefont {Hinton},
  \citenamefont {Ilie}, \citenamefont {Jolie}, \citenamefont {Petkov},\ and\
  \citenamefont {Radeck}}]{thomas2013}%
  \BibitemOpen
  \bibfield  {author} {\bibinfo {author} {\bibfnamefont {T.}~\bibnamefont
  {Thomas}}, \bibinfo {author} {\bibfnamefont {K.}~\bibnamefont {Nomura}},
  \bibinfo {author} {\bibfnamefont {V.}~\bibnamefont {Werner}}, \bibinfo
  {author} {\bibfnamefont {T.}~\bibnamefont {Ahn}}, \bibinfo {author}
  {\bibfnamefont {N.}~\bibnamefont {Cooper}}, \bibinfo {author} {\bibfnamefont
  {H.}~\bibnamefont {Duckwitz}}, \bibinfo {author} {\bibfnamefont
  {M.}~\bibnamefont {Hinton}}, \bibinfo {author} {\bibfnamefont
  {G.}~\bibnamefont {Ilie}}, \bibinfo {author} {\bibfnamefont {J.}~\bibnamefont
  {Jolie}}, \bibinfo {author} {\bibfnamefont {P.}~\bibnamefont {Petkov}},\ and\
  \bibinfo {author} {\bibfnamefont {D.}~\bibnamefont {Radeck}},\ }\href
  {https://doi.org/10.1103/PhysRevC.88.044305} {\bibfield  {journal} {\bibinfo
  {journal} {Phys. Rev. C}\ }\textbf {\bibinfo {volume} {88}},\ \bibinfo
  {pages} {044305} (\bibinfo {year} {2013})}\BibitemShut {NoStop}%
\bibitem [{\citenamefont {Togashi}\ \emph {et~al.}(2016)\citenamefont
  {Togashi}, \citenamefont {Tsunoda}, \citenamefont {Otsuka},\ and\
  \citenamefont {Shimizu}}]{togashi2016}%
  \BibitemOpen
  \bibfield  {author} {\bibinfo {author} {\bibfnamefont {T.}~\bibnamefont
  {Togashi}}, \bibinfo {author} {\bibfnamefont {Y.}~\bibnamefont {Tsunoda}},
  \bibinfo {author} {\bibfnamefont {T.}~\bibnamefont {Otsuka}},\ and\ \bibinfo
  {author} {\bibfnamefont {N.}~\bibnamefont {Shimizu}},\ }\href
  {https://doi.org/10.1103/PhysRevLett.117.172502} {\bibfield  {journal}
  {\bibinfo  {journal} {Phys. Rev. Lett.}\ }\textbf {\bibinfo {volume} {117}},\
  \bibinfo {pages} {172502} (\bibinfo {year} {2016})}\BibitemShut {NoStop}%
\end{thebibliography}%

\end{document}